\documentclass[fleqn,usenatbib]{mnras}

\usepackage{newtxtext,newtxmath}
\usepackage[T1]{fontenc}

\usepackage{float}
\usepackage{fleqn}
\usepackage{graphicx}
\usepackage{color}
\usepackage{hyperref}
\usepackage{epstopdf}
\usepackage{url}
\usepackage{subfloat}
\usepackage{caption}
\usepackage{gensymb}
\usepackage{multirow}
\usepackage{makecell}
\usepackage{esvect}
\usepackage{xcolor}
\usepackage{lscape}
\usepackage{pdflscape}
\usepackage{ae,aecompl}
\usepackage{wrapfig}
\usepackage{enumitem}

\usepackage{capt-of}
\usepackage{mhchem}
\def\apj{ApJ}

\def\araa{ARA\&A}
\def\aap{A\&A}
\def\apjl{ApJL}
\def\apjs{ApJS}

\DeclareRobustCommand{\VAN}[3]{#2}
\let\VANthebibliography\thebibliography
\def\thebibliography{\DeclareRobustCommand{\VAN}[3]{##3}\VANthebibliography}

\usepackage{graphicx}	% Including figure files
\usepackage{amsmath}	% Advanced maths commands
\title[The dark days are overcast]{The dark days are overcast: Iron-bearing clouds on HD~209458~b and WASP-43~b can explain low dayside albedos}

\author[K. L. Chubb and D. Samra et al.]{
	K. L. Chubb$^{1,2,*}$\thanks{E-mail: katy.chubb@bristol.ac.uk}
	and D. Samra,$^{3,*}$\thanks{E-mail: dominicsamra.uk@gmail.com}
	Ch. Helling,$^{3,4}$
	L. Carone,$^{3}$
	and D. M. Stam.$^{5}$
	\\
 $^{1}$University of Bristol, School of Physics, Tyndall Avenue, Bristol, BS8 1TL, UK \\
	$^{2}$Centre for Exoplanet Science, University of St Andrews, North Haugh, St Andrews, KY16 9SS, UK\\
	$^{3}$Space Research Institute, Austrian Academy of Sciences, Schmiedlstr. 6, A-8042, Graz, Austria \\
	$^{4}$Fakultät für Mathematik, Physik und Geodäsie, TU Graz, Petersgasse 16, Graz, A-8010, Austria \\
		$^{5}$Leiden Observatory, Niels Bohrweg 2, 2333 CA Leiden, Netherlands \\
		$^{*}$: these authors contributed equally to this work
}

\date{Accepted XXX. Received YYY; in original form ZZZ}

\pubyear{2023}

\begin{document}
	\label{firstpage}
	\pagerange{\pageref{firstpage}--\pageref{lastpage}}
	\maketitle
	
	\begin{abstract}
		%	This is a simple template for authors to write new MNRAS papers.
		%	The abstract should briefly describe the aims, methods, and main results of the paper.
		%		It should be a single paragraph not more than 250 words (200 words for Letters).
		%		No references should appear in the abstract. 
  We simulate the geometric albedo spectra of hot Jupiter exoplanets HD~209458~b and WASP-43~b, based on global climate model (GCMs) post-processed with kinetic cloud models. We predict WASP-43~b to be cloudy throughout its dayside, while HD~209458~b has a clear upper atmosphere around the hot sub-solar point, largely due to the inclusion of strong optical absorbers TiO and VO in the GCM for the latter causes a temperature inversion. In both cases our models find low geometric albedos - 0.026 for WASP-43b and 0.028 for HD~209458~b when averaged over the CHEOPS bandpass of $\sim$0.35~-~1.1~$\mu$m - indicating dark daysides, similar to the low albedos measured by observations. We demonstrate the strong impact of clouds that contain Fe-bearing species on the modelled geometric albedos; without Fe-bearing species forming in the clouds, the albedos of both planets would be much higher (0.518 for WASP-43~b, 1.37 for HD~209458~b). We conclude that a cloudy upper or mid-to-lower atmosphere that contains strongly absorbing Fe-bearing aerosol species, is an alternative to a cloud-free atmosphere in explaining the low dayside albedos of hot Jupiter atmospheres such as HD~209458~b and WASP-43~b.

	\end{abstract}
	
	
%-----------------------------------------------------------------------------
\section{Introduction}\label{sec:introduction}

The geometric albedo of a transiting exoplanet can be inferred by observing the star and the planet just before (or after) and during the planet's secondary eclipse. It is a measure of how reflective a planet is to incoming stellar radiation. Although this albedo is wavelength dependent, it is typically measured as an average over the bandpass of the instrument used for observation (see, for example, \cite{krenn_geometric_2023}).
There have been several observations to infer the geometric albedos of hot giant transiting exoplanets, and with a couple of exceptions, the observed albedos are generally low, indicating dark, poorly reflective daysides. For example, the geometric albedos of a population of around 20 hot gaseous exoplanets have been measured by studies such as \cite{15AnDeMo} and \cite{15EsDeJa}, with the finding that the majority have albedos typically less than 0.15 in the Kepler bandpass (0.42~-~0.91~$\mu$m). \cite{krenn_geometric_2023} confirm these findings by deriving a geometric albedo, $A_g$, of 0.076~$\pm$~0.016 for HD~189733~b from data of the CHaracterising ExOPlanet Satellite (CHEOPS) ($\sim$0.35~-~1.1~$\mu$m). This value is consistent with that previously inferred by \cite{evans_deep_2013} from their HST STIS secondary eclipse measurements of HD~189733~b; they found $A_g$~=~0.4~$\pm$~0.12 for the 0.29~-~0.45~$\mu$m region and $A_g$~$\leq$~0.12 for the 0.45~-~0.57~$\mu$m region. The decrease of this albedo towards longer wavelengths was interpreted as being due to absorption by sodium.
Some slightly higher values inferred from observations include $A_g$~=~0.2~$\pm$~0.04 for WASP-19b using TESS (0.6~-~1.0~$\mu$m)~\citep{24KoPa}, 0.225~$\pm$~0.004 for HAT-P-7b~\citep{13HeDe}, and $A_g$~$<$~0.33 for WASP-80~b using the WFC3 instrument onboard HST~\citep{23JaDeGa}.
For Kepler-7b $A_g = 0.25^{+0.01}_{-0.02}$~\citep{21HeMoKi}, and for WASP-178b 0.1~$<$~$A_g$~$<$~0.35 using CHEOPS~\citep{23PaScSi}. A noticeable exception with a much higher measured albedo is LTT~9779~b with $A_g$~=~0.8~$^{+0.10}_{-0.17}$ ~\citep{23HoJePa}. 
For comparison, the geometric albedo of Jupiter is around 0.44 from 0.4 - 17 $\mu{\rm m}$~\citep{18LiJiWe}.

HD~209458~b \citep[discovered by][]{Charbonneau2000} and WASP-43~b \citep[discovered by][]{Hellier2011} are both hot Jupiters with similar equilibrium temperatures of approximately 1400~K \citep{14BlHaMa,Zellem2014}. However, they orbit very different host stars: WASP-43~b orbits a K-type star every 0.813 days \citep{Gillon2012}, whilst HD~209458~b orbits a G0V star every 3.52 days \citep{Torres2008}. The planets also differ substantially in surface gravity with HD~209458~b having $\log_{10}(g\,{\rm [cms^{-2}]})=2.9$, whereas WASP-43~b has a surface gravity closer to 4, as $\log_{10}(g\,{\rm [cms^{-2}]})=3.7$ \citep{helling_cloud_2021}. HD~209458~b's and WASP-43~b's atmospheres have both been well studied theoretically as well as observationally. \cite{helling_mineral_2020} and \cite{22RoKaBa} both predict WASP-43~b to be cloudy based on global climate models (GCMs) combined with kinetic cloud models, with \cite{22RoKaBa} finding their cloudy models fit observed HST spectra better than clear ones. \cite{20Barstow} find evidence for clouds in the terminator region of HD~209458~b by analysing transmission spectra. \cite{23TaPa} analyse eclipse spectra of both WASP-43~b and HD~209458~b taking scattering properties of different materials used to form clouds into account. They find no evidence for dayside clouds on HD~209458~b but do expect dayside clouds on WASP-43~b.  More recently, WASP-43~b has also been the subject of JWST MIRI phase curve observations by \cite{TaylorBell_ERS2024} (as part of the JWST ERS program for transiting exoplanets - program ID: \#1366), which has also shown strong evidence for clouds on the planet's nightside.

\cite{21FrMaSt} infer a low geometric albedo, namely $A_g$~$<$~0.06, for WASP-43b based on secondary eclipse observations using the Hubble Space Telescope's (HST) WFC3/UVIS instrument, observed between 0.2~-~0.8~$\mu$m. 
They compute model atmospheres that include various cloud species, and deduce a lack of reflective cloud species on the dayside of the planet in the pressure layers probed by eclipse observations (i.e.\ P~$\leq$~1~bar), which yields WASP-43b's dark dayside. \cite{22BlKaPi} also find a low geometric albedo for WASP-43b, namely $A_g$~$<$~0.16, based on TESS observations in the optical, and \cite{scandariato_phase_2022} find $A_g$~$<$~0.087 using CHEOPS, TESS, and HST. The geometric albedo of HD~209458~b has also been observed to be low; using CHEOPS it was measured as 0.096~$\pm$~0.016 by \cite{brandeker_cheops_2022}. The authors find this low value of $A_g$ to be consistent with a cloud-free atmosphere.

The presence and types of clouds in exoplanet atmospheres are an important component of exploring why these measured albedos are so low. The cloud materials thought to be important in hot giant exoplanet atmospheres can be predicted using global climate models (GCMs) post-processed with kinetic cloud models \citep[e.g.][]{01AcMa,08HeAcAl,20GaThLe,22RoBaGa,Helling_grid2023}. These predictions can be confirmed via observations, such as those of the 5~-~13~$\mu$m region via JWST's MIRI LRS instrument, which allow for observations of vibrational mode absorption features of cloud particles~\citep{23GrLeWa,24DyMiDe}. Recent transmission observations by JWST, which probe the terminator regions of exoplanets, have yielded evidence for silicate cloud species in the atmospheres of hot Jupiter WASP-17b by \cite{23GrLeWa} (in the form of quartz, SiO$_2$[s]) and warm Neptune WASP-107b by \cite{24DyMiDe} (in the form of MgSiO$_3$[s], SiO$_2$[s] and SiO[s]). Albedo models and reflected light observations offer a complimentary technique which can further explore the nature of cloud species on the dayside of such exoplanets.  

It is well known that the types of cloud particles in the atmospheres of these types of planets will influence their observed primary eclipse (transmission) and/or secondary eclipse (emission) spectra \citep{15WaSi,17PiMa,20MiOrCh.arcis,23MaItAl,23TaPa}, and their albedos~\citep{21FrMaSt,23GoMaLe,23ChStHe}, through their absorption and scattering properties. Efforts are being made to progress knowledge on the properties of a large number of species predicted to be important in hot gaseous exoplanet atmospheres; investigating not only the distribution, size, and composition of clouds, but also the impact of fractal/irregularly versus spherically particle shapes on observed spectra~\citep{19AdGaPa,21DoMiTa,23LoWaLe}, and of the porosity of the particles~\citep{07MiWaKo,20SaHeMi,20OhOkTa}.

This paper is structured as follows. In Sect.~\ref{sec:atm_models}, we first present the methods (Sect.~\ref{sec:atm_methods}) and results (Sect.~\ref{sec:atm_results}) of atmospheric models of WASP-43~b and HD~209458~b which post-process GCM outputs with microphysical cloud formation models. We then use these results to compute geometric albedos of the same planets in Sect.~\ref{sec:albedo_models}, with methods outlined in Sect.~\ref{sec:albedo_methods} and the results in Sect.~\ref{sec:albedo_results}. We give our conclusions in Sect.~\ref{sec:conclusion}.

%-----------------------------------------------------------------------------
\section{Atmospheric modelling}\label{sec:atm_models}

\subsection{Methods}\label{sec:atm_methods}
	
We use a hierarchical modelling approach to cloud formation in the atmospheres of WASP-43~b and HD~209453~b, as has been used in \cite{helling_cloud_2021}. We use 1D local temperatures, pressures and vertical wind speed profiles ($T,p,v_{z}$-profiles) from the cloud-free general circulation model (GCM) of ExpeRT/MITgcm~\citep{20CaBaMo,21BaDeCa} as input for our kinetic cloud formation model~\citep{W&H2003,W&H2004,H&W2006,HWT2008}. The ExpeRT/MITgcm is run for a simulation time of 1500 Earth days  with solar metallicity, and the temperature, pressure and vertical velocity profiles are computed using the time-averages of their profiles over the last 100 days~\citep{schneider_exploring_2022}. 

In the case of WASP~43b, the ExpeRT/MITgcm used is the same as in \cite{TaylorBell_ERS2024}. 
For both planets, solar abundances are assumed for the GCM and for post-processing of the clouds, in the later case assuming \cite{Asplund2009} abundances. 
TiO and VO were taken out of the GCM computations for WASP-43~b in order to prevent a temperature inversion in the upper atmosphere, which \cite{TaylorBell_ERS2024} found to be a better match to observed dayside emission and photometry. Previous GCM studies such as \cite{15KaShFo} have also taken out TiO and VO from their WASP-43~b models due to observations and analyses suggesting there is no temperature inversion~\citep[e.g.][]{14BlHaMa,14LiKnWo}. Suggestions of a temperature inversion in the atmosphere of HD~209458~b have previously been deduced by studies such as \cite{07BuHuBu,09MaSe}, based on HST and Spitzer measurements available at the time. Slightly more recent observations and analyses conclude no evidence for a temperature inversion~\citep{14DiStBe,16LiStBe}, with more observations on the way from JWST to ideally provide more conclusive evidence. Similar GCMs of HD~209458~b to \cite{schneider_exploring_2022} such as \cite{21LePaHa} also keep TiO and VO in, and therefore model a temperature inversion. As pointed out by \cite{schneider_exploring_2022} the dayside temperature
of HD~209458~b is very close to the TiO and VO condensation curves~\citep[see e.g.][]{08DeViLe}, which indicate their thermal stability, which is why it is not entirely obvious whether the species should be left in or out of the gas phase. Given this uncertainty, we chose to explore the theoretical impact of allowing TiO and VO and therefore a thermal inversion in the atmosphere of HD~209458~b, but not WASP-43~b.
This causes a difference in the outcomes of our models, allowing us to explore how subtle differences in the atmospheric chemistry of two planets with otherwise similar effective temperatures can impact observed planetary properties such as albedo. The other main difference between the two planets is that the atmosphere of HD-209458~b is more inflated than that of WASP-43~b~\citep{schneider_exploring_2022}.
In total, 120 1D profiles were extracted from the GCM, spaced at 15 degrees in longitude, and every 22.5 degrees in latitude. An additional `polar latitude' of $86\degree$ was also included. The full spatial resolution is used for the mapping of the model clouds, but only a select number of regions are used for computing the dayside albedo (see Sect.~\ref{sec:albedo_methods}).
 
With the approach of using the output of GCMs as input to our kinetic cloud model, we are able to consider a set of 16 condensates (\ce{TiO2}[s], \ce{SiO}[s], \ce{SiO2}[s], \ce{MgO}[s], \ce{MgSiO3}[s], \ce{Mg2SiO4}[s], \ce{Fe2SiO4}[s], \ce{Fe}[s], \ce{Fe2O3}[s], \ce{FeO}[s], \ce{FeS}[s], \ce{KCl}[s], \ce{NaCl}[s], \ce{Al2O3}[s], \ce{CaSiO3}[s], \ce{CaTiO3}[s]) simultaneously, which consistently condense as mixed-composition (heterogeneous) cloud particles.
\textit{Condensation} is considered kinetically through a set of 132 reactions. To start the process of cloud formation, cloud condensation nuclei are formed through \textit{nucleation}, treated through modified classical nucleation theory \citep{ElsieLeeMCNT2018}, for 4 species 
(\ce{TiO2}[s], \ce{SiO}[s], \ce{KCl}[s], \ce{NaCl}[s]). \textit{Gravitational settling} of cloud particles occurs according to the size of the cloud particles assuming equilibrium drift velocities \citep{W&H2003}.
 
Cloud formation is coupled to the gas-phase through \textit{element depletion}, the gas-phase is then solved for using the chemical equilibrium code GGChem \citep{Woitke2018}. \textit{Element replenishment} is treated using a relaxation timescale approach, local elemental abundances return towards the initial elemental abundances considering mixing from the deep atmosphere on a given timescale (see Eqs.~6 and~7 in \cite{W&H2004}). The mixing timescale is computed from the vertical velocity component of the GCM using the standard-deviation of vertically adjacent cells from the GCM solution, as described in Appendix~B in \cite{Helling_grid2023}. 
Following \cite{samra_clouds_2023}, we use this mixing timescale increased by a factor of 100. This has been found to more appropriately reflect the efficiency of mixing and better reproduce observations of cloud optical depth for WASP~96b and WASP~39b \citep{samra_clouds_2023,Espinoza_ERS}. Furthermore, GCM studies find a similar relationship between the root-mean squared velocity times the pressure scale height, and the eddy diffusion coefficient for ($K_{zz}$) for passive tracer hazes and cloud particles \citep{Steinrueck_3DHazes2021,Parmentier2013}.

In order to compute the geometric albedo $A_g$ of a cloudy planet, a particle size distribution is required. The kinetic cloud formation model used treats the cloud particle population though the series of moments ($\rho L_{j} = \int_{V_l}^{\infty} V^{j/3}f(V){\rm d}V$), so the size distribution ($f(V)$) is not explicitly known. However a particle size distribution that is consistent with the moments can be re-constructed, providing a functional form with suitably few parameters \citep{HWT2008}.
In this work we derive the parameters for a log-normal distribution motivated by its common use in modelling cloud particle size distributions in exoplanet atmosphere models~\citep[e.g.]{01AcMa,15WaSi,19MaLi,RomanRauscher2019,21MuBaMa}. The derivation of the parameters can be found in Appendix~\ref{sec:log-normal}.

%-----------------------------------------------------------------------------
\subsection{Results}\label{sec:atm_results}

%%%%%%%%%%%%%%%%%%%%%%%%%%%%%%%%%%%%%%%%%%%%%%%%%%%%%%%%%%%%%%%%%%%%%%%%%%%%%%

%%%%%%%%%%%%%%%%%%%%%%%%%%%%%%%%%%%%%%%%%%%%%%%%%%%%%%%%%%%%%%%%%%%%%%%%%%%%%%

\begin{figure*}
	\centering
	\includegraphics[width=0.9\textwidth]{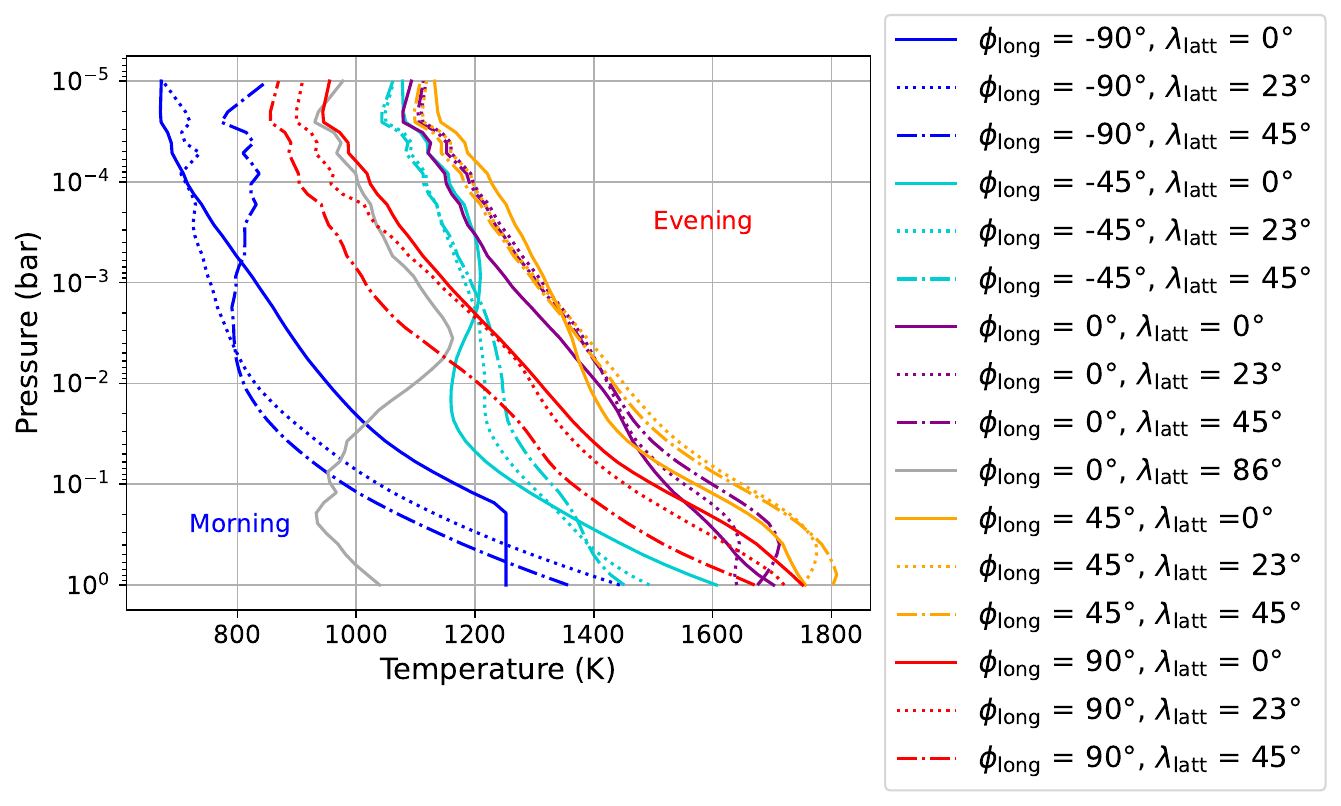}
 \includegraphics[width=0.6\textwidth]{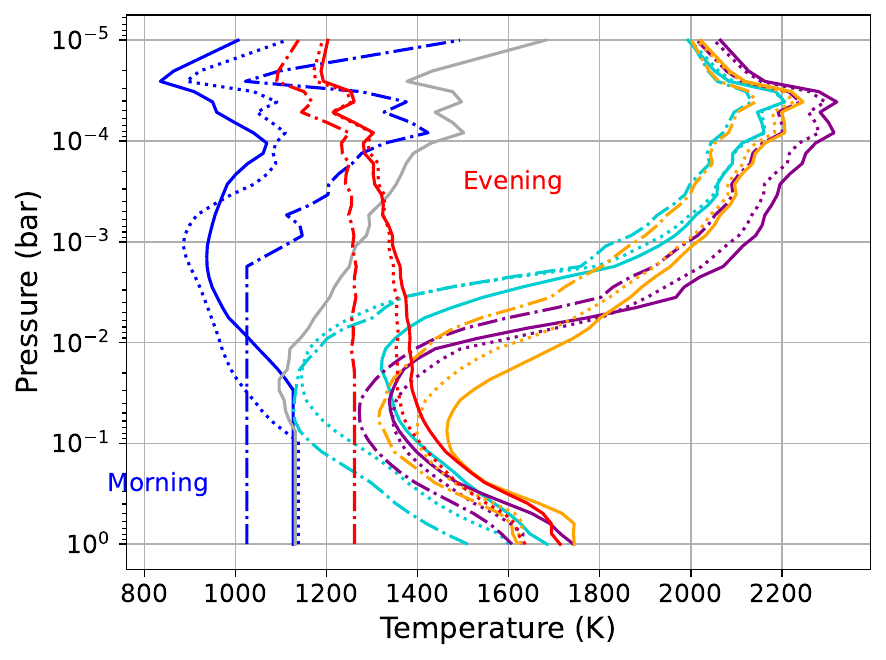}
 \caption{Pressure-temperature profiles for the various longitude-latitude regions across WASP-43~b (upper) and HD~209458~b (lower).}\label{fig:PT_16_regions_W43b}
\end{figure*}

\begin{figure}
    \centering    
    \includegraphics[width=0.49\textwidth]{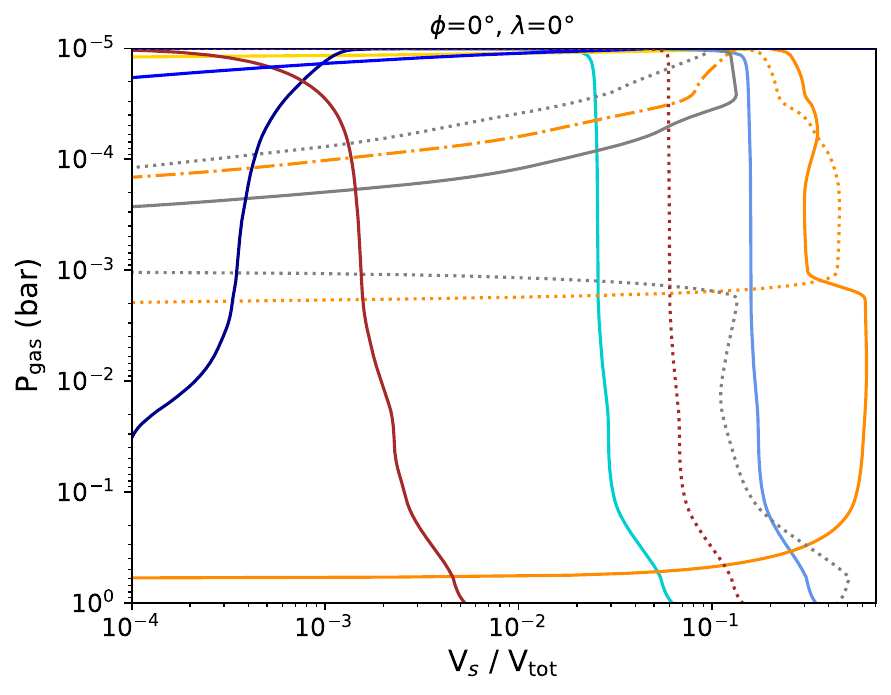}
    \includegraphics[width=0.49\textwidth]{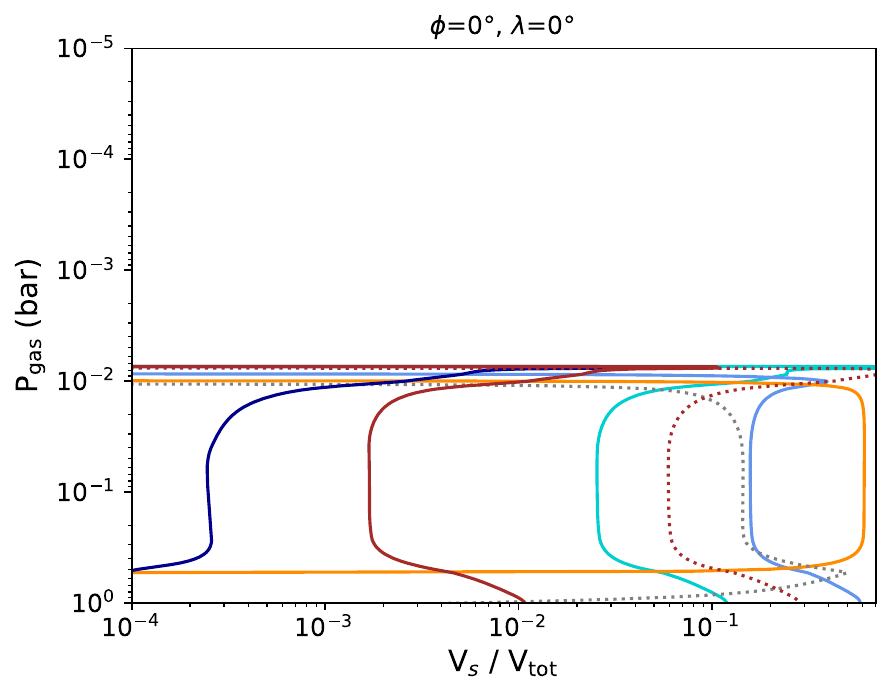}
    \includegraphics[width=0.49\textwidth]{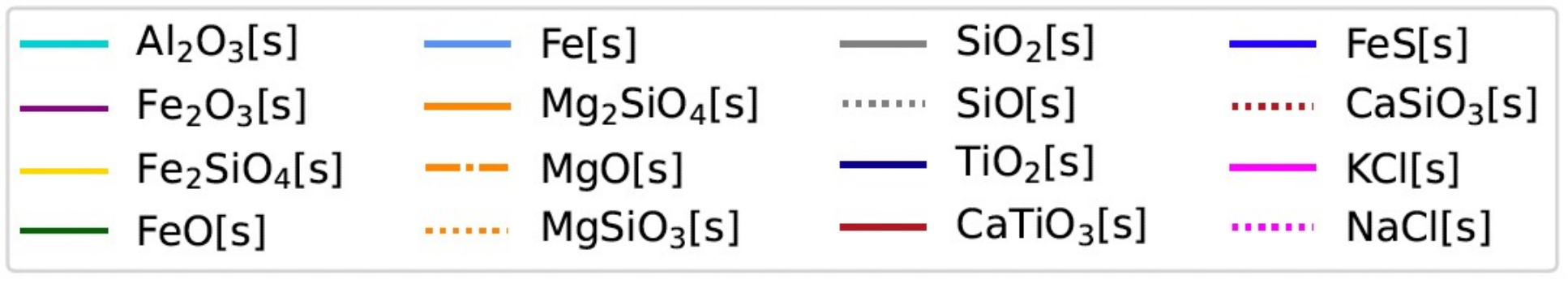}
    \caption{Composition $V_{s}$ of the cloud particles used in this work as fractions of the total cloud particle volume $V_{\rm tot}$, for the substellar regions on WASP-43~b (upper) and HD~209458~b (lower).}
    \label{fig:mat_comp_substellar}
\end{figure}

There is quite a difference between the pressure-temperature profiles of the two planets (seen in Fig.~\ref{fig:PT_16_regions_W43b}), with the region around the sub-solar point of HD~209458~b exhibiting a temperature inversion, due to the inclusion of TiO and VO in the GCM of HD~209458~b but not WASP-43~b (see discussion at the start of Sect.~\ref{sec:atm_methods}). 
As a consequence of these high temperatures, clouds do not form in the upper atmosphere (at pressures below $\sim$0.01~bar) around this region of HD~209458~b. In contrast, our models predict WASP-43b to be relatively cloudy throughout the dayside and down to low pressures, due to the relatively cooler upper-atmosphere temperatures. Fig.~\ref{fig:mat_comp_substellar} shows the composition $V_{s}$ of the cloud particles used in this work as fractions of the total cloud particle volume $V_{\rm tot}$, for the substellar regions on WASP-43~b (upper) and HD~209458~b (lower). The cloud particles are substantially mixed at all pressures, with the most abundant cloud species not exceeding 60\%. For the bulk of the cloud deck at the substellar point on both planets, the dominant cloud condensate is either \ce{Mg2SiO4}[s] or \ce{MgSiO3}[s], and \ce{Fe}[s] (the most abundant iron-bearing condensate) does not exceed around 20\%, except at the base of the cloud deck where silicate material evaporates.

%%%%%%%%%%%%%%%%%%%%%%%%%%%%%%%%%%%%%%%%%%%%%%%%%%%%%%%%%%%%%%%%%%%%%%%%%%%%%%
%%%%%%%%%%%%%%%%%%%%%%%%%%%%%%%%%%%%%%%%%%%%%%%%%%%%%%%%%%%%%%%%%%%%%%%%%%%%%%

\begin{figure}
    \centering    
        \includegraphics[width=0.49\textwidth]{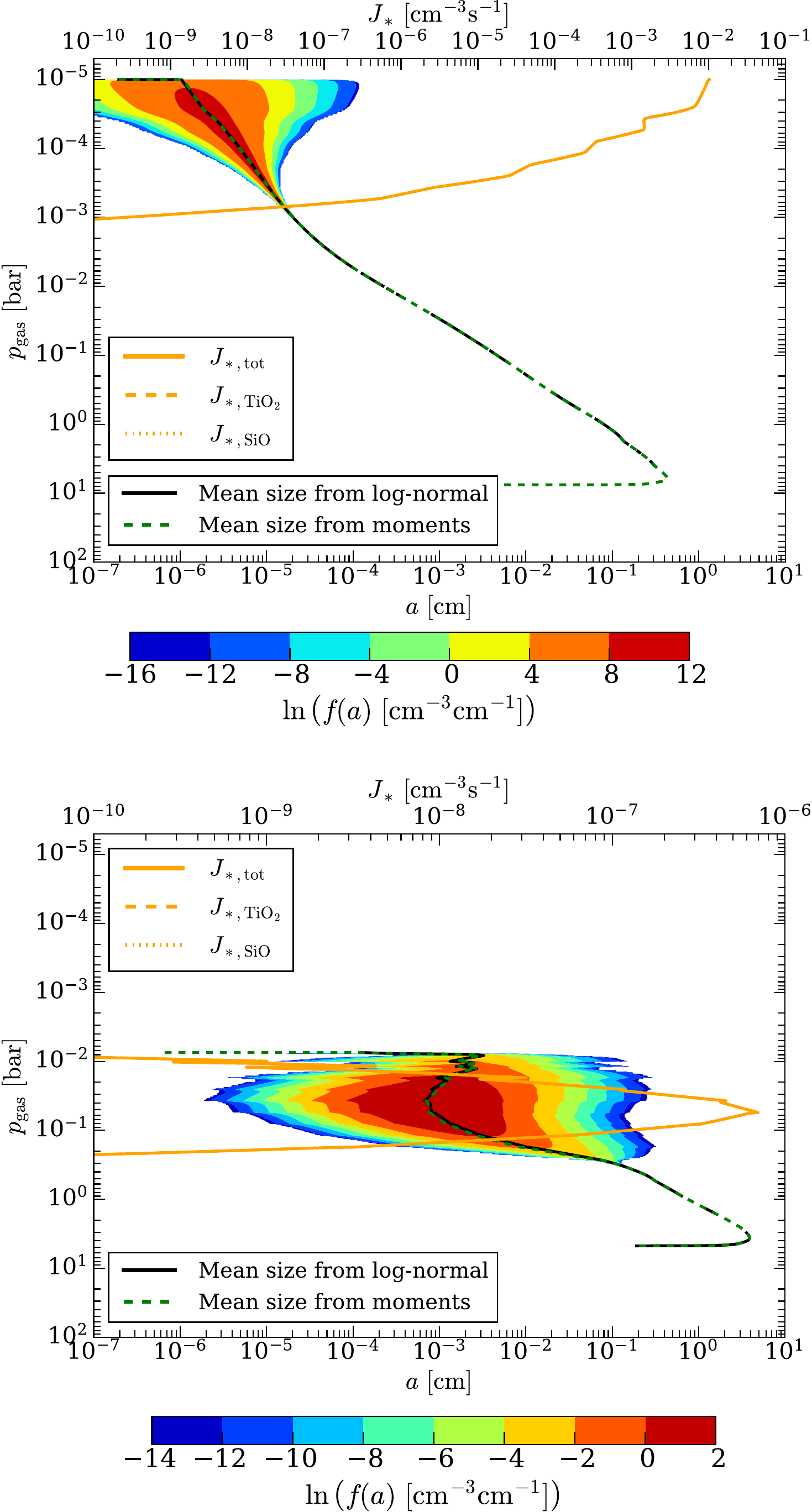}
    \caption{Log-normal cloud particle size distribution for the substellar regions on  WASP-43~b (upper) and HD209458~b (lower). The cloud particle size distribution is shown as filled contours. The average particle size from the log-normal distribution is shown as the black solid line. The average particle size derived from the moments is shown as the dashed green line. The orange lines indicate the nucleation rates.}
    \label{fig:nucleation_substellar}
\end{figure}

Figure~\ref{fig:nucleation_substellar} shows the derived log-normal size distributions for cloud particles at individual pressure levels throughout the atmosphere for both planets. In the case that the distribution becomes monodisperse (i.e.\ no valid solution for the log-normal parameters exists), only the average particle size is shown. The average particle size derived from the moments, i.e.\ $\langle a \rangle = (3/4\pi)^{1/3}L_1/L0$, is shown in dashed green, with a good agreement between the log-normal average particle size at all pressures.
Also shown are the nucleation rates, showing the broadness of the size distribution in regions with efficient nucleation: there is a significant population of small cloud particles resulting from `recently nucleated' particles that have not yet undergone substantial particle growth. In the deeper parts of the cloud deck, the cloud particle size distribution is close to monodisperse, because there nucleation has ceased to be an efficient process. Hence, the only processes that affect the particle size distribution are bulk growth/evaporation, or settling. As bulk growth is very efficient, the particles quickly grow to the maximum size for their pressure level to then rain out rapidly. Thus the only stable particle size distribution possible is close to monodisperse around this - large - particle size. See 
%sect blah in
\cite{W&H2003} for a discussion of the maximum particle size when considering equilibrium velocity rain-out of cloud particles. 

%%%%%%%%%%%%%%%%%%%%%%%%%%%%%%%%%%%%%%%%%%%%%%%%%%%%%%%%%%%%%%%%%%%%%%%%%%%%%%
%%%%%%%%%%%%%%%%%%%%%%%%%%%%%%%%%%%%%%%%%%%%%%%%%%%%%%%%%%%%%%%%%%%%%%%%%%%%%%

\begin{figure*}
      \centering
      \includegraphics[width=0.49\textwidth]{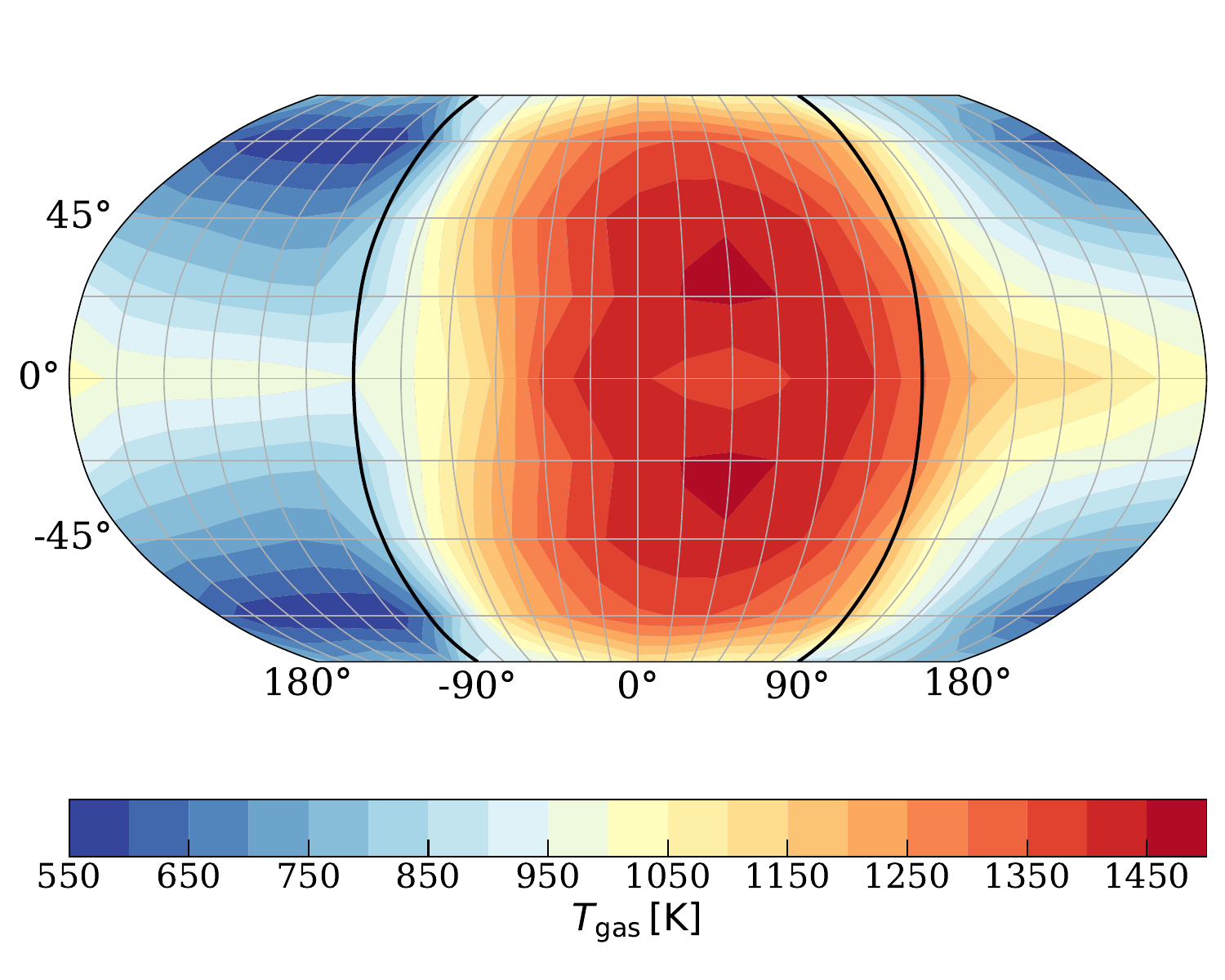}
      \includegraphics[width=0.49\textwidth]{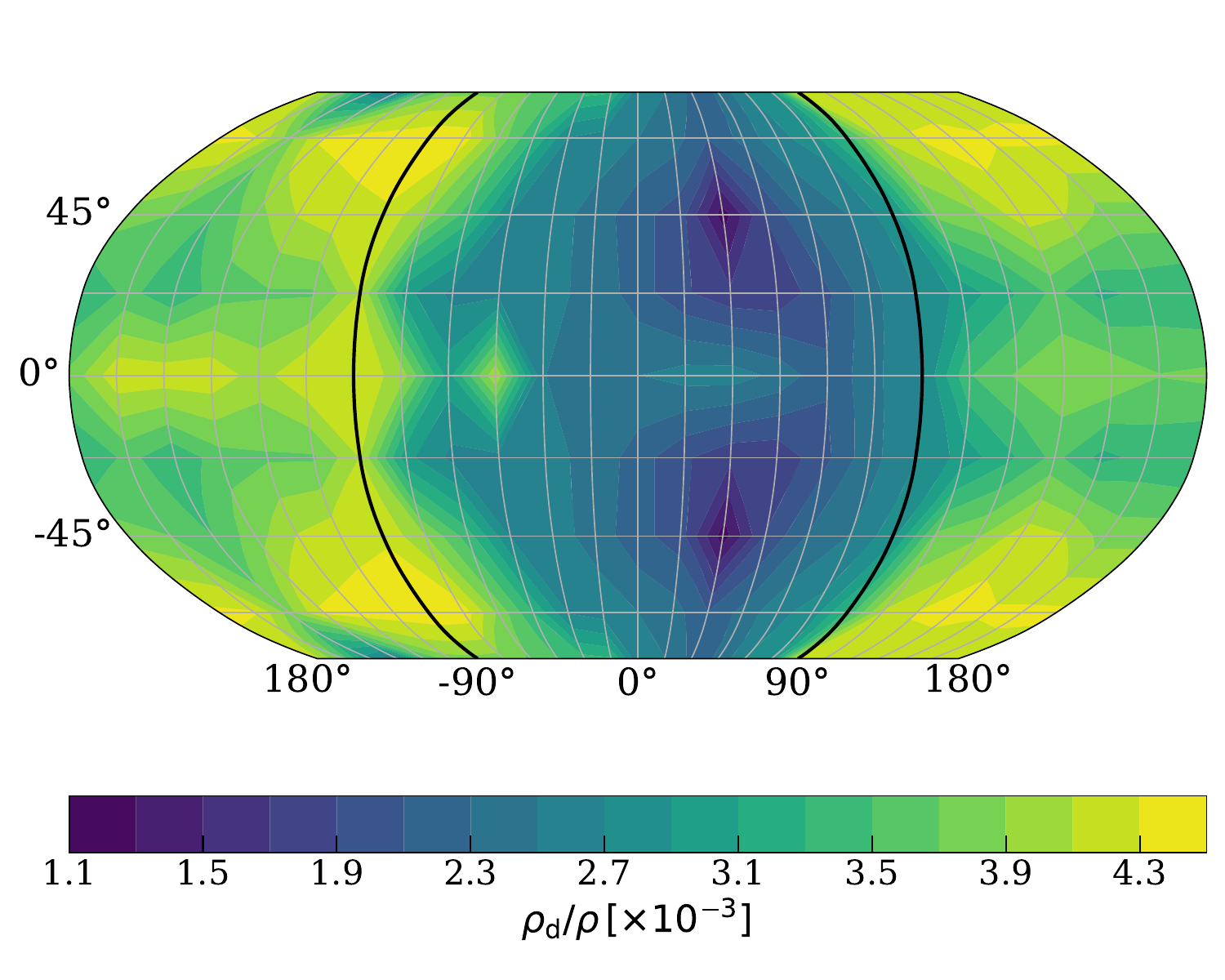}
      \includegraphics[width=0.49\textwidth]{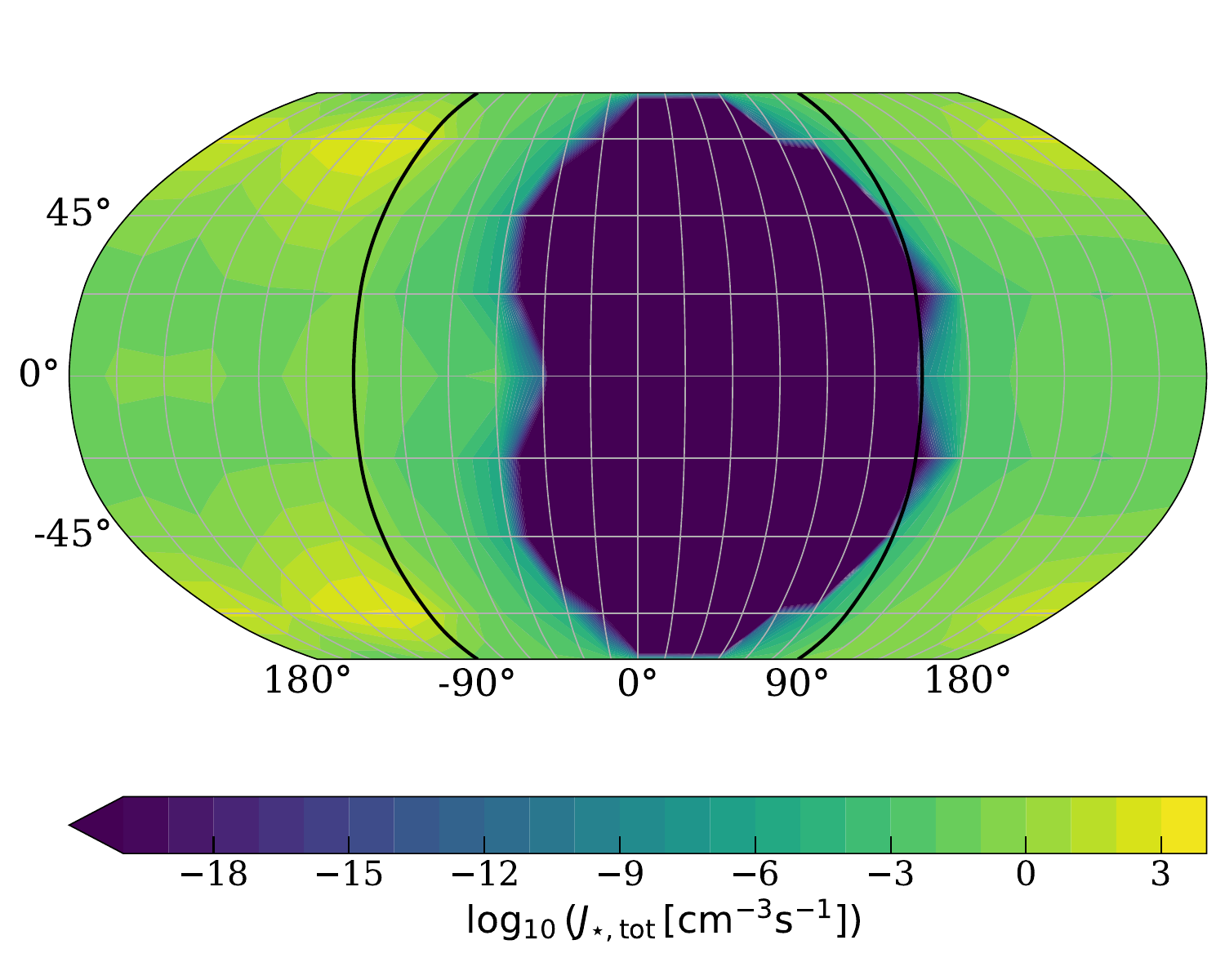}
      \includegraphics[width=0.49\textwidth]{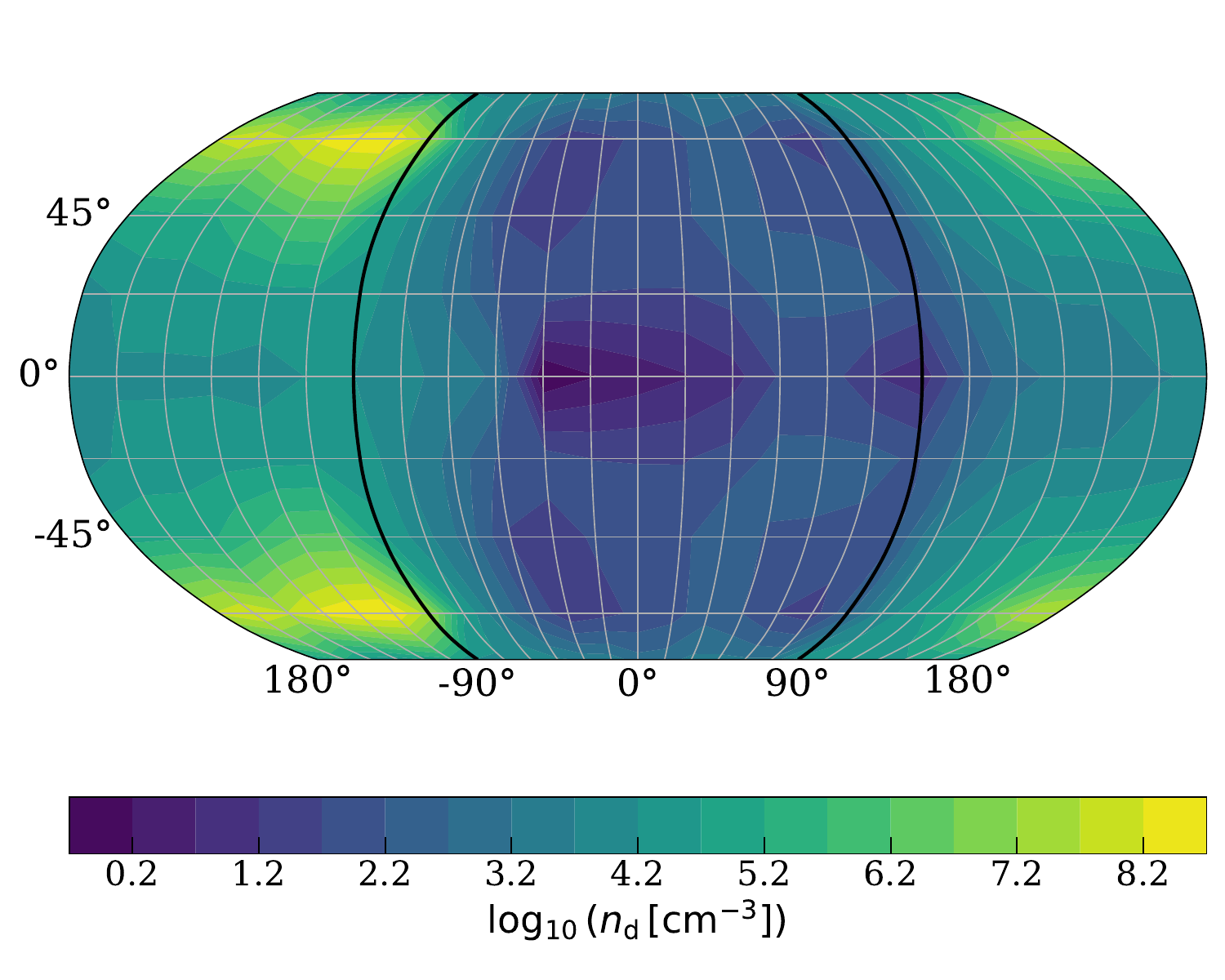}\\
      \includegraphics[width=0.49\textwidth]{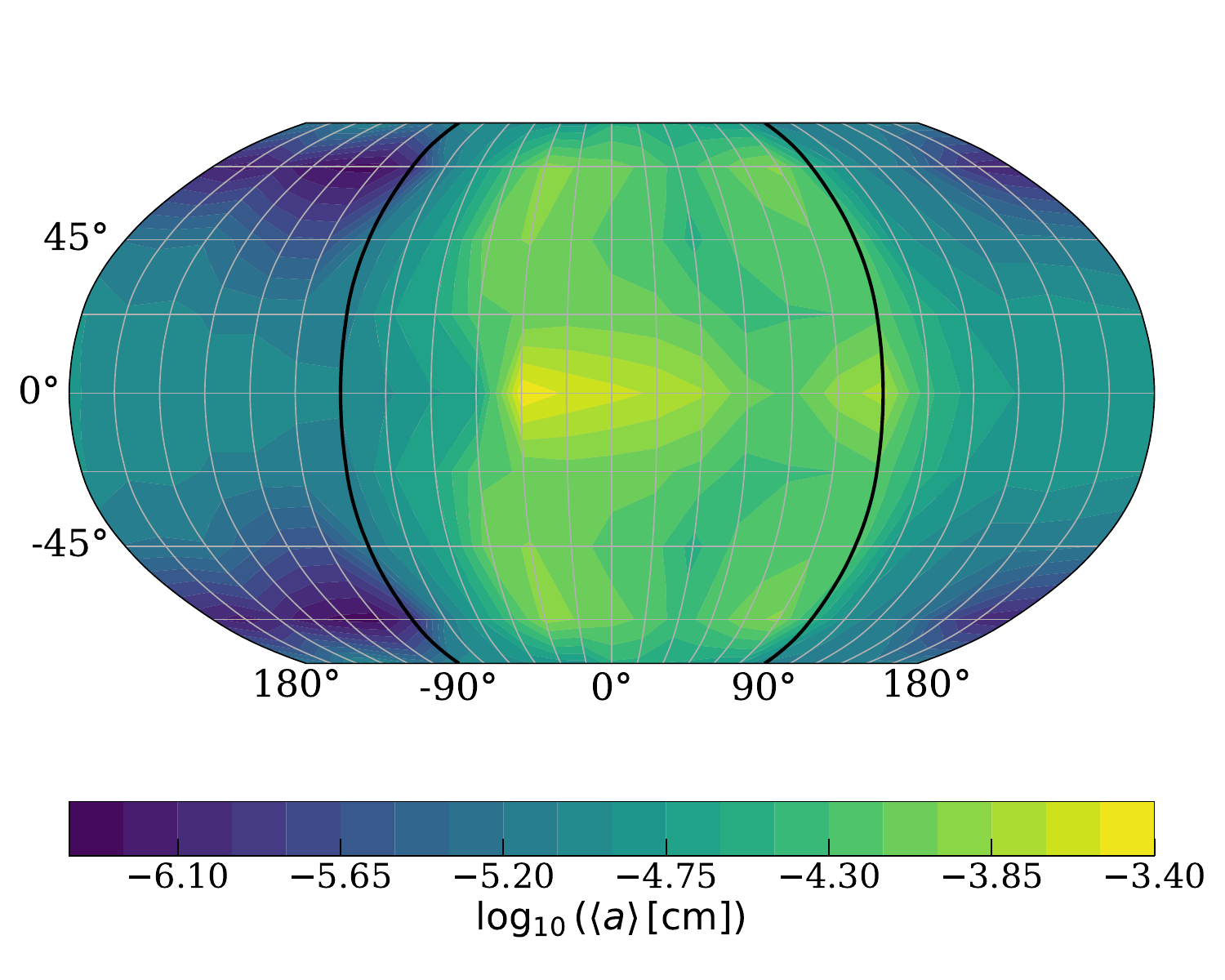}
      \includegraphics[width=0.49\textwidth]{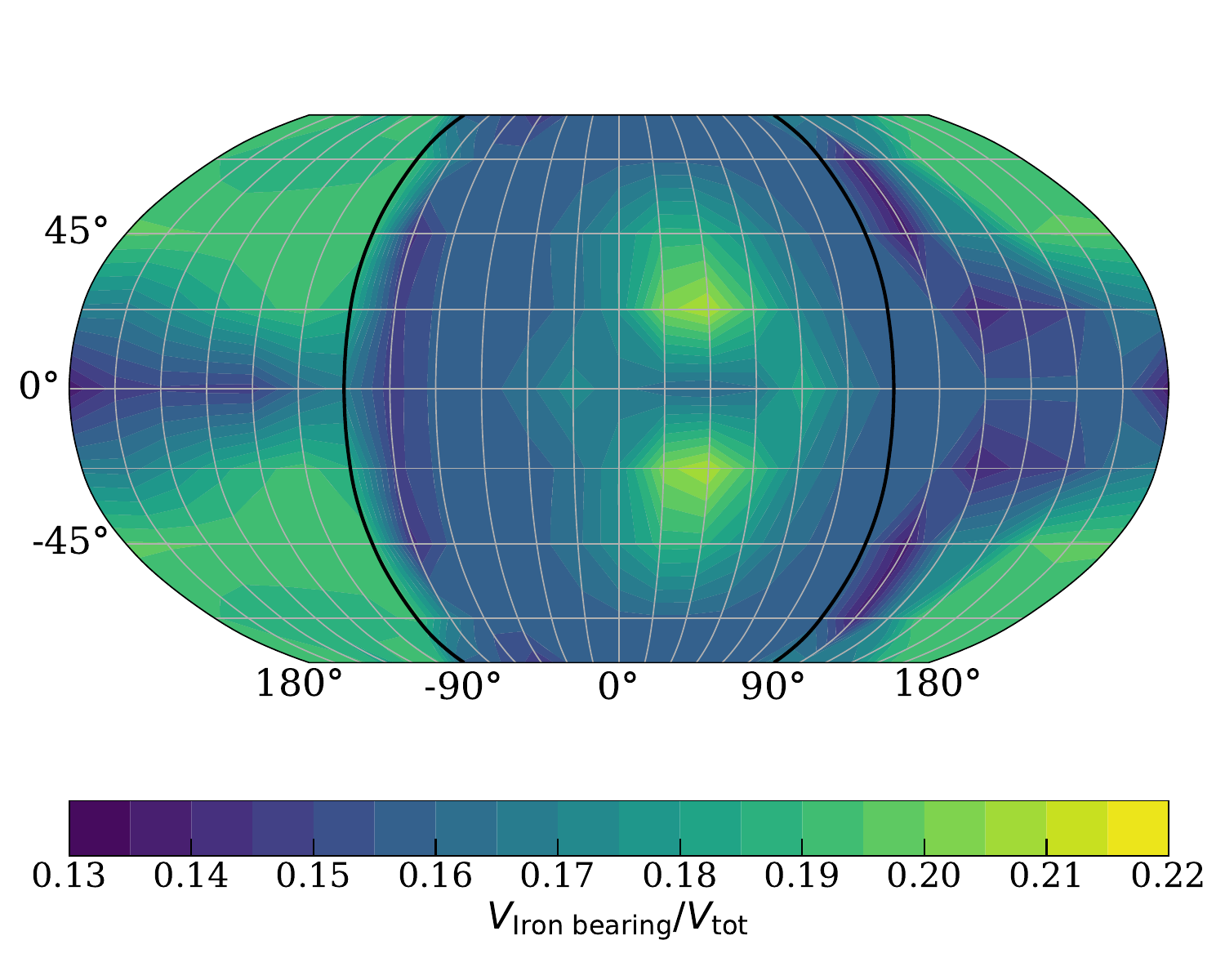}\\
      \caption{GCM and cloud formation results for WASP-43~b at $p_{\rm gas} = 10\,{\rm mbar}$. The maps are centred on the substellar point ($\phi_{\rm long} = 0\degree,\, \lambda_{\rm latt} = 0\degree$), with the terminator limbs ($\phi_{\rm long}=\pm 90\degree$) shown in bold black lines. \textbf{Top Left:} The local gas temperature from the GCM solution. \textbf{Top Right:} Cloud particle mass load in the atmosphere as the ratio of the density of cloud particle mass to the gas density. \textbf{Middle Left:} Total nucleation rate. \textbf{Middle Right:} Cloud particle number density. \textbf{Bottom Left:} Average cloud particle size. \textbf{Bottom Right:} Volume fraction of the cloud particles composed of iron-bearing condensate species. The iron-bearing condensates volume fraction is the sum of the volume fractions for \ce{Fe}[s], \ce{FeS}[s], \ce{FeO}[s], \ce{Fe2O3}[s], \ce{Fe2SiO4}[s].}
      \label{fig:W43b_10mbar_maps}
  \end{figure*}

\begin{figure*}
      \centering
      \includegraphics[width=0.49\textwidth]{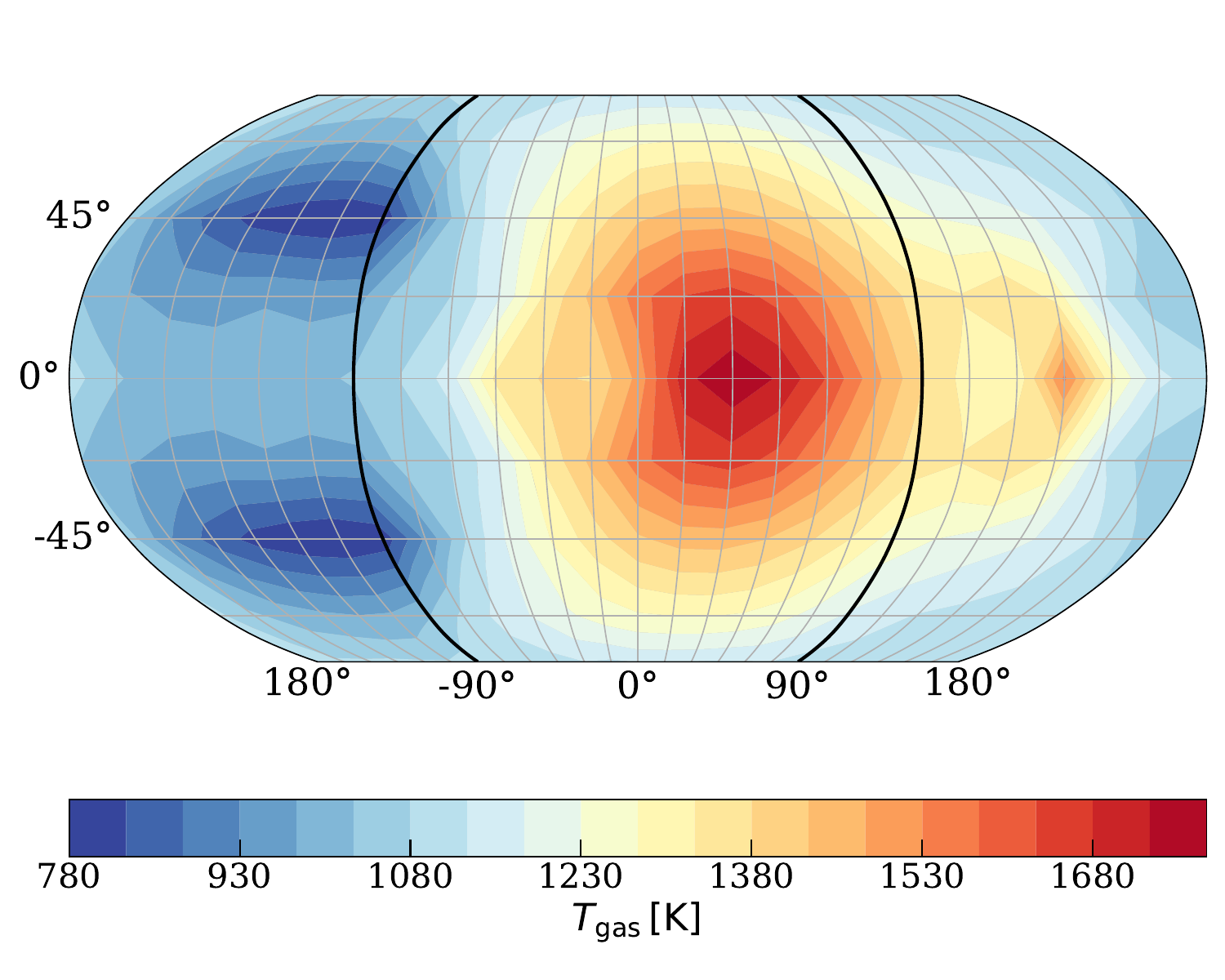}
      \includegraphics[width=0.49\textwidth]{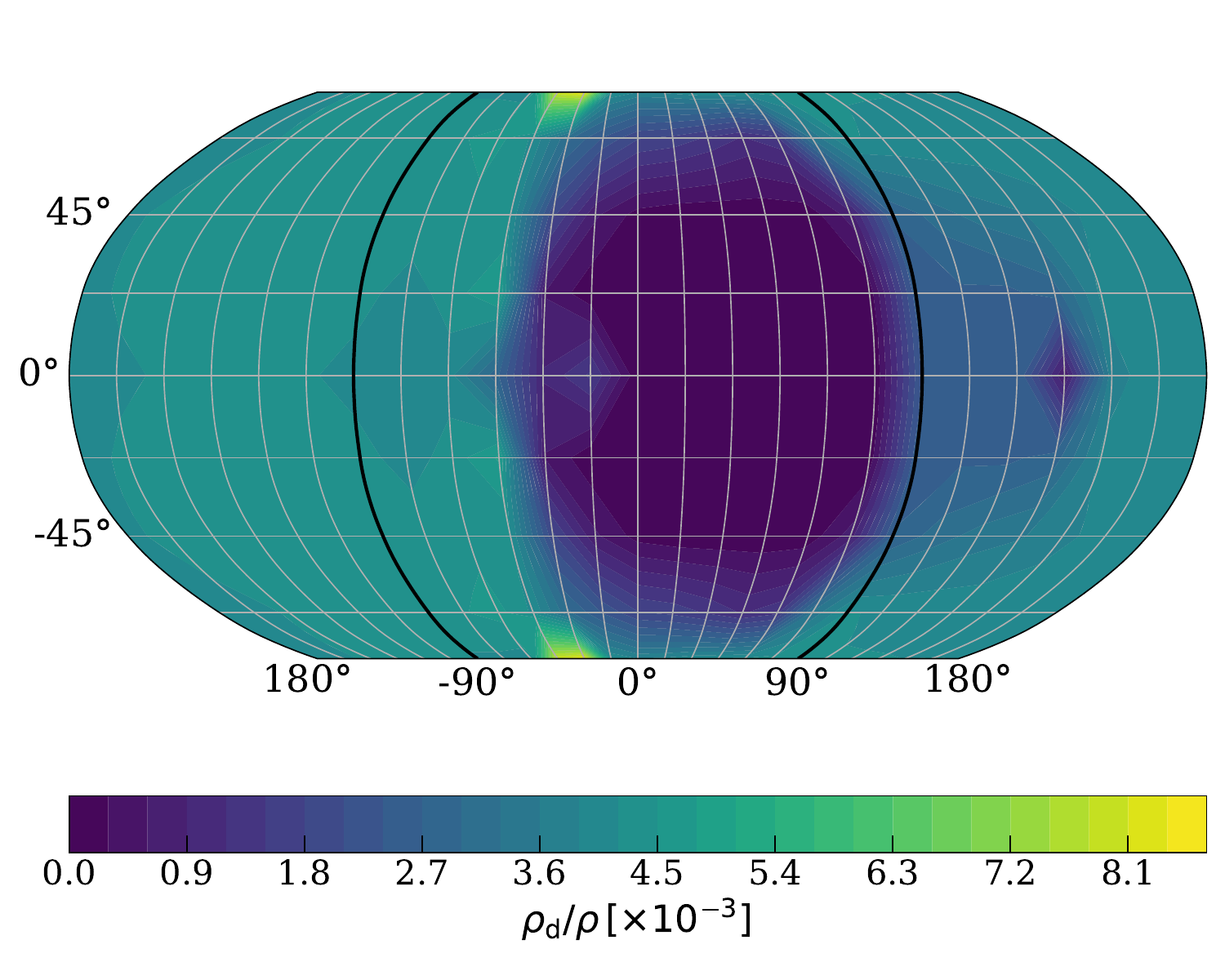}\\
      \includegraphics[width=0.49\textwidth]{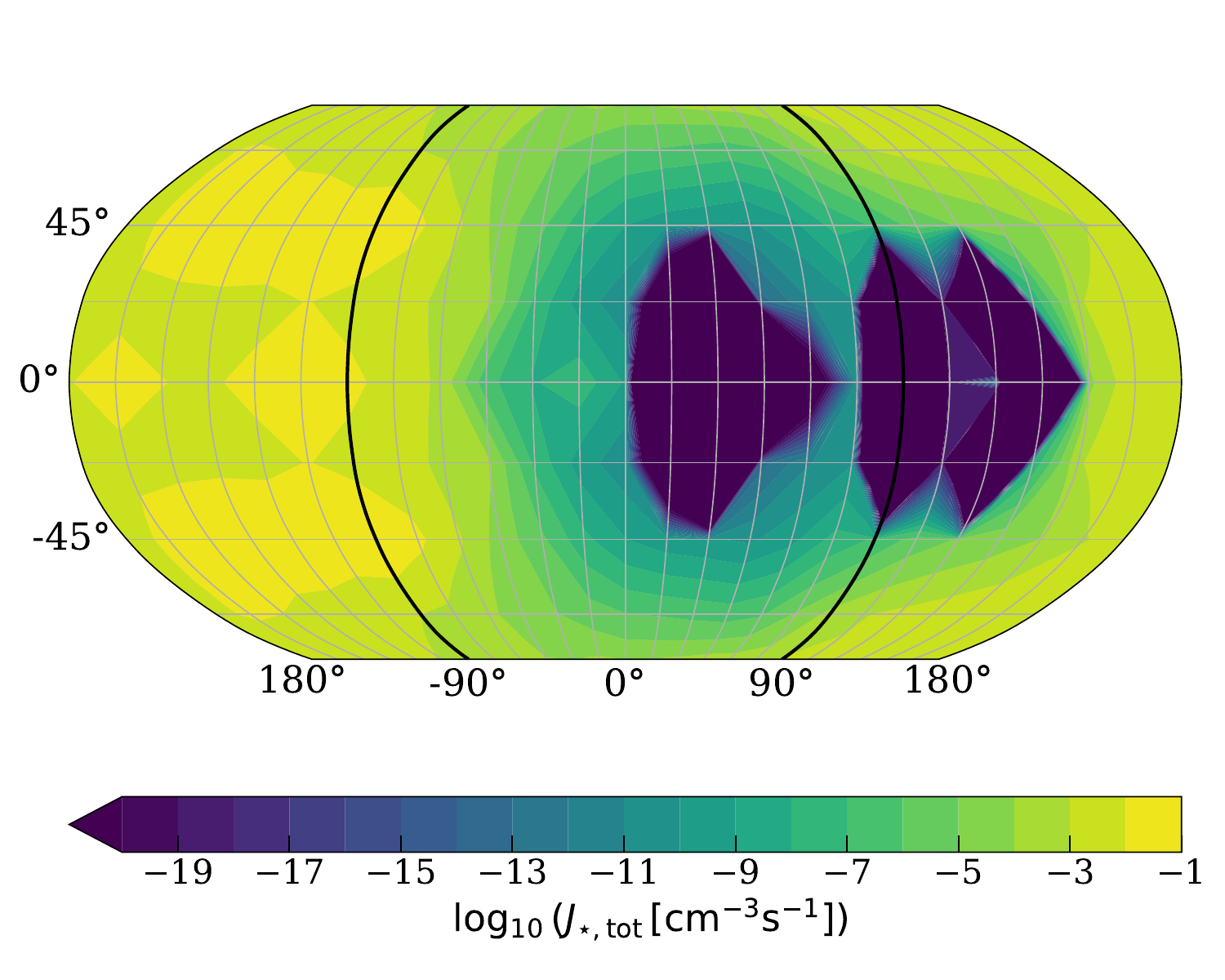}
      \includegraphics[width=0.49\textwidth]{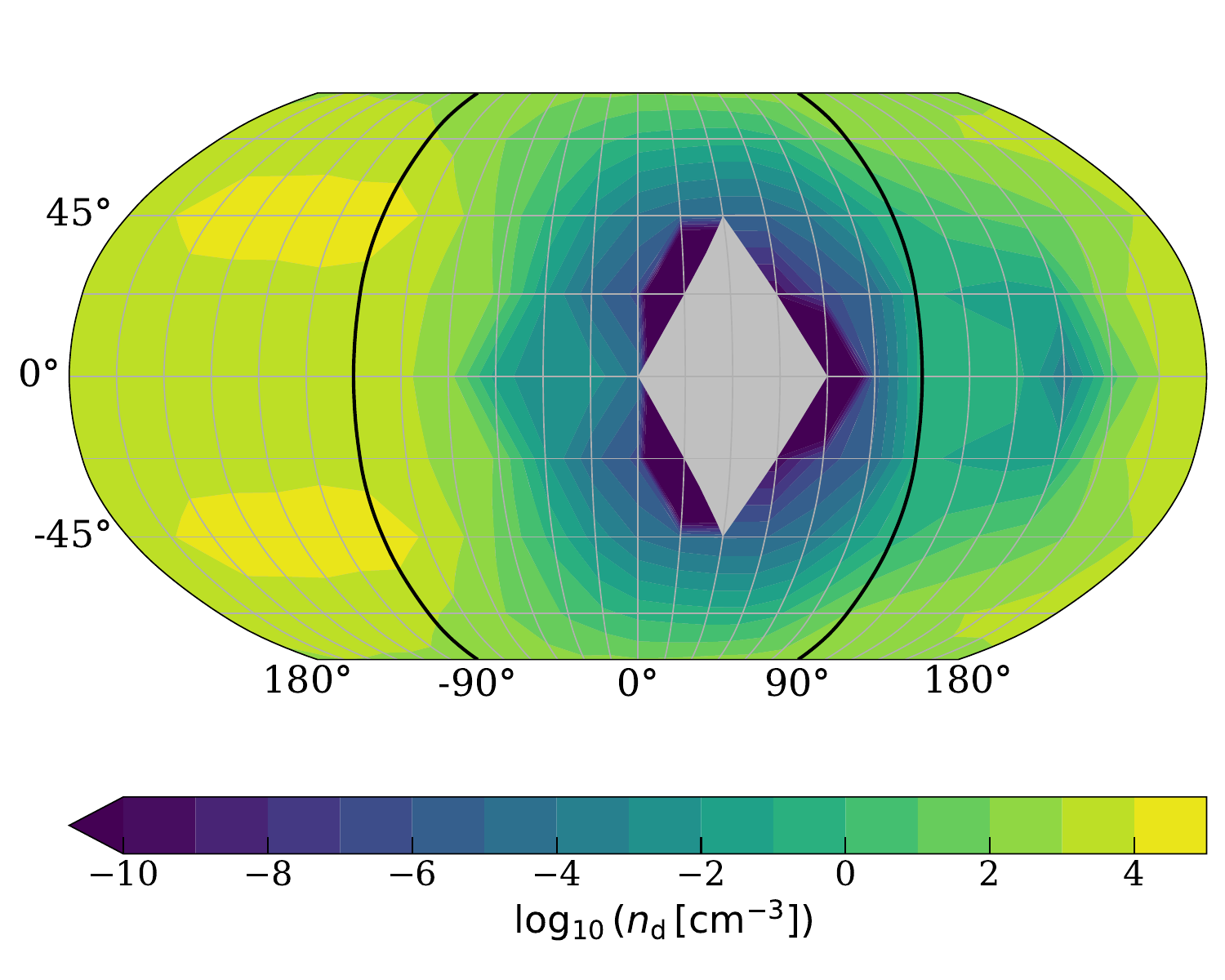}\\
      \includegraphics[width=0.49\textwidth]{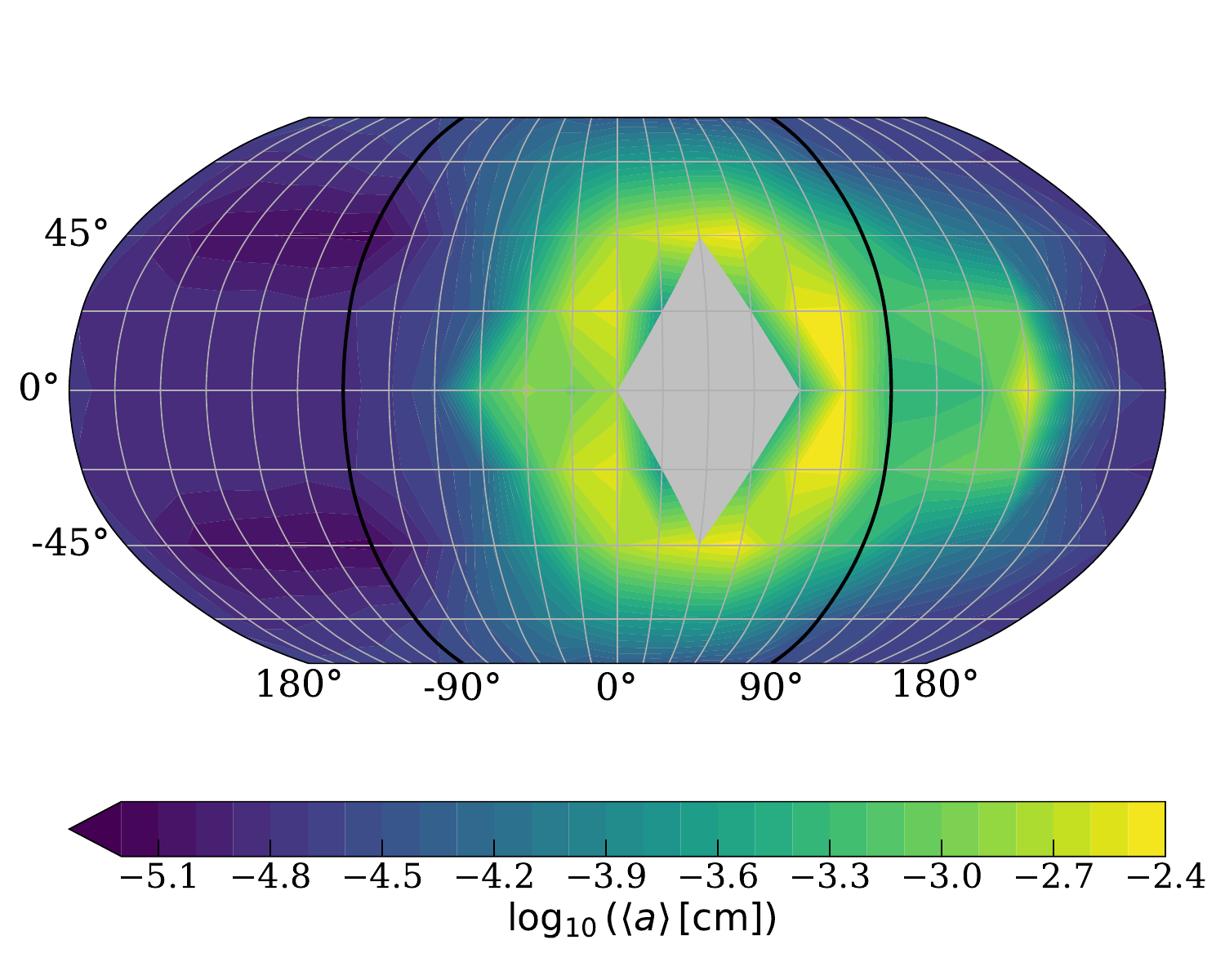}
    \includegraphics[width=0.49\textwidth]{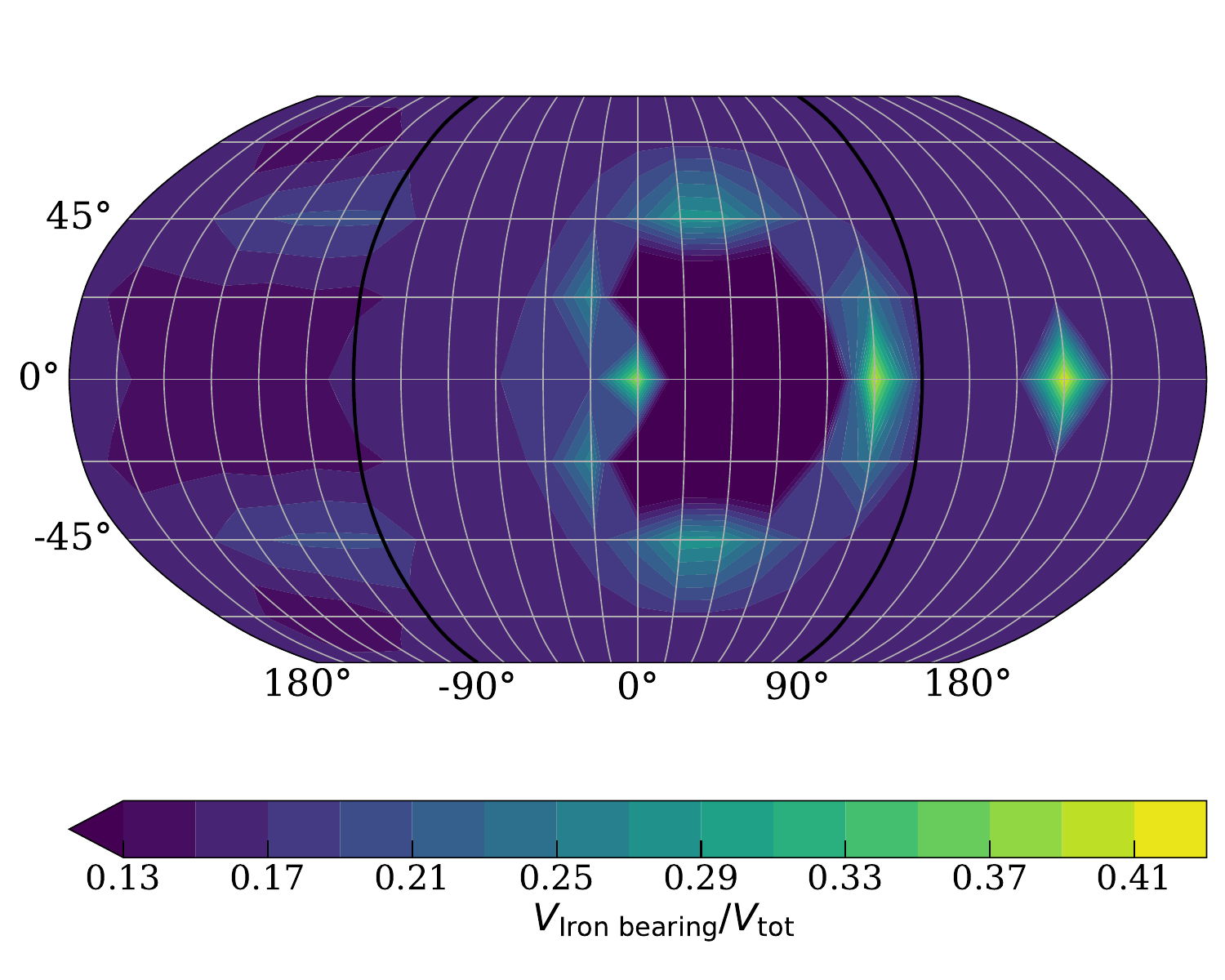}
      \caption{GCM and cloud formation results for HD~209458~b at $p_{\rm gas} = 10\,{\rm mbar}$. The maps are centred on the substellar point ($\phi_{\rm long} = 0\degree,\, \lambda_{\rm latt} = 0\degree$), with the terminator limbs ($\phi_{\rm long}=\pm 90\degree$) shown in bold black lines. \textbf{Top Left:} The local gas temperature from the GCM solution. \textbf{Top Right:} Cloud particle mass load in the atmosphere as the ratio of the density of cloud particle mass to the gas density. \textbf{Middle Left:} Total nucleation rate. \textbf{Middle Right:} Cloud particle number density. \textbf{Bottom Left:} Average cloud particle size. \textbf{Bottom Right:} Volume fraction of the cloud particles composed of iron-bearing condensate species.}
      \label{fig:HD209458b_10mbar_maps}
  \end{figure*}

As our aim is to investigate the dayside geometric albedo, it is also instructive to look at isobaric maps of various cloud properties for the dayside of these exoplanets. Isobaric maps are shown for WASP~43~b and HD~209458~b at 10 mbar in Figs.~\ref{fig:W43b_10mbar_maps} and~\ref{fig:HD209458b_10mbar_maps}, respectively. The grey regions indicate where there are no cloud particles. Additional maps at other pressure levels can be found in %Appendix~\ref{sec:additional}, 
Figs.~\ref{fig:W43b_solar_01mbar_maps}~-~\ref{fig:HD209458b_mbar_maps}. The effect of the temperature inversion at the substellar point is also nicely seen in maps of the cloud formation results, which is in particular evident in Fig.~\ref{fig:HD209458b_mbar_maps} as almost the entire dayside of HD~209458~b is devoid of cloud formation at that pressure level of 1~mbar (the grey area).

%%%%%%%%%%%%%%%%%%%%%%%%%%%%%%%%%%%%%%%%%%%%%%%%%%%%%%%%%%%%%%%%%%%%%%%%%%%%%%
%%%%%%%%%%%%%%%%%%%%%%%%%%%%%%%%%%%%%%%%%%%%%%%%%%%%%%%%%%%%%%%%%%%%%%%%%%%%%%

%%%%%%%%%%%%%%%%%%%%%%%%%%%%%%%%%%%%%%%%%%%%%%%%%%%%%%%%%%%%%%%%%%%%%%%%%%%%%%
%%%%%%%%%%%%%%%%%%%%%%%%%%%%%%%%%%%%%%%%%%%%%%%%%%%%%%%%%%%%%%%%%%%%%%%%%%%%%%

Where the temperature is higher, silicate materials are less thermally stable than iron-bearing oxides, metallic iron and iron sulphide, hence the volume fraction of iron bearing species increases, this coincides with a reduction in cloud particle mass load ($\rho_{\rm d}/\rho$). For example this is illustrated nicely for WASP-43~b in Fig.~\ref{fig:W43b_10mbar_maps}, comparing the top left and right plots, and the bottom right plot. This is more difficult to see for HD~209458~b at 10 mbar in Fig.~\ref{fig:HD209458b_10mbar_maps} because of the absence of cloud material at the hot-spot on the dayside.

The grey regions indicate where there is no cloud material: for particular quantities (number density $n_{\rm d}$, and average particle size $\langle a \rangle$) there is no data across these regions. For quantities that are only meaningful to discuss above a certain limit, but their values nonetheless remain non-zero, extended colour-bar minimums are used. For example, we truncate the colour maps for all nucleation rates at $J_{\rm tot} = 10^{-20}\, [{\rm cm^{-3}s^{-1}}]$. This is equivalent to approximately one new particle per ${\rm cm^{3}}$ every 1000 current universe lifetimes, a very conservative lower limit for meaningful nucleation. Hence any nucleation rate values below this limit are shown as part of the last colour level. The reason for the sudden drop-off in nucleation rate seen in the maps (Figs.~\ref{fig:W43b_10mbar_maps}~-~\ref{fig:HD209458b_10mbar_maps}), is because when nucleation becomes thermally unfavourable it tends to reduce rapidly (see Fig.~\ref{fig:nucleation_substellar}).

%-----------------------------------------------------------------------------
\section{Geometric albedo modelling}\label{sec:albedo_models}

%-----------------------------------------------------------------------------
 \subsection{Methods}\label{sec:albedo_methods}

The geometric albedo $A_g$ of a transiting exoplanet is a property that can be inferred by observing the star and the planet just before/after and during the planet's secondary eclipse.
It is defined as the ratio of reflected flux at $\alpha$~=~0$\degree$, when the planet is directly behind the host star as seen by an observer, compared to a Lambertian (i.e.\ white and isotropically reflecting) flat disk of the same cross-sectional area. For gaseous exoplanets of the temperatures considered in this paper, we expect reflected light to dominate in the visible and UV where thermal emission is negligible, 
before thermal emission starts to dominate in the infrared. Instruments such as CHEOPS therefore cover the reflected light dominated part of the spectra.

In this work we use the Polarisation of Hot Exoplanets (PolHEx) code~\citep{23ChStHe} to simulate the flux that is reflected by the model atmospheres of hot gas giant exoplanets HD~209458~b and WASP-43~b, as described in Sect.~\ref{sec:atm_results}. PolHEx is based on the adding-doubling radiative transfer algorithm of \cite{87HaBoHo}, adapted for modelling polarised reflected light spectra of exoplanets by \cite{06StRoCo, 08Stam}, and recently further adapted for inhomogeneous transiting exoplanets by \cite{23ChStHe}. Multiple scattering effects are fully included in the radiative transfer part of the code~\citep{87HaBoHo}. We use a planetary scattering matrix for the model atmosphere which is normalised on the planet's geometric albedo $A_g$ when the 
reflected flux is integrated over the visible part of the planet at a phase angle of $\alpha$~=~0$\degree$, i.e. when a transiting planet is directly behind the star from the point of view of the observer. We compute $A_g$ as a function of the wavelength, and integrate over the CHEOPS bandpass (0.35~-~1.1~$\mu$m) in order to be able to compare simulations against observations of WASP-43~b and HD~209458~b.
	
It has been shown that errors of several percent can be introduced into computations of the geometric albedo when the linear polarisation state of light is not taken into account~\citep{05StHo} because the polarisation state influences the scattering processes.
We therefore use a 3~$\times$~3 matrix to compute the reflected flux and the geometric albedo. See \cite{23ChStHe} for details. 

Figure~\ref{fig:PT_16_regions_W43b} gives the pressure-temperature profiles of the different atmospheric regions as computed using the GCM and kinetic cloud models. 
We chose to model the reflection (geometric albedo) spectra using the regions represented by solid lines in these figures only, as it can be seen that the latitudinal variation is not as strong as the longitudinal variation. 
As a consequence of the higher temperatures in the upper atmosphere of HD~209458~b, due to the inclusion of TiO and VO in the GCM, clouds do not form in the upper atmosphere (at pressures below $\sim$0.01~bar) around this region of HD~209458~b. In contrast, our models predict WASP-43b to be relatively cloudy throughout the dayside and up to low pressures, due to the relatively cooler upper-atmosphere temperatures, with no TiO and VO included in those GCMs~\citep{TaylorBell_ERS2024}. 

 For the different cloud species predicted by the kinetic cloud models of the two planets, we combine the refractive indices together for each cloud layer to form inhomogenous clouds using effective medium theory~\citep[e.g.][]{16MiDiLi}. The refractive indices of individual materials that we assume are given in Table~\ref{t:ref_index}. The material volume fractions V$_{s}$~/~V$_{\rm tot}$ of the different materials forming the mixed-material cloud particles as a function of pressure are illustrated in Fig.~\ref{fig:cloud_W43b} for WASP-43~b and Fig.~\ref{fig:cloud_HD209} for HD~209458~b, for the longitude and latitude regions used in this work. It can be seen that while WASP-43~b is expected to retain clouds up to high altitudes across the dayside, the clouds are limited to higher pressure layers, $>$~0.01~bar, around the hotspot region of HD~209458~b according to our model atmospheres. We do not include refractive indices for NaCl or KCl in Table~\ref{t:ref_index} as these do not contribute to the material volume fractions of either of our target planets. The only refractive index data we could find for CaTiO$_3$ (perovskite) is from \cite{03PoKeFa}, but it is only available at wavelengths greater than 2~$\mu$m which is higher than we need for this study, so we also do not use it here. 

\begin{figure*}
	\centering
	\includegraphics[width=0.49\textwidth]{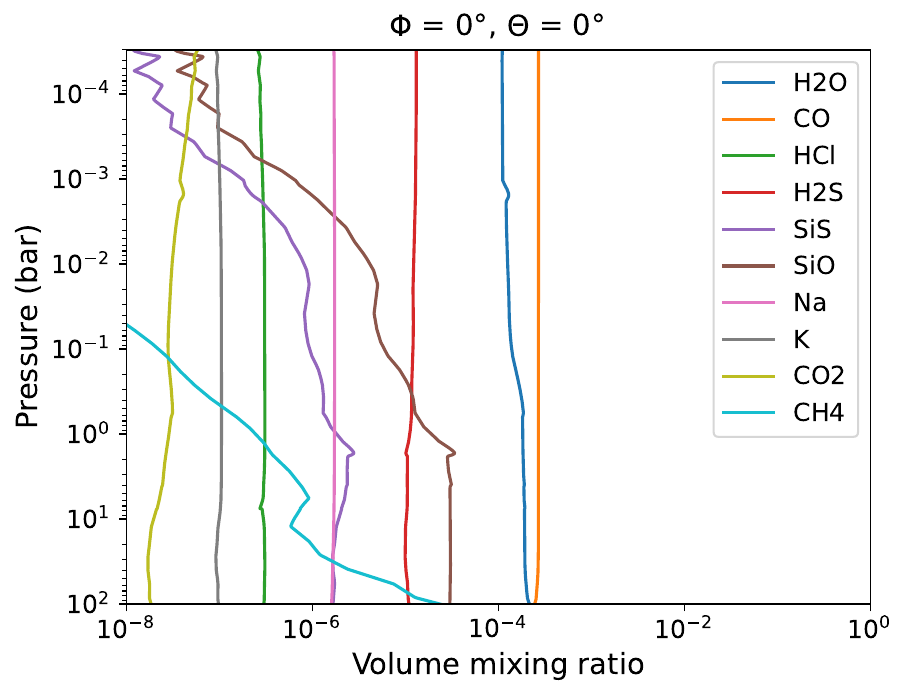}
 \includegraphics[width=0.49\textwidth]{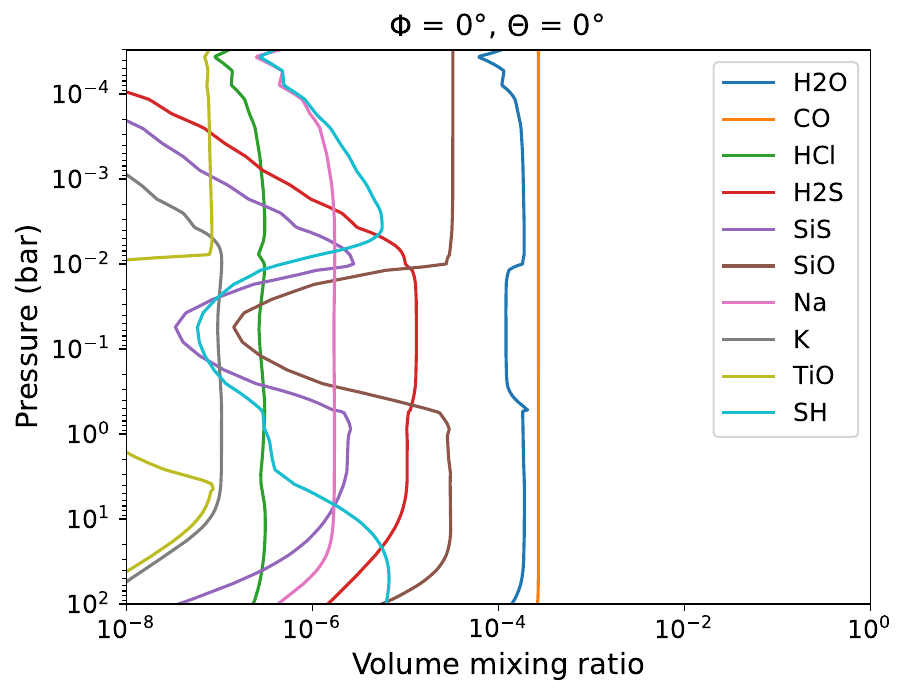}
 \caption{Volume mixing ratios (VMRs) of the most abundant gas-phase molecular and atomic species at the sub-stellar points of WASP-43~b (left) and HD~209458~b (right), based on our model atmospheres. {\bf The VMR of VO is approximately 3~$\times$10$^{-9}$ throughout the atmospheres and is thus not visible in these plots.}}\label{fig:mixing_ratios}
\end{figure*}

 The most abundant gas-phase molecular and atomic species for the sub-stellar point for the two planets are shown in Fig.~\ref{fig:mixing_ratios} (left) for WASP-43~b and Fig.~\ref{fig:mixing_ratios} (right) for HD~209458~b. We include the species shown in our models at the volume mixing ratios (VMRs) specified, using opacities which were computed for constructing the ExoMolOP database~\citep{20ChRoAl.exo}, and line lists for each species as follows. H$_2$O~\citep{ExoMol_H2O}, CO~\citep{15LiGoRo.CO}, HCl~\citep{22GoRoHa}, H$_2$S~\citep{ExoMol_H2S}, SiS~\citep{18UpCoTe.SiS}, SiO~\citep{21YuTeSy.SiO}, Na~\citep{19AlSpLe.broad,KURonline}, K~\citep{16AlSpKi.broad,KURonline}, TiO~\citep{19McMaHo.TiO}, SH~\citep{ExoMol_SH}, CO$_2$~\citep{ExoMol_CO2}, CH$_4$~\citep{ExoMol_CH4}, VO~\citep{ExoMol_VO}. 
 We note that the VMR of VO lies just below the lower limit of the x-axis of Fig.~\ref{fig:mixing_ratios} (right), at approximately 3~$\times$10$^{-9}$ throughout the atmosphere.

%%%%%%%%%%%%%%%%%%%%%%%%%%%%%%%%%%%%%%%%%%%%%%%%%%%%%%%%%%%%%%%%%%%%%%%%%%%%%%
%%%%%%%%%%%%%%%%%%%%%%%%%%%%%%%%%%%%%%%%%%%%%%%%%%%%%%%%%%%%%%%%%%%%%%%%%%%%%%

%%%%%%%%%%%%%%%%%%%%%%%%%%%%%%%%%%%%%%%%%%%%%%%%%%%%%%%%%%%%%%%%%%%%%%%%%%%%%%
\begin{table*}
	\centering
	\begin{tabular}{lll} 
		\hline
		\rule{0pt}{3ex}Species & Name  & Source \\
		\hline
		\rule{0pt}{3ex}Al$_2$O$_3$[s] (crystalline) & Corundum &  \cite{12Palik} \\
    \rule{0pt}{3ex}CaSiO$_3$[s] (crystalline) & Wollastonite  &   \cite{13He} \\
		\rule{0pt}{3ex}Fe$_2$O$_3$[s] (solid) & Hematite &  \cite{05Triaud}$^a$\\
		\rule{0pt}{3ex}Fe$_2$SiO$_4$[s] (crystalline) & Fayalite &   Unpublished$^b$\\
		\rule{0pt}{3ex}FeO[s] (amorphous) & Wustite &  \cite{95HeBeMu}\\
  \rule{0pt}{3ex}FeS[s] (crystalline) & Troilite &\cite{94PoHoBe}\\
		\rule{0pt}{3ex}Fe[s] (metallic) & Iron & \cite{12Palik} \\
		\rule{0pt}{3ex}Mg$_2$SiO$_4$[s] (amorphous) & Forsterite & \cite{03JaDoMu}\\ 
		\rule{0pt}{3ex}MgO[s] (cubic) & Magnesium oxide & \cite{12Palik} \\
		\rule{0pt}{3ex}MgSiO$_3$[s] (amorphous) & Enstatite  &  \cite{95DoBeHe}\\
		\rule{0pt}{3ex}SiO$_2$[s] (crystalline) & Quartz &   \cite{12Palik} \\
		\rule{0pt}{3ex}SiO[s] (non-crystalline) & Silicon oxide &  \cite{12Palik} \\
		\rule{0pt}{3ex}TiO$_2$[s] (rutile) & Rutile & \cite{11ZePoMu}\\
	\end{tabular}
	\caption{Sources used for the real ($n$) and imaginary ($k$) parts of the refractive indices of the species that form the cloud particles in this work. All species are in the solid phase, indicated by [s].}
	%\end{sidewaystable*}
	\label{t:ref_index}
	%\rule{0pt}{0.2ex}
% \vspace{-0.1ex}
	\flushleft{\textit{$^a$ Downloaded via the 
			Aerosol Refractive Index Archive (ARIA) at \url{http://eodg.atm.ox.ac.uk/ARIA/}}}\\
	\flushleft{\textit{$^b$ Accessed via the Database of Optical Constants for Cosmic Dust at  \url{https://www.astro.uni-jena.de/Laboratory/OCDB/crsilicates.html}}} \\
\end{table*}
%%%%%%%%%%%%%%%%%%%%%%%%%%%%%%%%%%%%%%%%%%%%%%%%%%%%%%%%%%%%%%%%%%%%%%%%%%%%%%	
	
For planets that orbit very close to their host star, such that the planet-star distance is not much greater than the planetary radius, it is no longer a good approximation to consider incident starlight to be parallel. We follow the methods of \cite{19Palmer} and introduce a stellar grid to represent the contribution from different regions of the star to each planetary grid point, and include the planet-star distance, stellar radius and planetary radius into our computations. We compute limb darkening using the quadratic law~\citep{50Kopal,11Howarth} with coefficients of $a$~=~0.67679, $b$~=~0.07477 for WASP-43 and $a$~=~0.44756, $b$~=~0.27543 for HD~209458 taken from Table~2 of \cite{21Claret}.
In Fig.~\ref{fig:closein}, we demonstrate the difference made by implementing the stellar grid and limb darkening effects to the close-in model exoplanet atmosphere of WASP-43b, by plotting the reflected flux $F$
%the geometric albedo $A_g$ 
at $\lambda$~=~0.35~$\mu$m as a function of the planetary phase angle $\alpha$. 
%The geometric albedo is $F$ at a phase angle of 0$^\circ$).
%(although this albedo is defined at a phase angle of 0$^\circ$). 
The differences are largest at $\alpha=0^\circ$, the phase angle that we are focused on in this study, showing the importance of accounting for the non--parallel incident light.

%-----------------------------------------------------------------------------
 \subsection{Results}\label{sec:albedo_results}

\begin{figure*}
	\centering
    \includegraphics[width=0.49\textwidth]{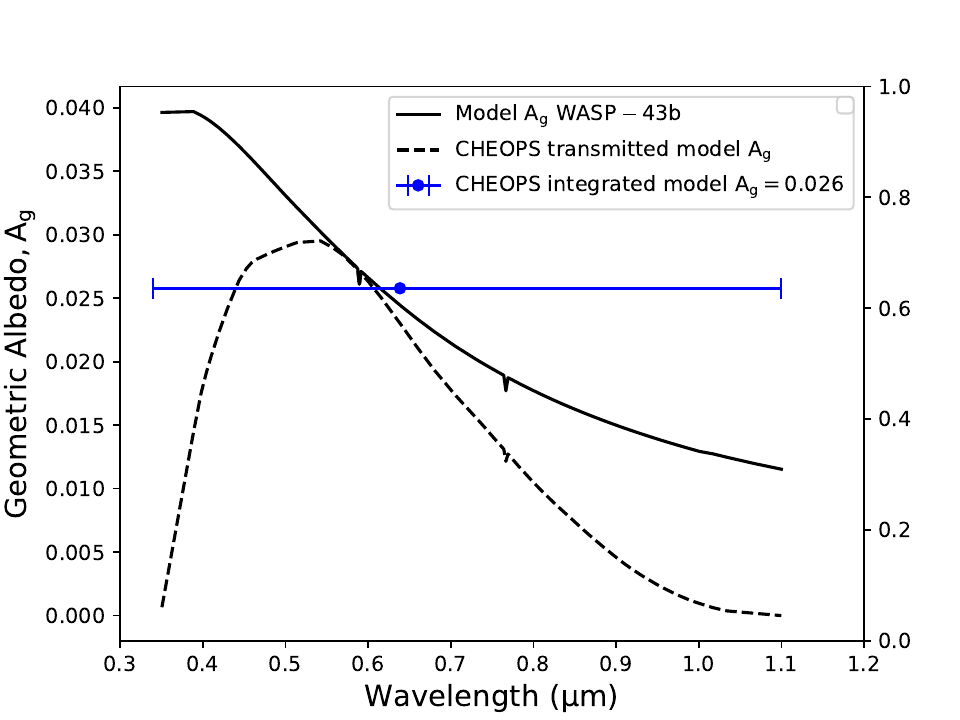}
        \includegraphics[width=0.49\textwidth]{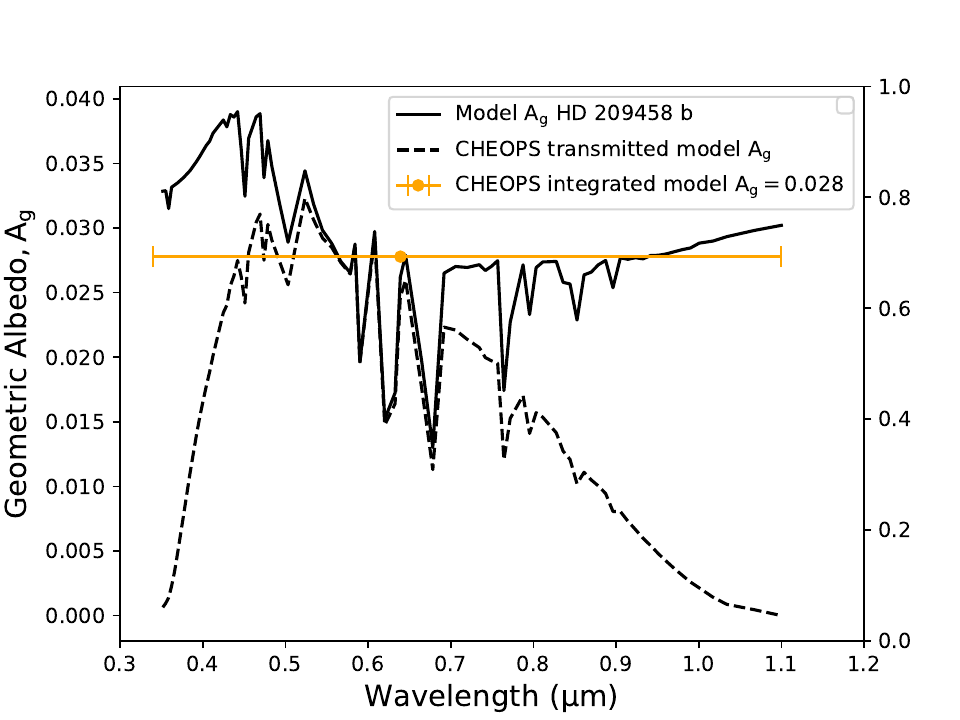}
    \caption{Geometric albedos $A_g$ of WASP-43~b (left) and HD~209458~b (right) for our model horizontally inhomogeneous atmospheres, and integrated over the CHEOPS bandpass. Here `transmitted geometric albedo' is the signal multiplied by the CHEOPS filter throughput function and CCD sensitivity curve (transmission fraction).}\label{fig:cheops_albedos}
\end{figure*}

\begin{figure*}
	\centering
    \includegraphics[width=0.49\textwidth]{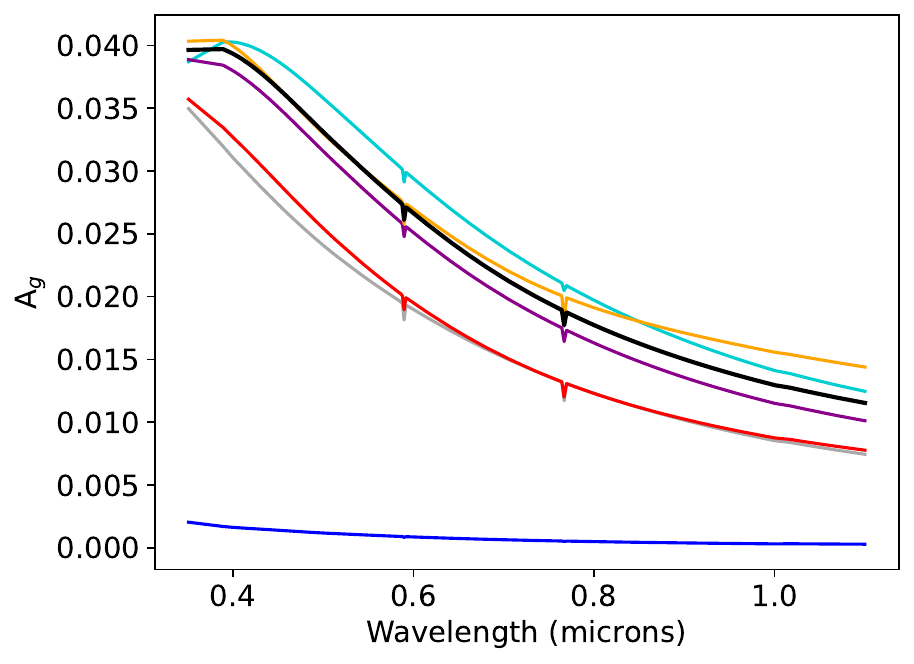}
        \includegraphics[width=0.49\textwidth]{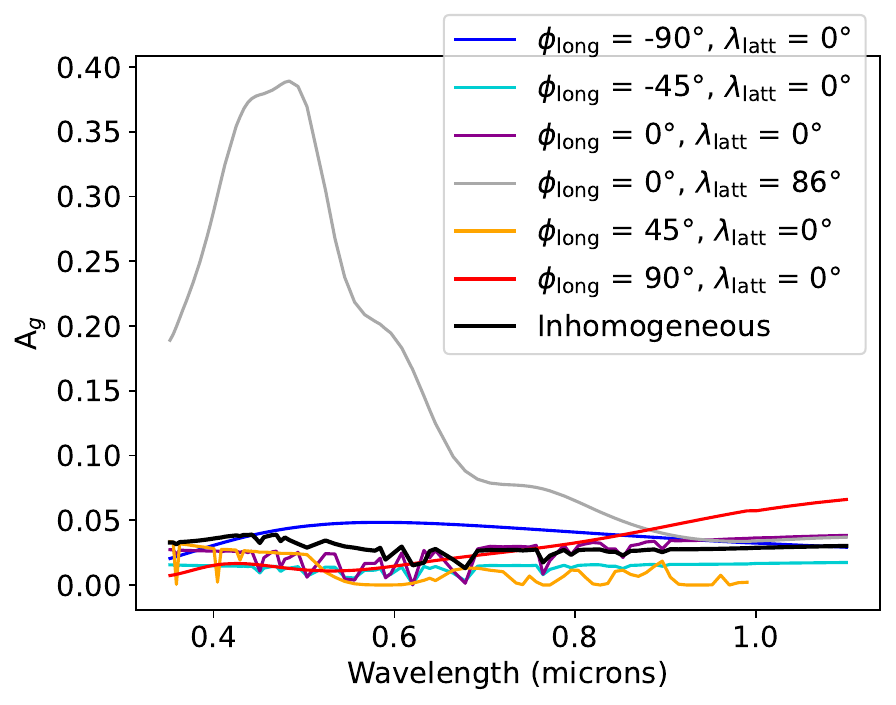}
    \caption{Geometric albedos $A_g$ of WASP-43~b (left) and HD~209458~b (right) for selected atmospheric regions, and (in black) a combination of the shown regions for the assumption of a mainly horizontally inhomogeneous atmosphere.}\label{fig:tau_A_F_W43b}
\end{figure*}

Figure~\ref{fig:cheops_albedos} shows the main results of our study: the
geometric albedo $A_g$ computed wavelength-dependent and integrated over the CHEOPS bandpass for WASP-43~b and HD~209458~b. The `transmitted geometric albedo' refers to the geometric albedo weighted by the CHEOPS filter throughput function and the CCD sensitivity, as taken from \url{http://svo2.cab.inta-csic.es/theory/fps/}.
Although the pressure-temperature profiles of the atmospheres show marked differences (Fig.~\ref{fig:PT_16_regions_W43b}), resulting in different levels of cloudiness across the daysides, our models predict both planets to exhibit relatively similar geometric albedos across the CHEOPS bandpass. The geometric albedos of WASP-43b and HD~209458~b as observed by CHEOPS alone are only $\sim$~0.14~\citep{scandariato_phase_2022}\footnote{estimated based on parameters and figures provided in the paper} and 0.096~$\pm$~0.016~\citep{brandeker_cheops_2022}, respectively. 
If we compare our model results to these by computing over the CHEOPS bandpass, we find 0.026 (compared to $\sim$~0.14) for WASP-43~b and 0.028 (compared to 0.096~$\pm$~0.016) for HD~209458~b. 
Figure~\ref{fig:tau_A_F_W43b} shows the wavelength-dependent geometric albedos for each of the atmospheric regions we included in our models, along with the horizontally inhomogeneous atmospheres for comparison, for WASP-43~b and HD~209458~b. 

It is interesting to see in Fig.~\ref{fig:tau_A_F_W43b} that the high-latitude region ($\lambda_{\rm latt} = 86^\circ$) of HD~209458~b has a much larger albedo than the equatorial region. This is due to the higher proportion of reflective materials as compared to Fe-bearing species in this region's upper atmospheric clouds; see Fig.~\ref{fig:cloud_HD209}. The reason for this can be seen in Fig.~\ref{fig:PT_16_regions_W43b}: qualitatively, the pressure-temperature structure of the high-latitude region is similar to that of the morning terminator. However, it is hotter overall due to it being on the dayside. Comparing the pressure-temperature structure of the high-latitude region with Fig.~14 in \cite{helling_cloud_2021} (which shows Solar-abundance supersaturation curves), it is clear that the temperature inversion for the high-latitude region exceeds $1400\,{\rm K}$ at $\sim 10^{-4}\,{\rm bar}$. For this upper part of the high-latitude region atmospheres, the temperature is above the thermal stability for all the condensates except \ce{TiO2}[s], \ce{Al2O3}[s], \ce{CaTiO3}[s] and \ce{CaSiO3}[s]. Crucially, this does not include any of the Fe-bearing species necessary for dark clouds. However, the temperatures in the upper atmosphere are still low enough for cloud formation to occur, through \ce{TiO2}[s] nucleation at low pressures (until above $9\times10^{-4}\,{\rm bar}$). 

%%%%%%%%%%%%%%%%%%%%%%%%%%%%%%%%%%%%%%%%%%%%%%%%%%%%%%%%%%%%%%%%%%%%%%%%%%%%%%
%%%%%%%%%%%%%%%%%%%%%%%%%%%%%%%%%%%%%%%%%%%%%%%%%%%%%%%%%%%%%%%%%%%%%%%%%%%%%%

%%%%%%%%%%%%%%%%%%%%%%%%%%%%%%%%%%%%%%%%%%%%%%%%%%%%%%%%%%%%%%%%%%%%%%%%%%%%%%
%%%%%%%%%%%%%%%%%%%%%%%%%%%%%%%%%%%%%%%%%%%%%%%%%%%%%%%%%%%%%%%%%%%%%%%%%%%%%%

\begin{figure*}
	\centering
    \includegraphics[width=0.49\textwidth]{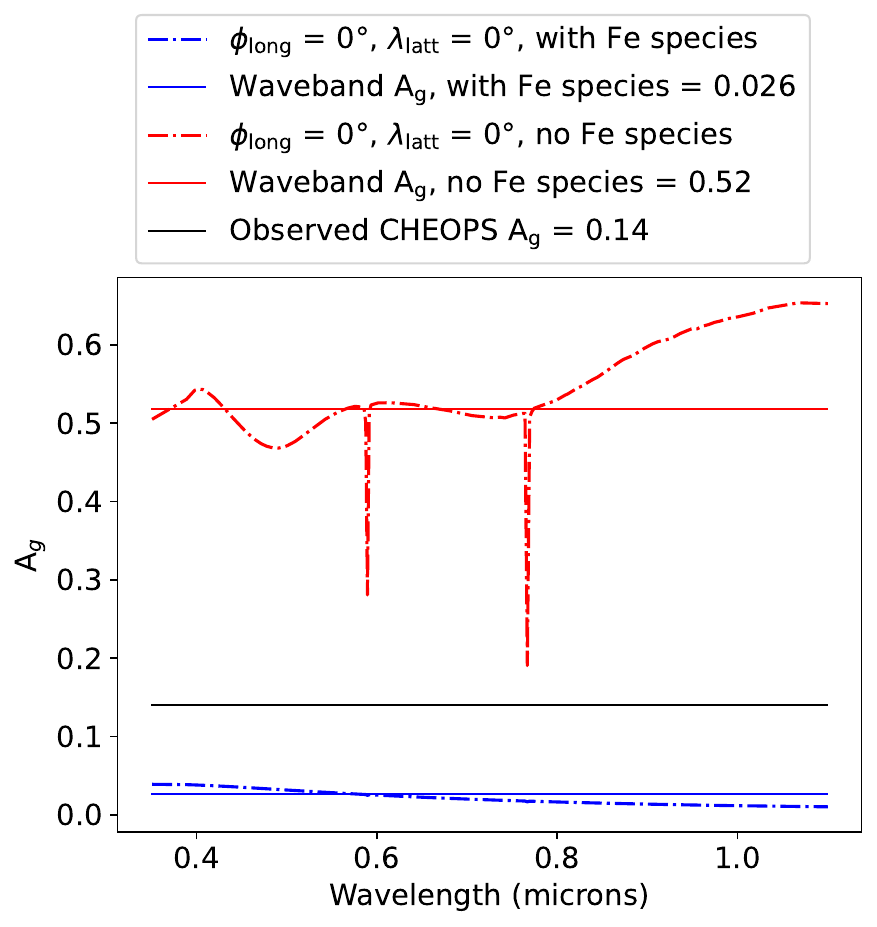}
        \includegraphics[width=0.49\textwidth]{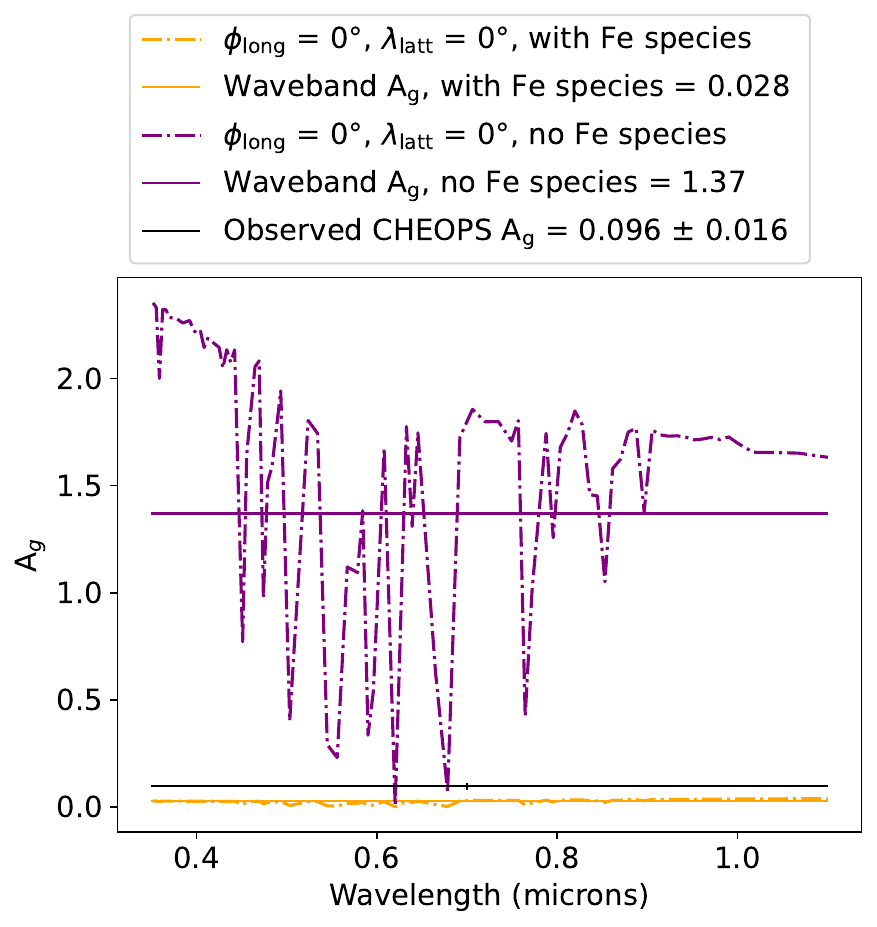}
  \caption{Geometric albedos $A_g$ of WASP-43~b (left) and HD~209458~b (right) for an atmospheric region on the middle of the planetary disk, comparing with and without including Fe-bearing species. Values of $A_g$ integrated over the CHEOPS bandpass for our model atmospheres are also shown, along with values deduced from CHEOPS observations of $A_g$~$\sim$~0.14 (WASP-43~b)~\citep{scandariato_phase_2022} and 0.096~$\pm$~0.016 (HD~209458~b)~\citep{brandeker_cheops_2022}. 
  }\label{fig:tau_A_F_W43b_Fe}
\end{figure*}

It can be seen in Fig.~\ref{fig:tau_A_F_W43b_Fe} that models using the same input parameters for both WASP-43b (left) and HD~209458~b (right) but leaving out strongly absorbing Fe-bearing species such as Fe[s], FeO[s], FeS[s], and Fe$_2$O$_3$[s] in forming the cloud layers yield a much higher geometric albedo than when all species are included. Although \ce{Fe2SiO4}[s] is also an Fe-bearing species, it does not have the same strongly absorbing properties as the others mentioned, and behaves more like a reflecting silicate. In the case of HD~209458~b, $A_g$ can reach very high in the instance that we remove the Fe-bearing species, due to the lack of absorbing clouds high in the atmosphere and the now highly-reflective clouds further down. Strong molecular and atomic absorption features can be seen due to the clear upper atmosphere. 
 
In general, our predicted geometric albedos from GCM and kinetic cloud model predictions are slightly lower than those inferred from observations. We might expect that an atmosphere which has a slightly lower proportion of Fe-bearing species than assumed by us to match the observed data better. The reduced albedo features just under 0.6 and 0.8~$\mu$m in Fig.~\ref{fig:tau_A_F_W43b_Fe} are due to the strongly absorbing Na and K doublets. 

%%%%%%%%%%%%%%%%%%%%%%%%%%%%%%%%%%%%%%%%%%%%%%%%%%%%%%%%%%%%%%%%%%%%%%%%%%%%%%
%%%%%%%%%%%%%%%%%%%%%%%%%%%%%%%%%%%%%%%%%%%%%%%%%%%%%%%%%%%%%%%%%%%%%%%%%%%%%%

%-----------------------------------------------------------------------------		
\section{Conclusion}\label{sec:conclusion}

Based on our model atmospheres and using a radiative transfer code that fully includes multiple scattering, polarization and non-parallel incident starlight, we find geometric albedos $A_g$ of 0.026 for WASP-43b and 0.028 for HD~209458~b in the CHEOPS bandpass, which compare to $A_g$~$\sim$~0.14~\citep{scandariato_phase_2022} and 0.096~$\pm$~0.016~\citep{brandeker_cheops_2022} measured geometric albedos for these planets, respectively. 
%\kc{Due to Fe-bearing species which are expected throughout both atmospheres (point out any differences). Talk about studies which find evidence for clouds or clear daysides on these planets, and other modelling studies.} 
We compare to a model atmosphere which does not contain any Fe-bearing species to demonstrate the dramatic impact these particles have on the albedo. We find that $A_g$ increases from 0.024 to 0.518 when no strongly absorbing Fe-bearing species are included in our WASP-43b model atmosphere, and from 0.021 (with Fe-bearing species) to 1.37 (without Fe-bearing species) for HD~209458~b. 

Our models predict Fe-bearing species to be abundant enough in these atmospheres to reduce $A_g$ to a value consistent with, or lower than, observations, in contrast to highly reflective clouds formed from silicate species alone. We predict that these atmospheres contain a combination of cloud species whose combined scattering and absorption properties lead to a relatively low observed albedo. We demonstrate that the contribution of Fe-bearing species to cloud particles, even at relatively small abundances, can reduce the overall geometric albedo to the same extent as a clear atmosphere. 

It is important to note that our model predictions must also match observed transmission (primary eclipse) and emission (secondary eclipse) spectra of exoplanets, which typically do not provide evidence for very dense cloud cover. As discussed in \cite{samra_clouds_2023}, there are various mechanisms which can explain an atmosphere which is relatively cloudy but still transparent enough that clear molecular and atomic signatures can be seen; for example, increased atmospheric mixing or increased porosity of cloud particles. It would be of interest to explore further the impact of other cloud properties such as different particle size distributions or cloud particles with a fractal nature \citep[see, e.g.,][]{23LoWaLe} on the computed geometric albedo. The assumptions and choices made throughout our models, such as including TiO and VO in the GCM of HD~209458~b but not in that of WASP-43b (see discussion at the start of Sect.~\ref{sec:atm_methods}), have an impact on our results which may not be indicative of the true nature of the planetary atmospheres. We do, however, demonstrate how these different choices impact our final models, and find that despite these differences, both our model atmospheres yield relatively low geometric albedos, largely due to the impact that including Fe-bearing species in the formation of clouds has on the atmospheres. 

We conclude that the geometric albedo alone cannot be used as conclusive evidence for a clear dayside, particularly for the types of exoplanet explored in this study, and must be used in combination with other observational techniques in order to deduce the most likely nature of the atmosphere. We stress that albedo models should be used in combination with other spectra derived observational techniques, such as transmission~\citep{23GrLeWa,24DyMiDe}, emission~\citep{23TaPa}, or even polarised reflection spectra~\citep{18BaKeBo,23ChStHe}, to gain the maximum amount of information possible about the cloud particles in these exotic exoplanet atmospheres.

%-----------------------------------------------------------------------------  
		\section*{Acknowledgements}
		
		This project has received funding from STFC, under project number  ST/V000861/1. Ch.H. and L.C. acknowledges funding from the European Union H2020-MSCA-ITN-2019 under grant agreement no. 860470 (CHAMELEON). 
  % does not belong here
  D.S. acknowledges financial support and use of the computational facilities of the Space Research Institute of the Austrian Academy of Sciences. K.L.C further acknowledges funding by UK Research and Innovation (UKRI) under the UK government’s Horizon Europe funding guarantee as part of an ERC Starter Grant [grant number EP/Y006313/1].

		%%%%%%%%%%%%%%%%%%%%%%%%%%%%%%%%%%%%%%%%%%%%%%%%%%
		\section*{Data Availability}
		
		The data underlying this article will be shared on reasonable request to the corresponding author(s).
		
		%	The inclusion of a Data Availability Statement is a requirement for articles published in MNRAS. Data Availability Statements provide a standardised format for readers to understand the availability of data underlying the research results described in the article. The statement may refer to original data generated in the course of the study or to third-party data analysed in the article. The statement should describe and provide means of access, where possible, by linking to the data or providing the required accession numbers for the relevant databases or DOIs.

		
		
		\bibliographystyle{mnras}
        \bibliography{references}		
		
		%\begin{thebibliography}{99}
		%\end{thebibliography}
		
		%%%%%%%%%%%%%%%%%%%%%%%%%%%%%%%%%%%%%%%%%%%%%%%%%%
		
		%%%%%%%%%%%%%%%%% APPENDICES %%%%%%%%%%%%%%%%%%%%%
		
		%	\newpage
		\appendix

        \section{Log-Normal Derivation}\label{sec:log-normal}
        
     We are going to consider the log-normal distribution written as a function of the particle radius $a$, the mean of the logarithm of particle size $\mu$, the standard deviation of the logarithm of particle size $\sigma$, and particle number density $n_d$:
     \begin{align}
         f(a;\mu,\sigma,n_{\rm d}) = \frac{n_{\rm d}}{a\sigma\sqrt{2\pi}}\exp\left(-\frac{\left(\ln(a)-\mu\right)^2}{2\sigma^2}\right).
     \end{align}
     Now we need to compute the moments of the distribution in radius-space of the cloud particles:
     \begin{align}
         K_j = \int_{a_{\rm l}}^{\infty} a^j f(a) {\rm d}a
     \end{align}
     where $f(a)$ is the distribution function of the cloud particles in radius $a$. We note that the units of this are ${\rm [cm^{-4}]}$, where ${\rm [cm^{-1}]}$ is from the distribution in radius-space, and ${\rm [cm^{-3}]}$ is the number density of cloud particles ($n_{\rm d}$) in the atmosphere. In other words, the integral of the distribution function is not normalised to 1. Hence, the moments in radius space have units of ${\rm cm}^{j-3}$. For the log-normal distribution this is therefore:
     \begin{align}
         K_j = \int_{a_{\rm l}}^{\infty} a^j \frac{n_{\rm d}}{a\sigma\sqrt{2\pi}}\exp\left(-\frac{\left(\ln(a)-\mu\right)^2}{2\sigma^2}\right) {\rm d}a.
     \end{align}
    To do this, we let $x=\ln(a)$, we also make the assumption $a_{\rm l} \rightarrow 0$. We note this is different from the assumption made in \cite{HWT2008} for the Gaussian of $a_{\rm l} \rightarrow -\infty$, however the extension of the integral domain does not introduce a large error as, for both distributions, the function rapidly diminishes in these regions. Thus now with our change of units to $x$, we have limits of $\pm \infty$ and the integral becomes
     \begin{align}
         K_j = \int_{-\infty}^{\infty}  \frac{n_{\rm d}}{\sigma\sqrt{2\pi}}\exp\left(jx-\frac{\left(x-\mu\right)^2}{2\sigma^2}\right) {\rm d}x.
         \label{Integral_in_x}
     \end{align}
     Next we focus on re-arranging the exponent in Eq.~\ref{Integral_in_x}:
     \begin{align}
         jx-\frac{\left(x-\mu\right)^2}{2\sigma^2} = \frac{1}{2\sigma^2}\left(2\sigma^2jx-x^2+2x\mu-\mu^2\right).
     \end{align}
     First, we do the clever trick of introducing $(\mu +j\sigma^2)^2-(\mu +j\sigma^2)^2$ inside the brackets. Then expanding the second part of this term, grouping like-terms of $x$, and cleaning up, we arrive at:
     \begin{align}
     \begin{split}
         jx- & \frac{\left(x-\mu\right)^2}{2\sigma^2} = \frac{1}{2\sigma^2} \cdot \\
         & \left( -x^2 +2(\mu +j\sigma^2)x - (\mu + j\sigma^2)^2 +j\sigma^2(2\mu+j\sigma^2)\right).
     \end{split}
     \end{align}
     We recognise that by defining $\mu' = \mu+j\sigma^2$, the first three terms of the bracket reduce to simply $-(x^2-\mu')^2$. Thus now substituting back into Eq.~\ref{Integral_in_x}, we arrive at:
     \begin{align}
         K_j = \exp\left(\frac{j(2\mu+j\sigma^2)}{2}\right)\int_{-\infty}^{\infty} \frac{n_{\rm d}}{\sigma\sqrt{2\pi}}\exp\left(-\frac{(x-\mu')^2}{2\sigma^2}\right) {\rm d}x,
     \end{align}
     where the integral is now just the integral over a Gaussian distribution (with the addition of the normalisation factor of $n_{\rm d}$). Using the standard result:
     \begin{align}
        \int_{-\infty}^{\infty}\exp\left(-c(x+b)^2\right) {\rm d}x = \sqrt{\frac{\pi}{c}},
     \end{align}
     where we let:
     \begin{align}
         c &= \frac{1}{2\sigma^2}\\
         b & = - \mu'
     \end{align}
     thus just the integral simplifies to:
     \begin{align}
         \frac{n_{\rm d}}{\sigma\sqrt{2\pi}} \int_{-\infty}^{\infty} \exp\left(-\frac{(x-\mu')^2}{2\sigma^2}\right) {\rm d}x = \frac{n_{\rm d}}{\sigma\sqrt{2\pi}} \sqrt{2\pi}\sigma = n_{\rm d}.
     \end{align}
     This is a standard and expected result, given the normalisation of the moments $K_j$ as discussed before. We therefore arrive (after far too much detail) at the expression for the $K_j$ moment for a log-normal distribution:
     \begin{align}\label{eq:KJ}
         K_j = n_{\rm d}\exp\left(\frac{j(2\mu+j\sigma^2)}{2}\right).
     \end{align}

     Finally we derive a set of equations for the parameters $n_{\rm d},\mu,\sigma^2$, using the moments $K_1,K_2,K_3$. We avoid the zeroth moment, $K_0$, as it is used in the closure conditions of the moment equations, and therefore actually assumes a different form of the size distribution \citep{HWT2008}.
     Taking the logarithm of Eq.~\ref{eq:KJ}:
     \begin{align}
         \ln(K_j) = \ln(n_{\rm d})+\frac{j}{2}(2\mu+j\sigma^2),
     \end{align}
     and taking the difference of the logarithm of adjacent moments, $K_2,\,K_3$:    
     \begin{align}
        \begin{split}
             \ln(K_2)-\ln(K_1) &= (2\mu+2\sigma^2)-\frac{1}{2}(2\mu+\sigma^2) \\
         &= \mu + \frac{3}{2}\sigma^2 \label{A15} \\
        \end{split}
     \end{align}
     and
     \begin{align}
        \begin{split}
             \ln(K_3)-\ln(K_2) &= \frac{3}{2}(2\mu+3\sigma^2)-(2\mu+2\sigma^2) \\
         &= \mu + \frac{5}{2}\sigma^2 \label{A17}.
        \end{split}
     \end{align}
     Then subtracting Eq.~\ref{A15} from Eq.~\ref{A17}, we arrive at:    
     \begin{align}
         \sigma^2 = \ln(K_3)-2\ln(K_2)+\ln(K_1). \label{A18}
     \end{align}
     Substituting this result back into Eq.~\ref{A15} we get:
     \begin{align}
         \ln(K_2)-\ln(K_1) &= \mu + \frac{3}{2}\left(\ln(K_3)-2\ln(K_2)+\ln(K_1)\right)
     \end{align}
     therefore
     \begin{align}
         \mu &= -\frac{3}{2}\ln(K_3)+4\ln(K_2)-\frac{5}{2}\ln(K_1). \label{A20}
     \end{align}

     Finally, for $n_{\rm d}$ we take Eq.~\ref{A18} and Eq.~\ref{A20} for $\mu$ and $\sigma^2$ respectively, and substitute them into the formula for $\ln(K_2)$:
     \begin{align}
     \begin{split}
         \ln(K_2) &= \ln(n_{\rm d}) + 2\mu +2\sigma^2 \\
          &= \ln(n_{\rm d})\\
          &\ \ \ -3\ln(K_3)+8\ln(K_2)-5\ln(K_1)\\
          &\ \ \ +2\ln(K_3)-4\ln(K_2)+2\ln(K_1),
     \end{split}
     \end{align}
     simplifying, therefore:
     \begin{align}
          \ln(n_{\rm d}) = \ln(K_3)-3\ln(K_2)+3\ln(K_1).
     \end{align}
     Exponentiating this gives:
     \begin{align}
         n_{\rm d} = \frac{K_3K_1^3}{K_2^3}.
     \end{align}

     So the final solution for the parameters for the log-normal distribution, in terms of radius-space moments $K_1,\,K_2,\,K_3$ are:
     \begin{align}
         n_{\rm d} &= \frac{K_3K_1^3}{K_2^3} \\
         \mu &= \ln\left(\frac{K_2^4}{\sqrt{K_3^{3}K_1^{5}}}\right) \\
         \sigma^2 &= \ln\left(\frac{K_3K_1}{K_2^2}\right).
     \end{align}

     %Finally, we want to verify the results, and thus re-arrange the parameters into expressions for $K_1,K_2,K_3$. These can be derived easily, from the above by dimensional analysis (I have done this as a check, but its just re-arrangement), but also using Eq 41. So all the derivations should now be ok.

\section{Additional Figures}\label{sec:additional}

  \begin{figure*}
      \centering
      \includegraphics[width=0.49\textwidth]{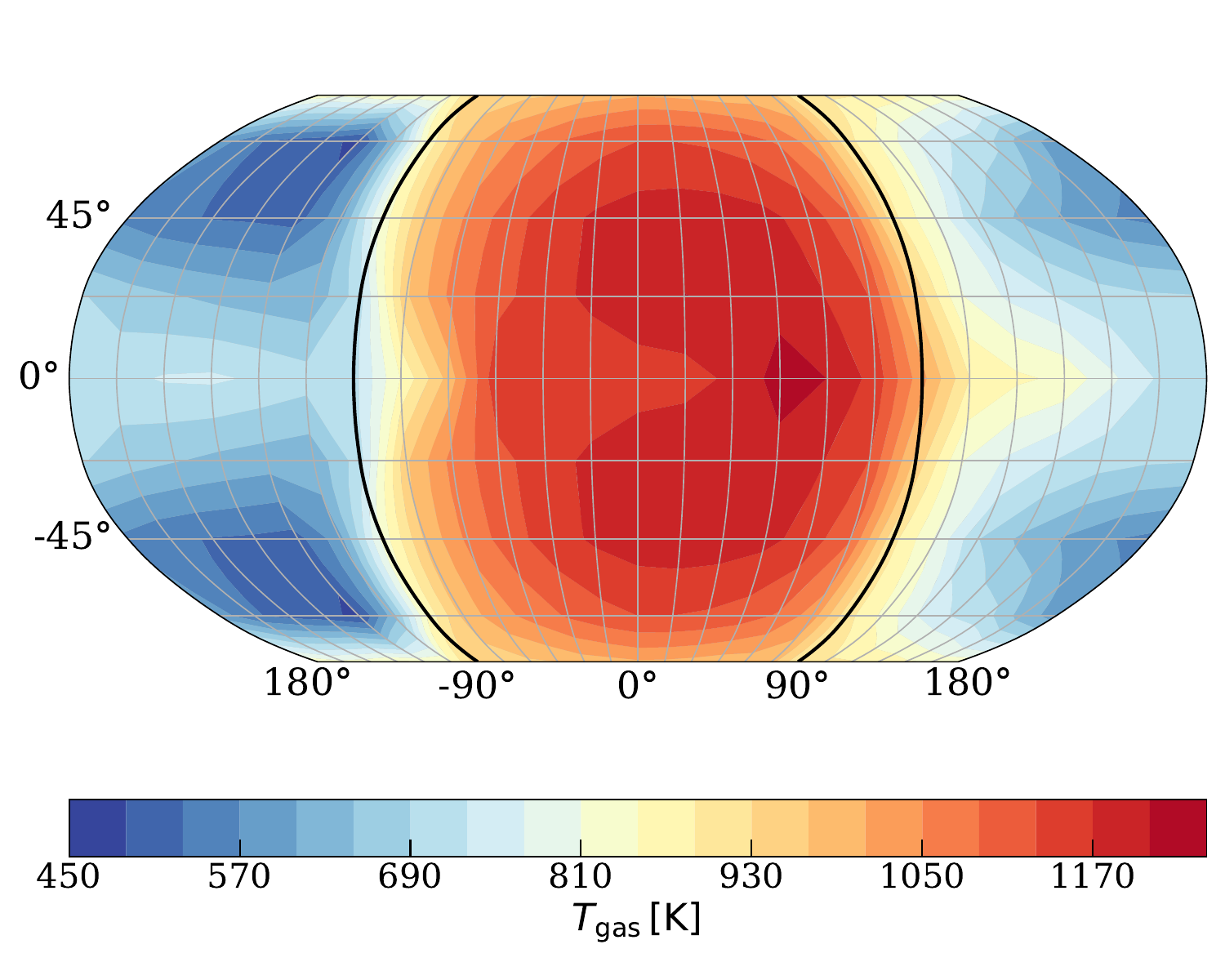}
      \includegraphics[width=0.49\textwidth]{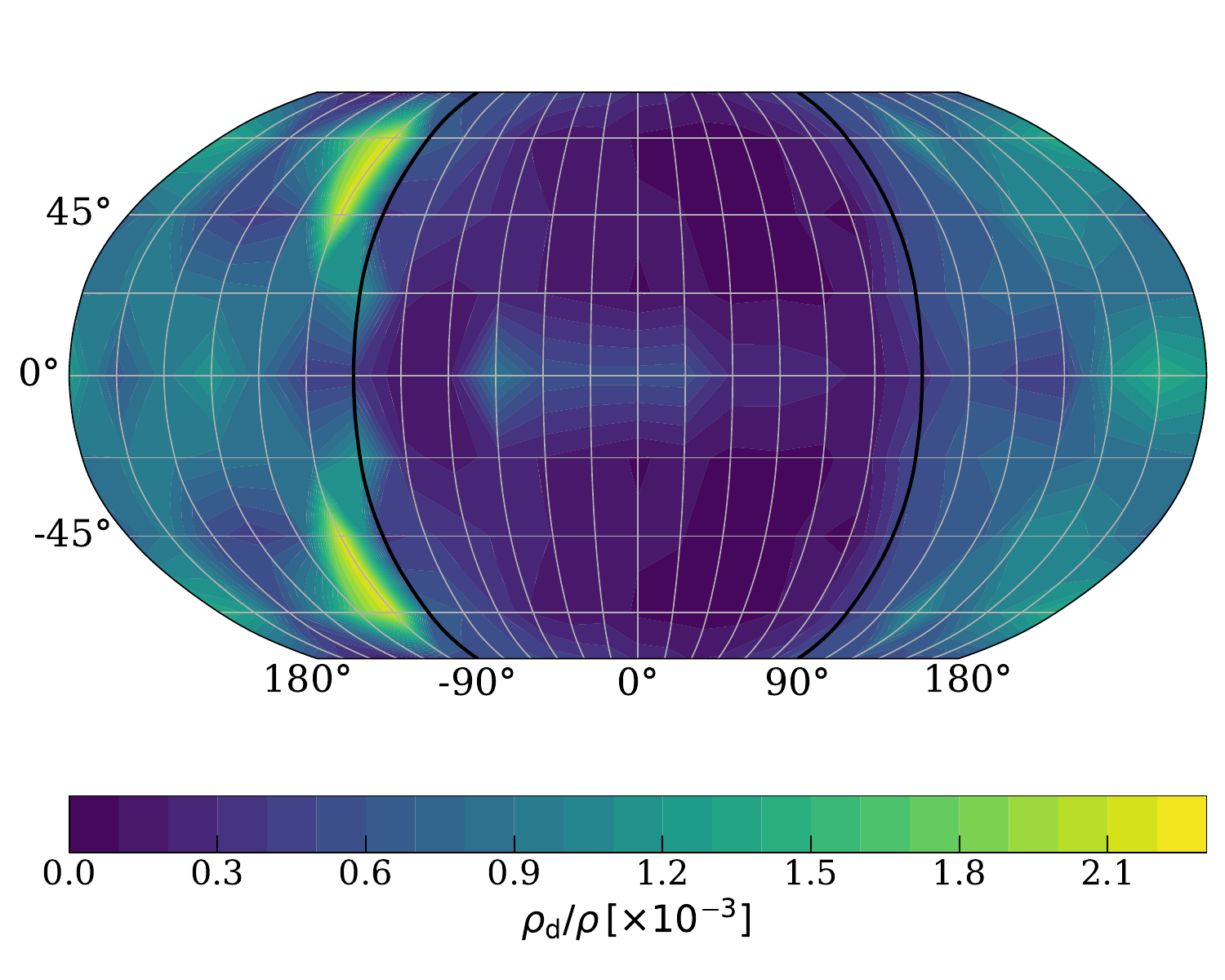}
      \includegraphics[width=0.49\textwidth]{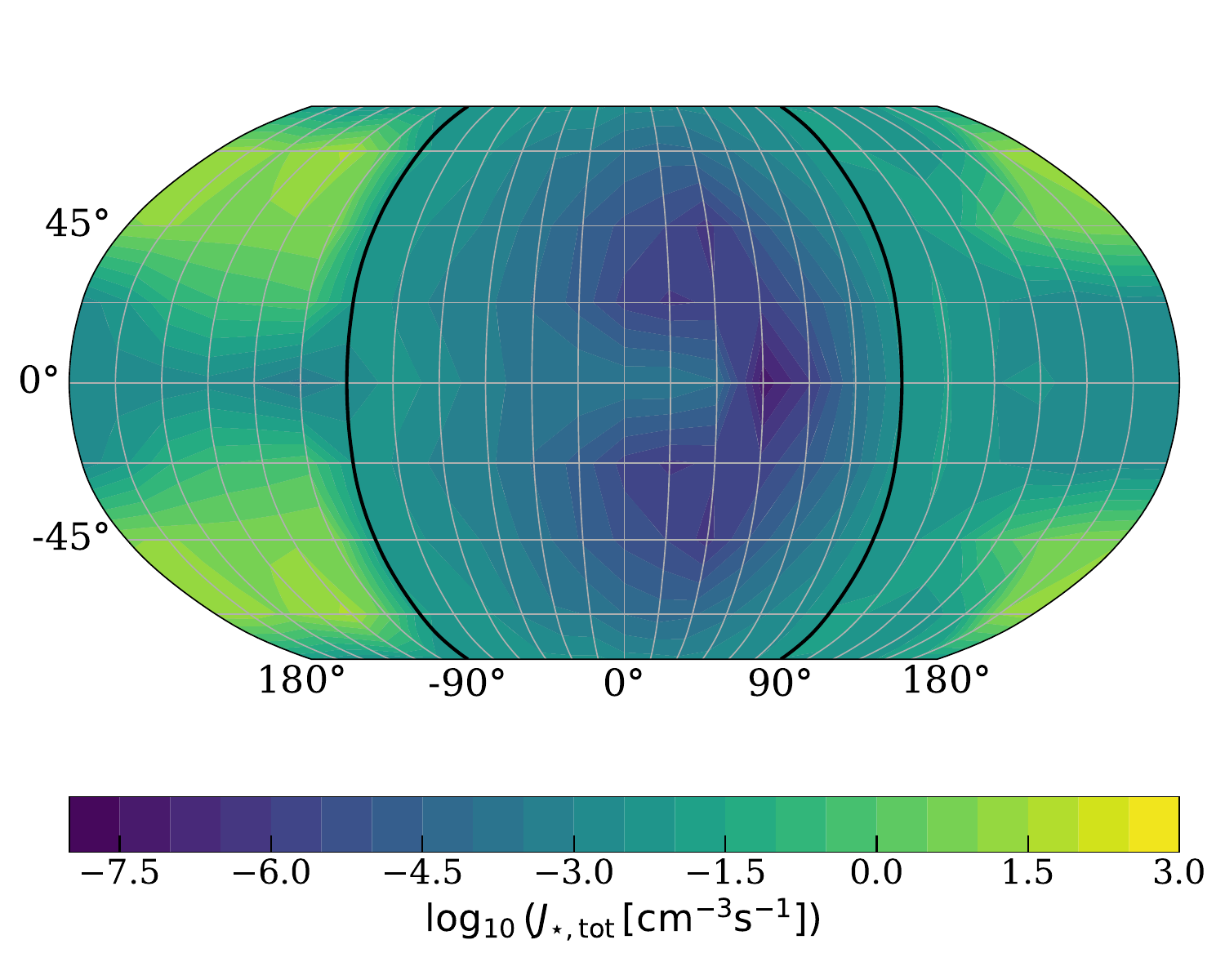}
      \includegraphics[width=0.49\textwidth]{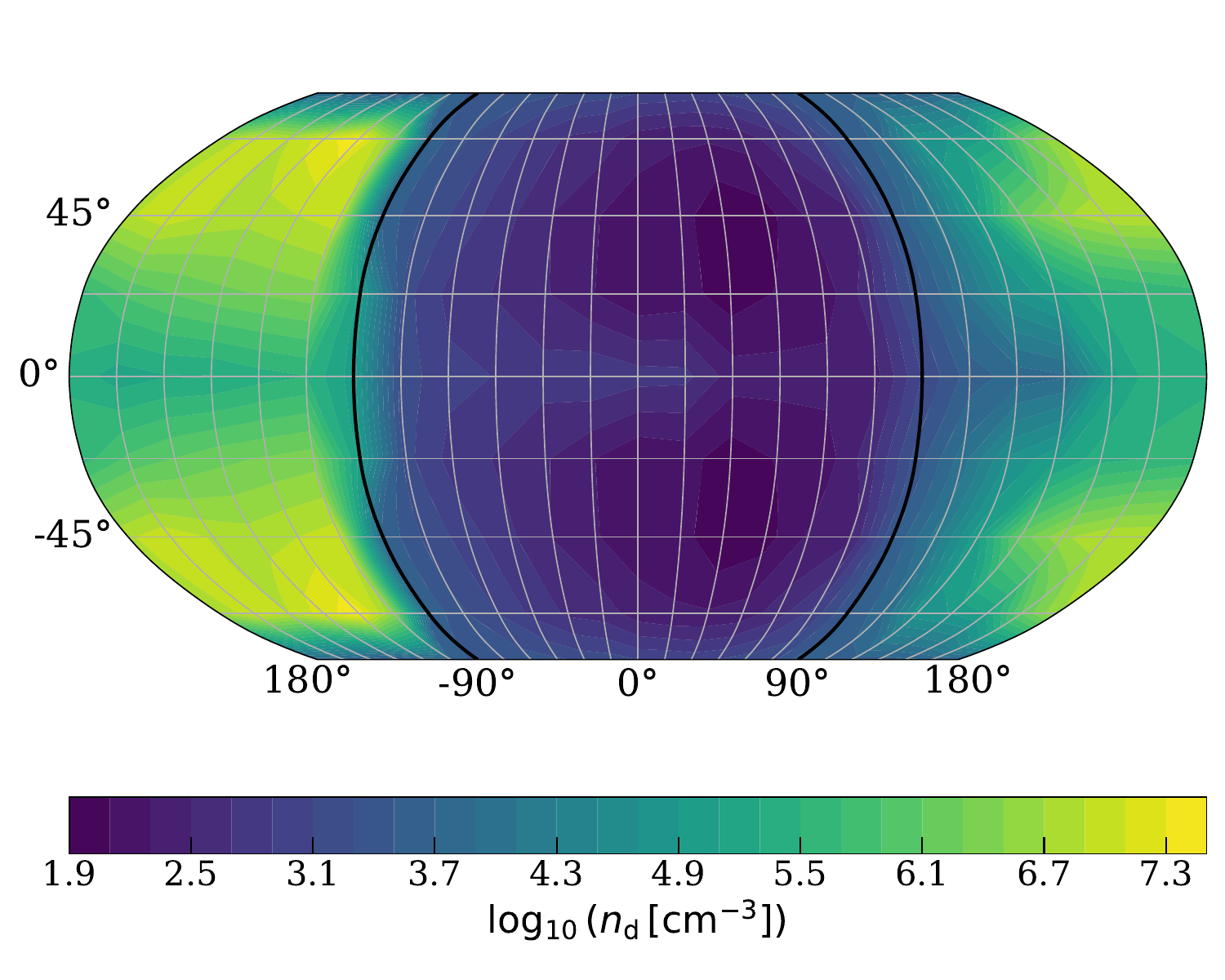}\\
      \includegraphics[width=0.49\textwidth]{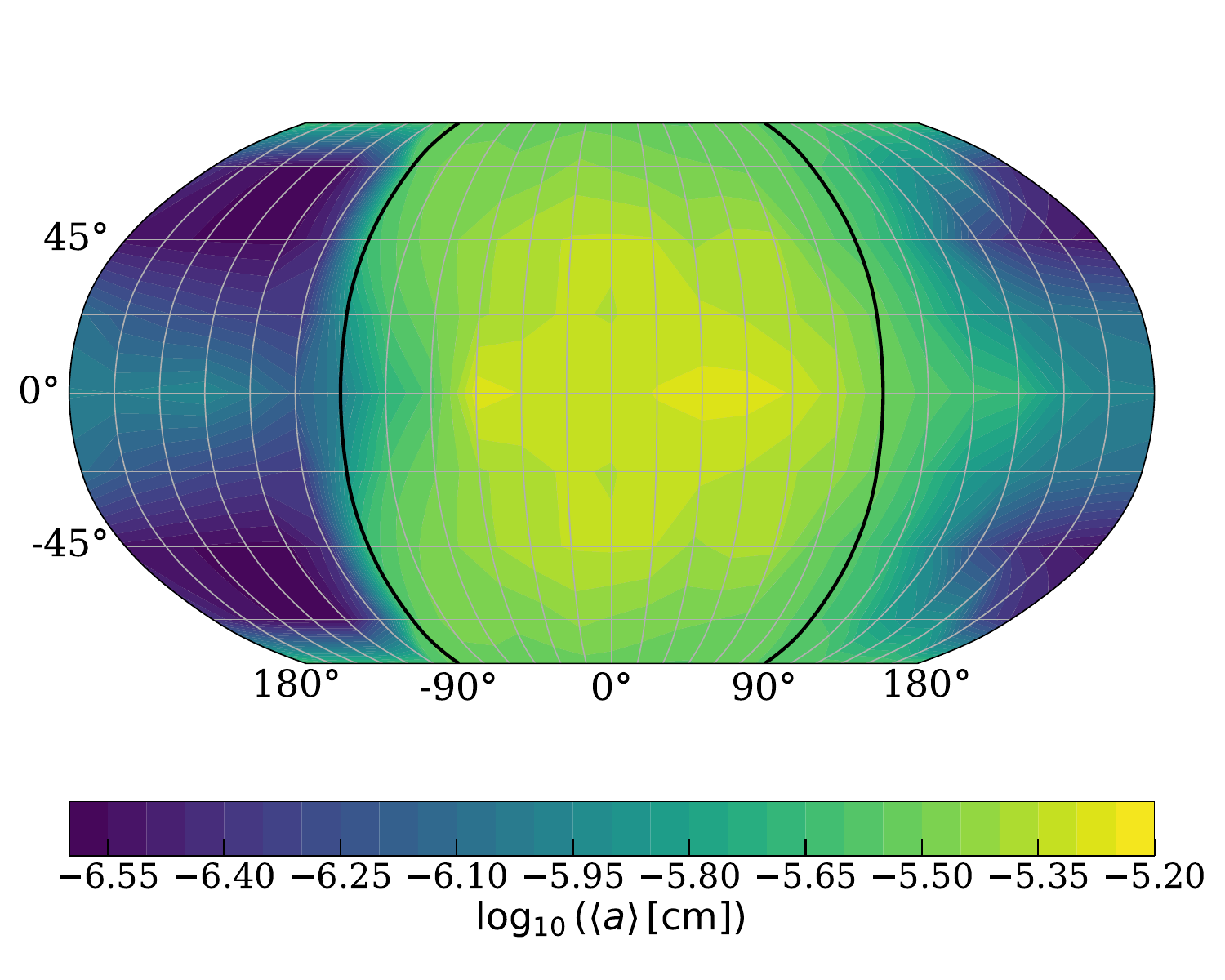}
      \includegraphics[width=0.49\textwidth]{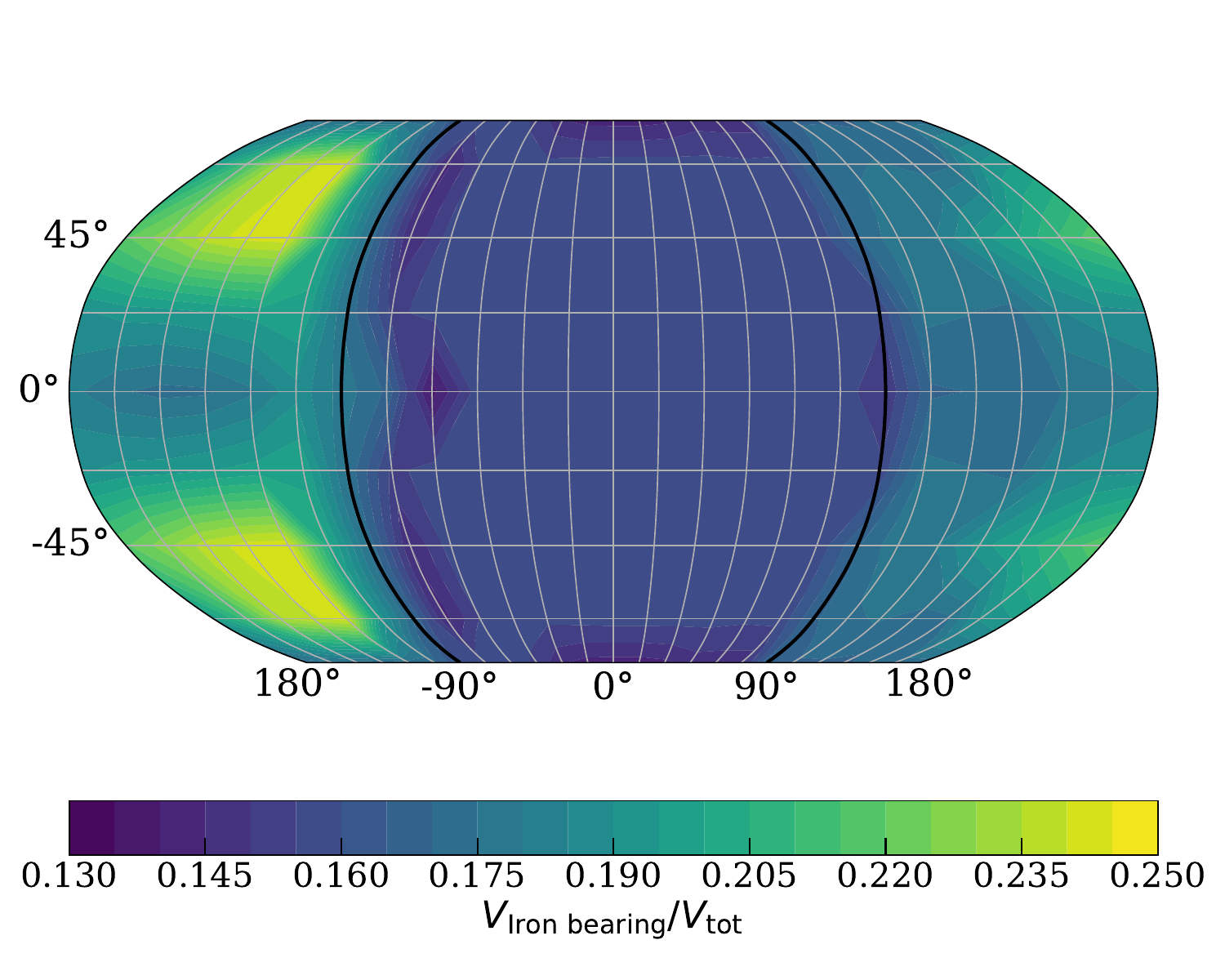}\\
      \caption{GCM and cloud formation results for WASP-43~b at $p_{\rm gas} = 0.1\,{\rm mbar}$. The maps are centred on the substellar point ($\phi_{\rm long} = 0\degree,\, \lambda_{\rm latt} = 0\degree$), with the terminator limbs ($\phi_{\rm long}=\pm 90\degree$) shown in bold black lines. \textbf{Top Left:} The local gas temperature from the GCM solution. \textbf{Top Right:} Cloud particle mass load in the atmosphere as the ratio of the density of cloud particle mass to the gas density. \textbf{Middle Left:} Total nucleation rate. \textbf{Middle Right:} Cloud particle number density. \textbf{Bottom Left:} Average cloud particle size. \textbf{Bottom Right:} Volume fraction of the cloud particles composed of iron-bearing condensate species.}
      \label{fig:W43b_solar_01mbar_maps}
  \end{figure*}

  \begin{figure*}
      \centering
      \includegraphics[width=0.49\textwidth]{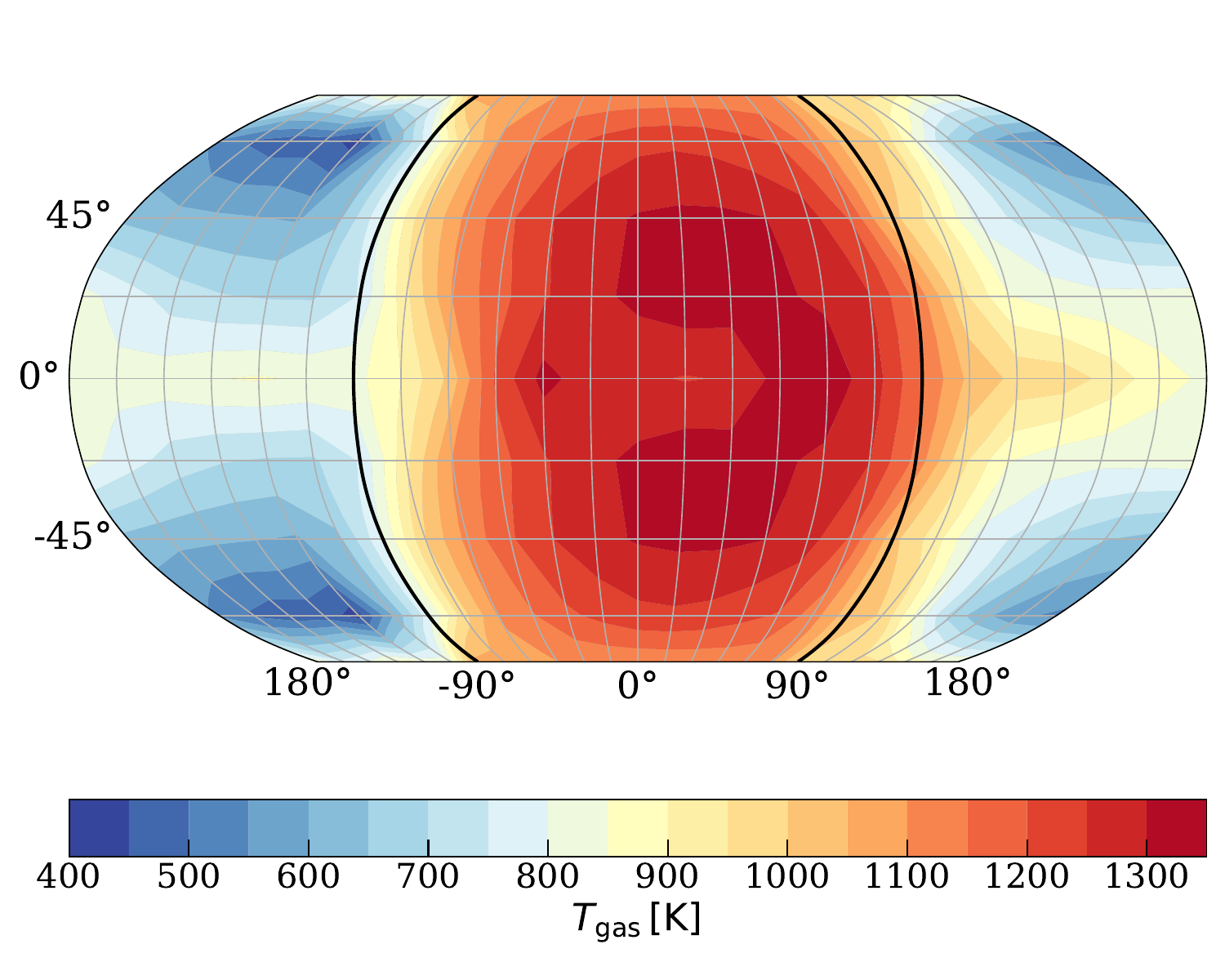}
      \includegraphics[width=0.49\textwidth]{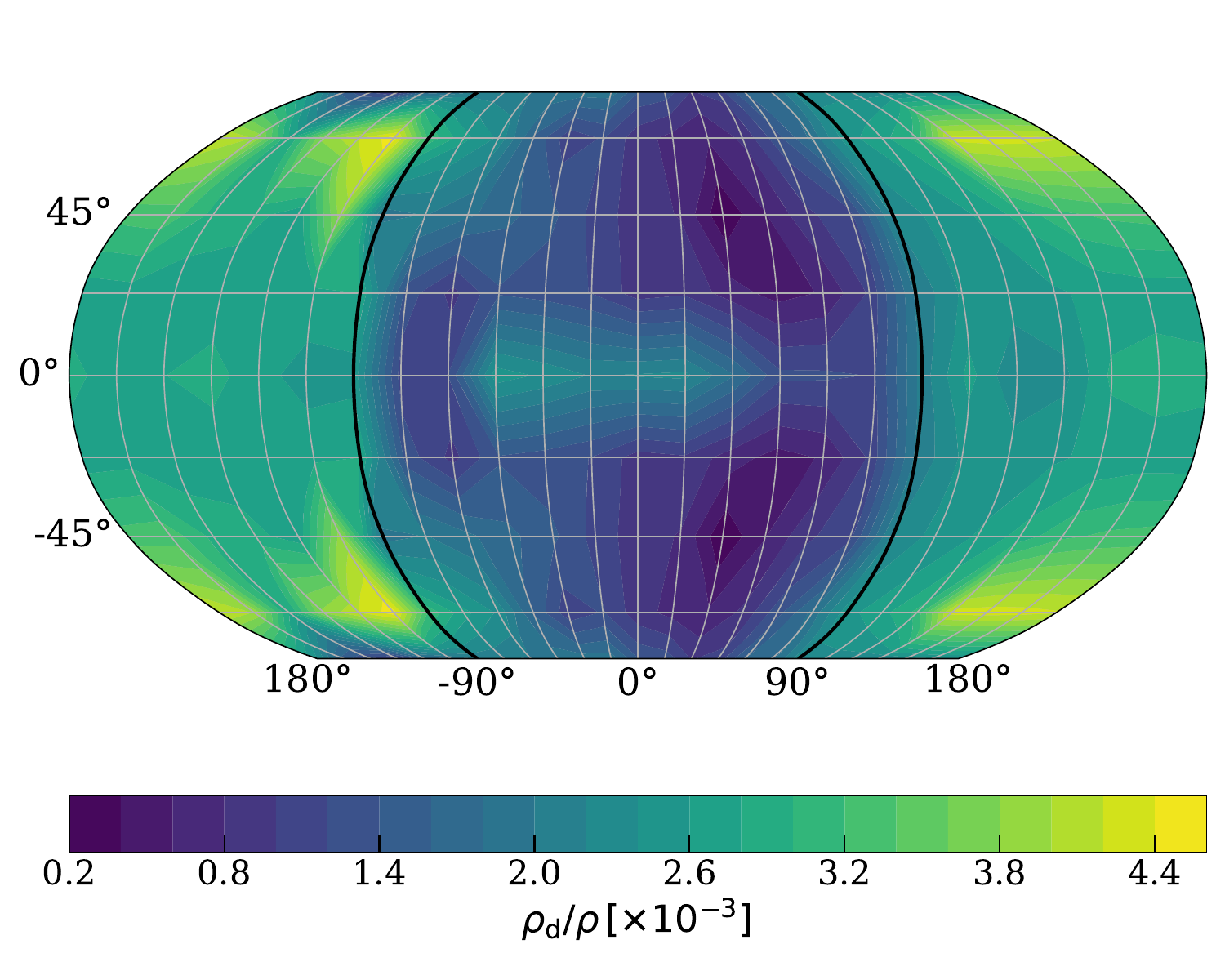}
      \includegraphics[width=0.49\textwidth]{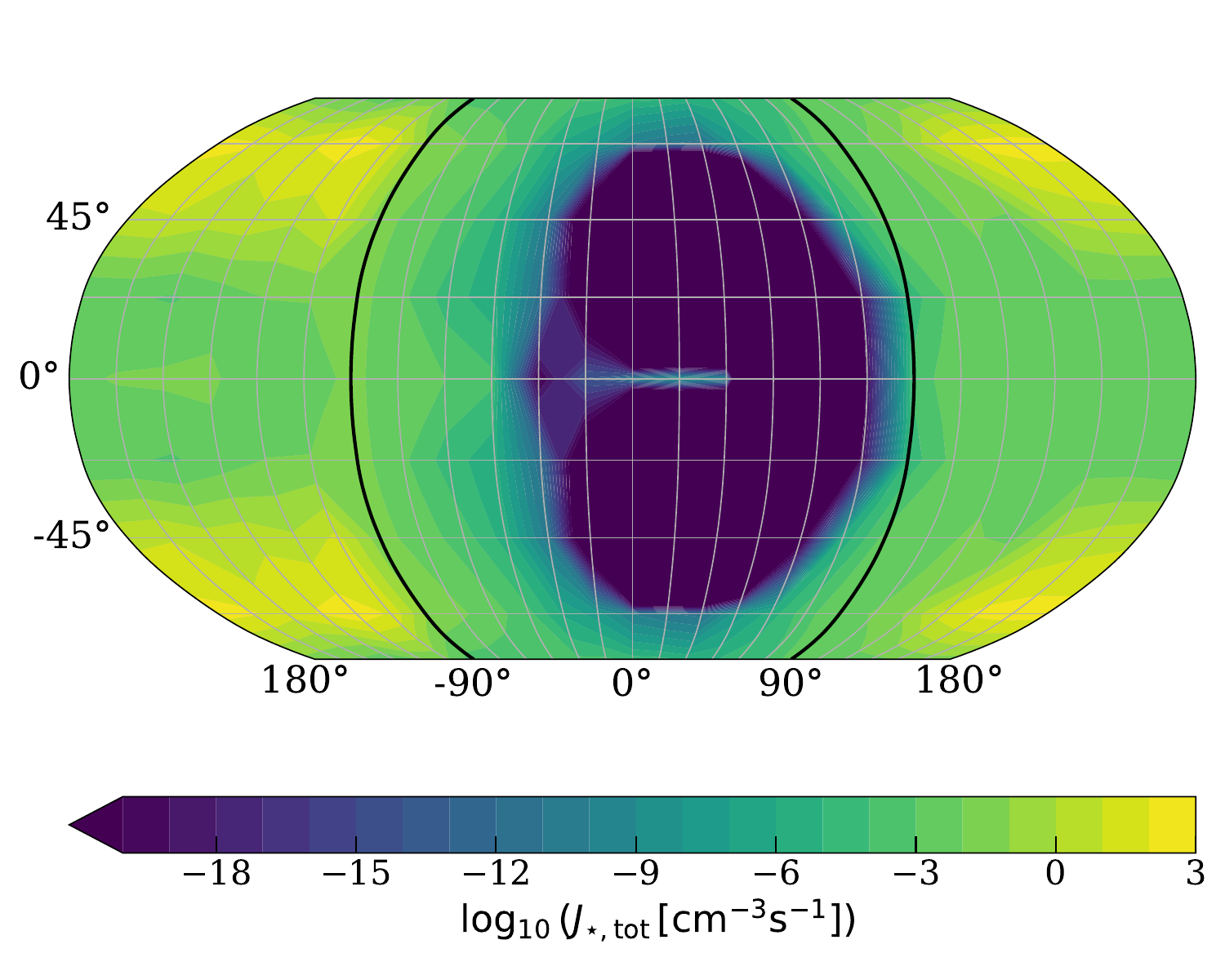}
      \includegraphics[width=0.49\textwidth]{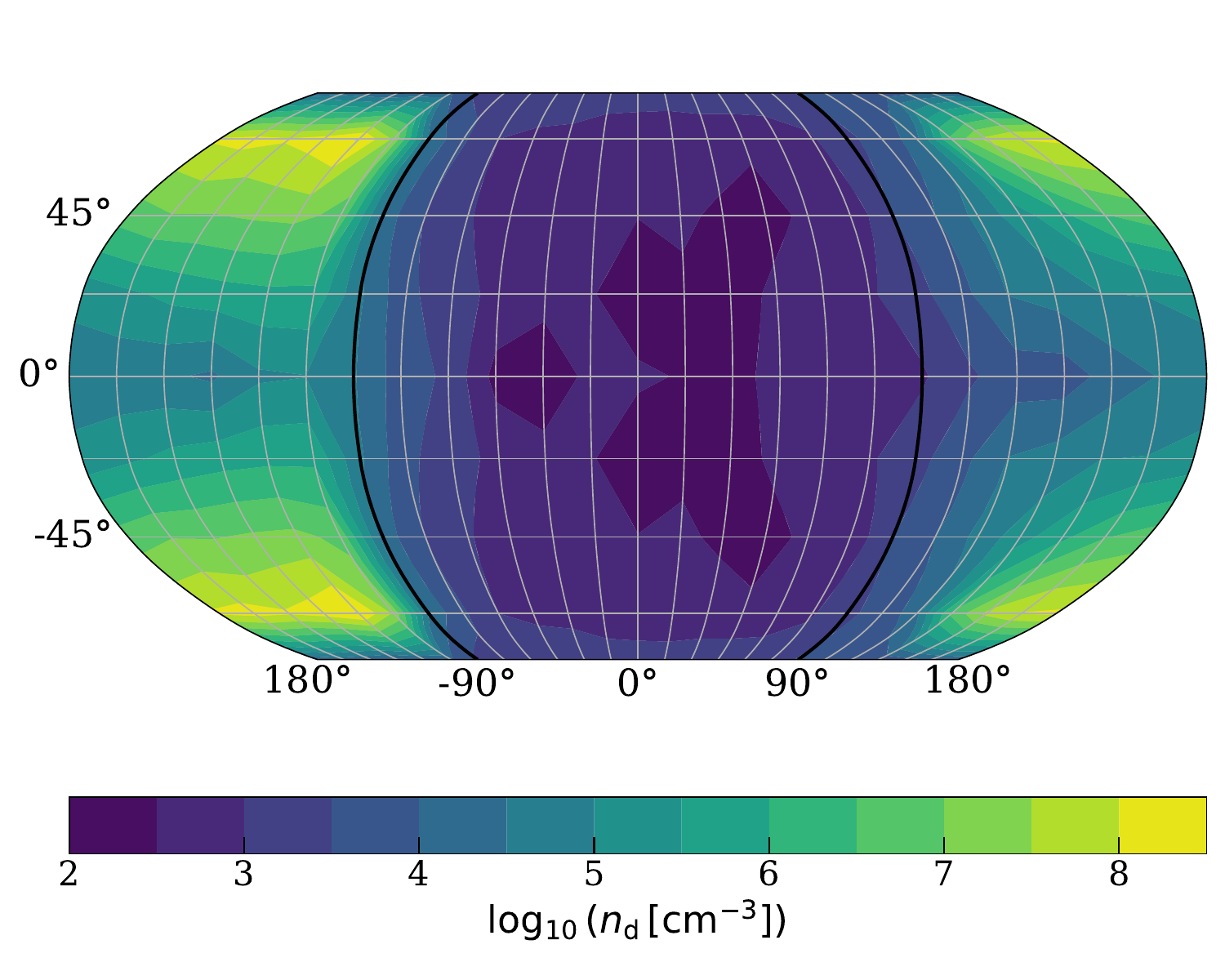}\\
      \includegraphics[width=0.49\textwidth]{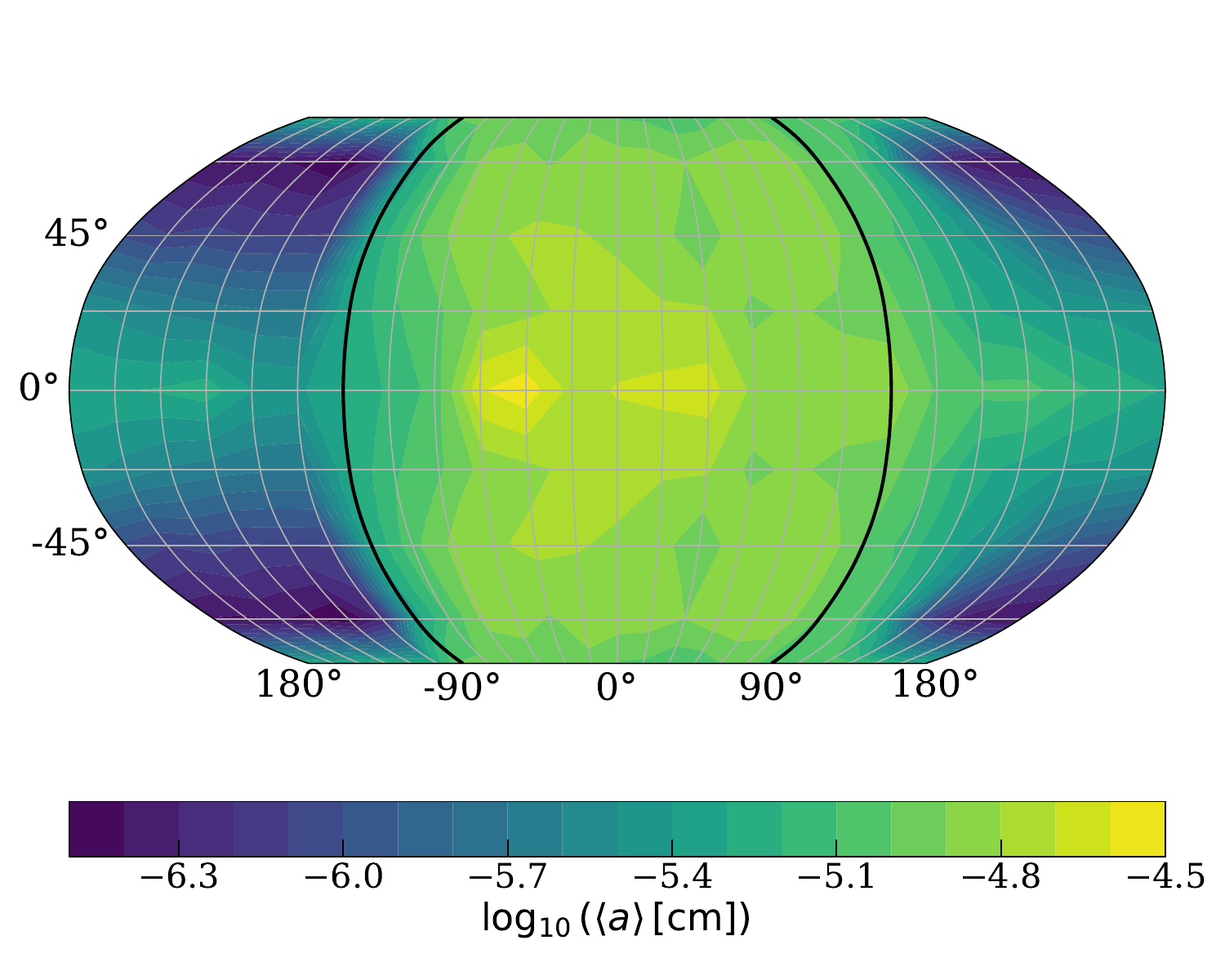}
      \includegraphics[width=0.49\textwidth]{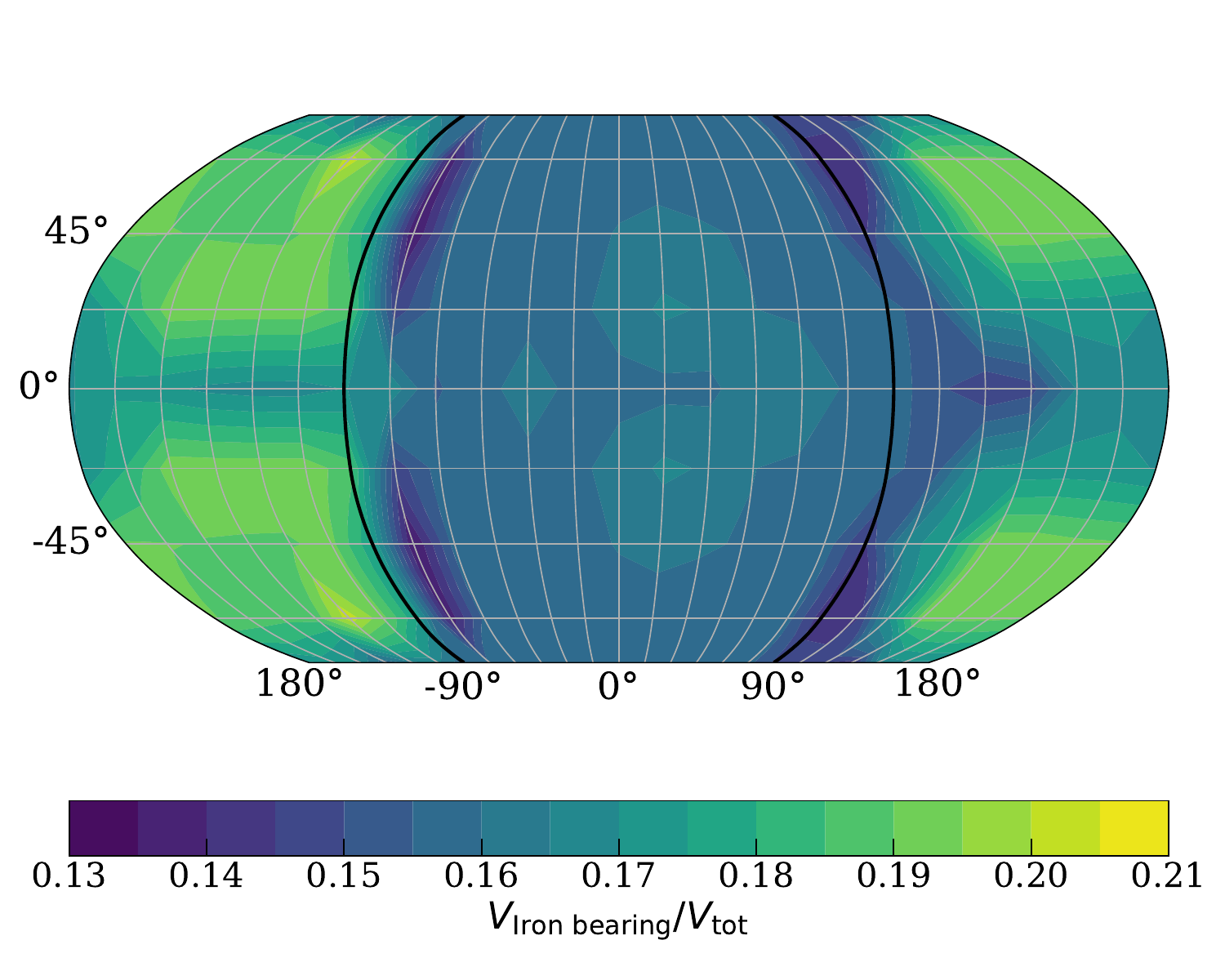}\\
      \caption{GCM and cloud formation results for WASP-43~b at $p_{\rm gas} = 1\,{\rm mbar}$. The maps are centred on the substellar point ($\phi_{\rm long} = 0\degree,\, \lambda_{\rm latt} = 0\degree$), with the terminator limbs ($\phi_{\rm long}=\pm 90\degree$) shown in bold black lines. \textbf{Top Left:} The local gas temperature from the GCM solution. \textbf{Top Right:} Cloud particle mass load in the atmosphere as the ratio of the density of cloud particle mass to the gas density. \textbf{Middle Left:} Total nucleation rate. \textbf{Middle Right:} Cloud particle number density. \textbf{Bottom Left:} Average cloud particle size. \textbf{Bottom Right:} Volume fraction of the cloud particles composed of iron-bearing condensate species.}
      \label{fig:W43b_solar_mbar_maps}
  \end{figure*}

\begin{figure*}
      \centering
      \includegraphics[width=0.49\textwidth]{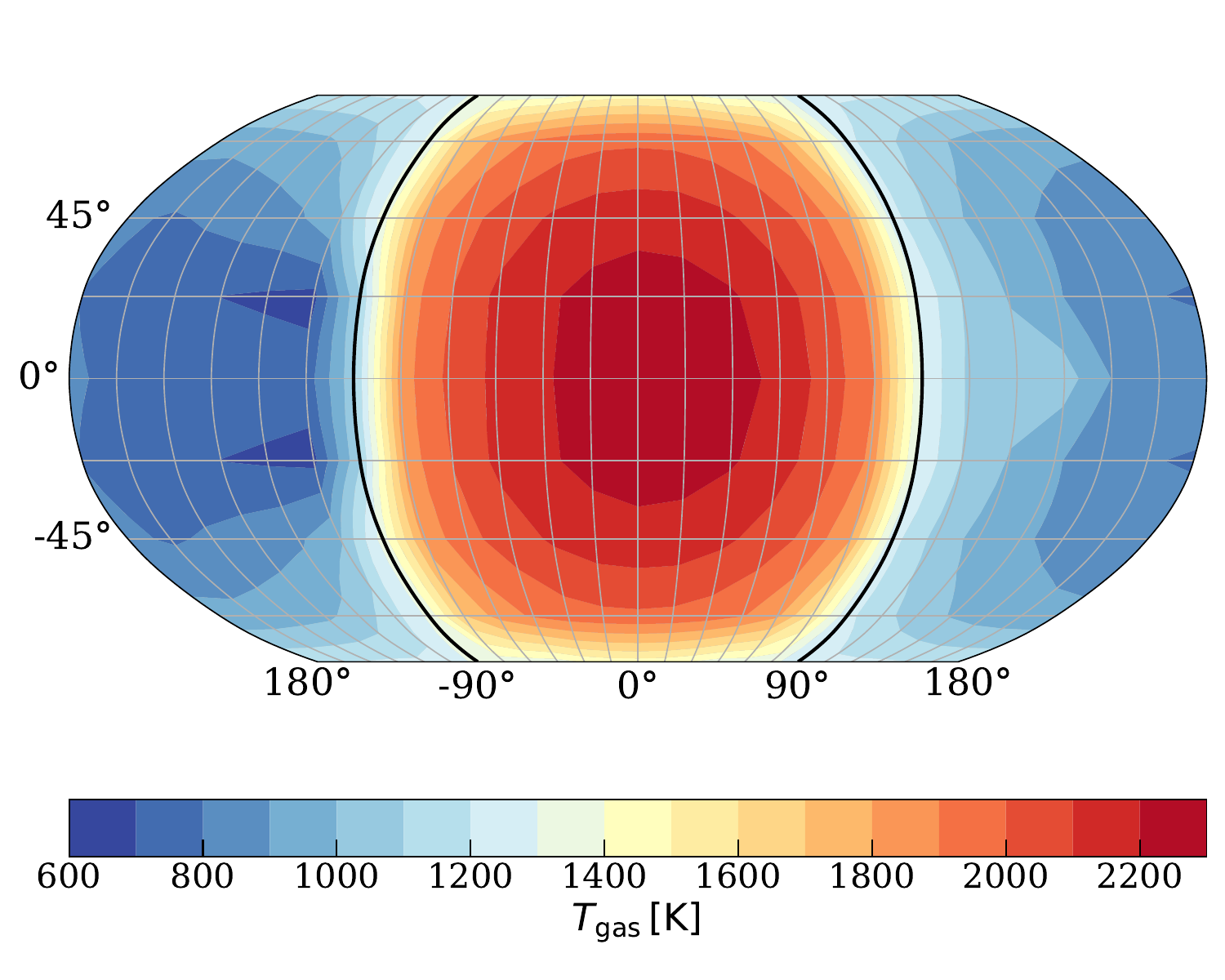}
      \includegraphics[width=0.49\textwidth]{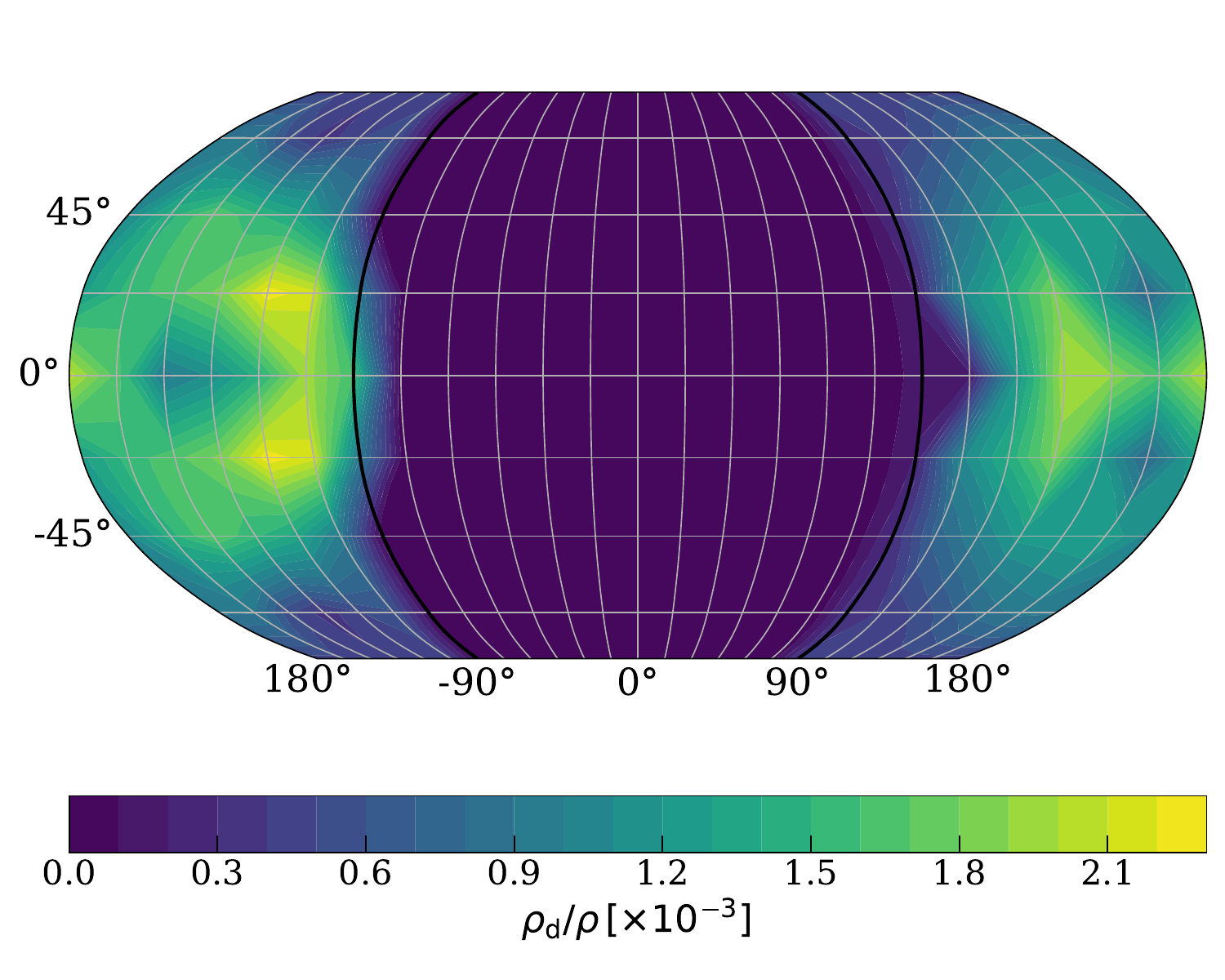}\\
      \includegraphics[width=0.49\textwidth]{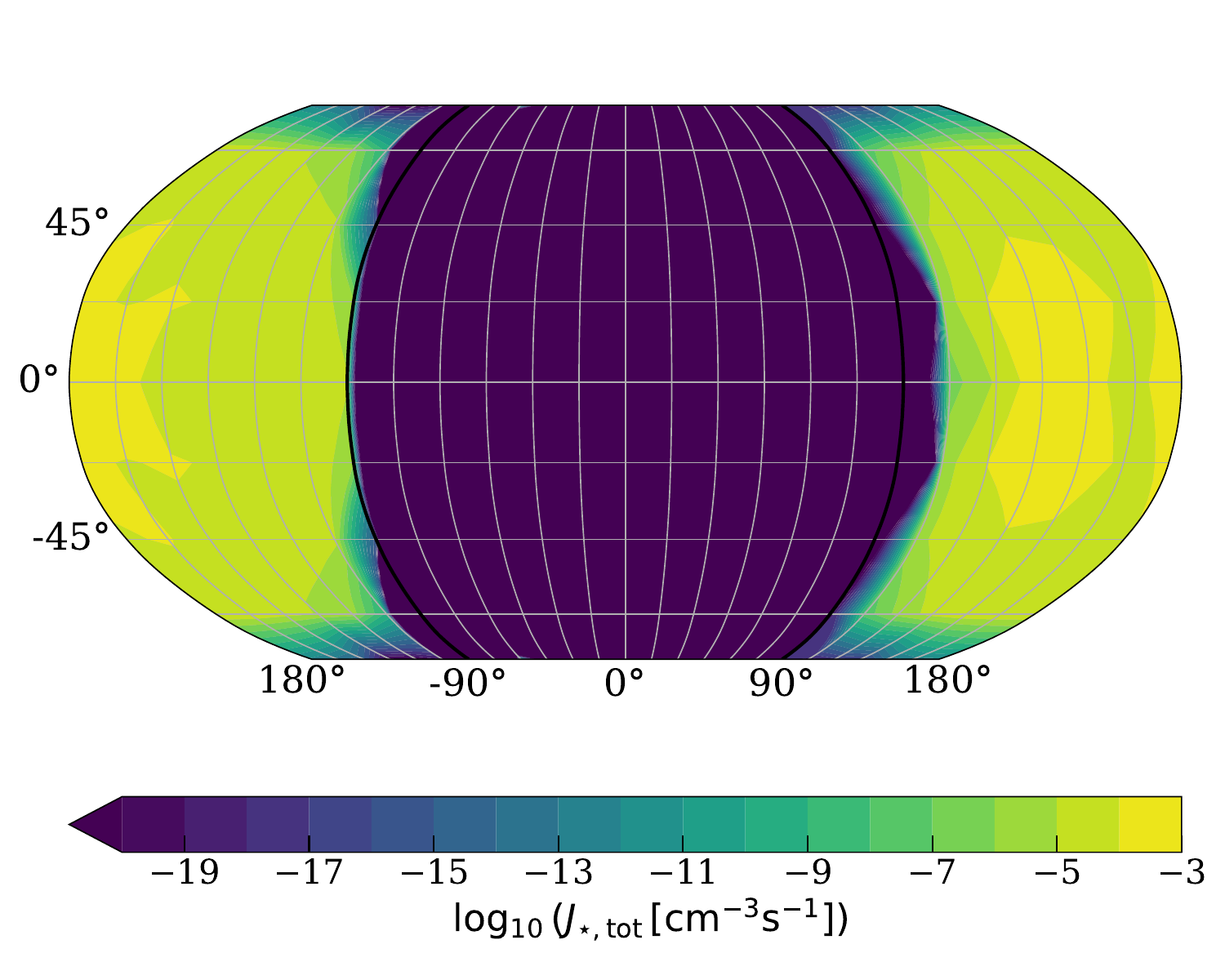}
      \includegraphics[width=0.49\textwidth]{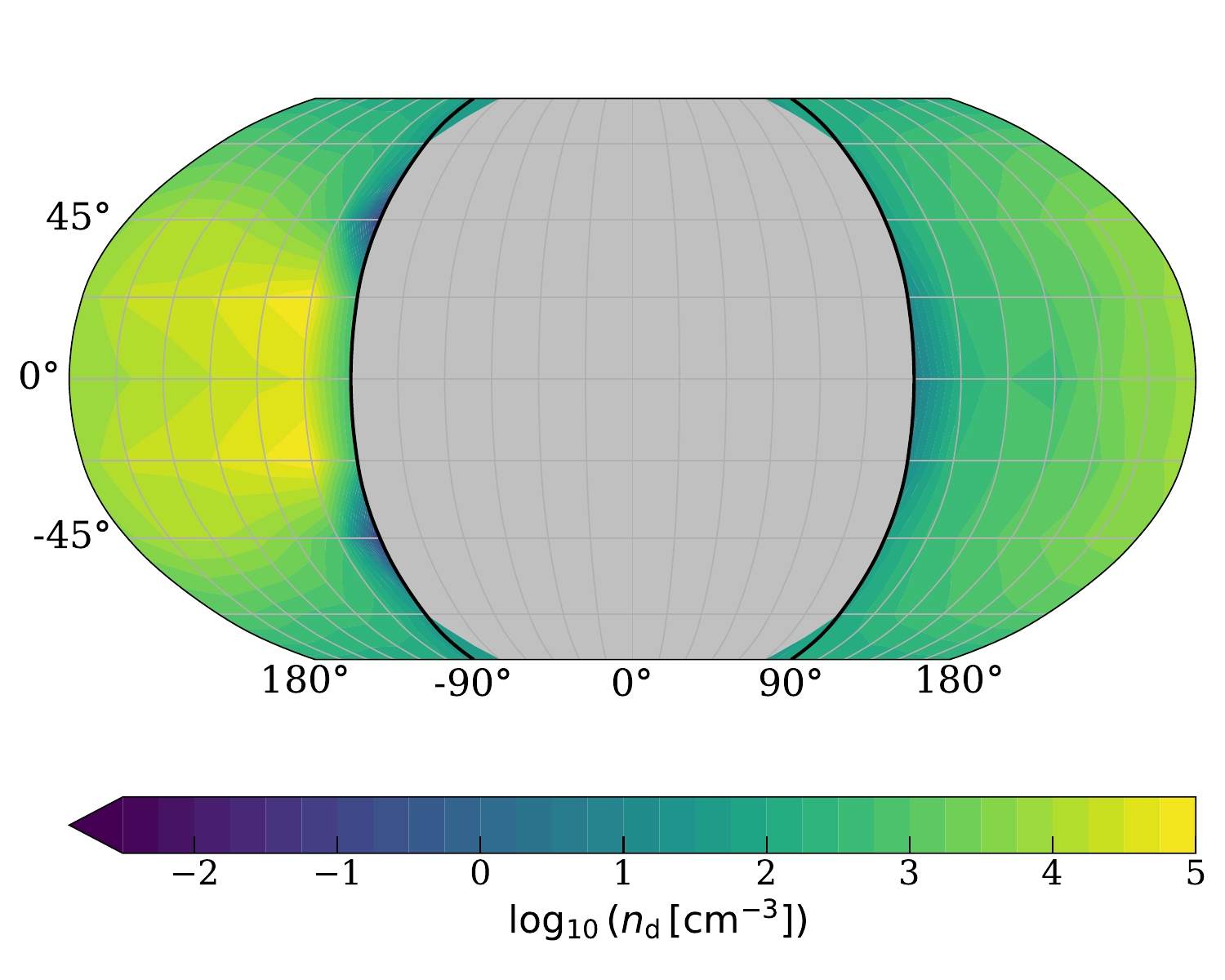}\\
      \includegraphics[width=0.49\textwidth]{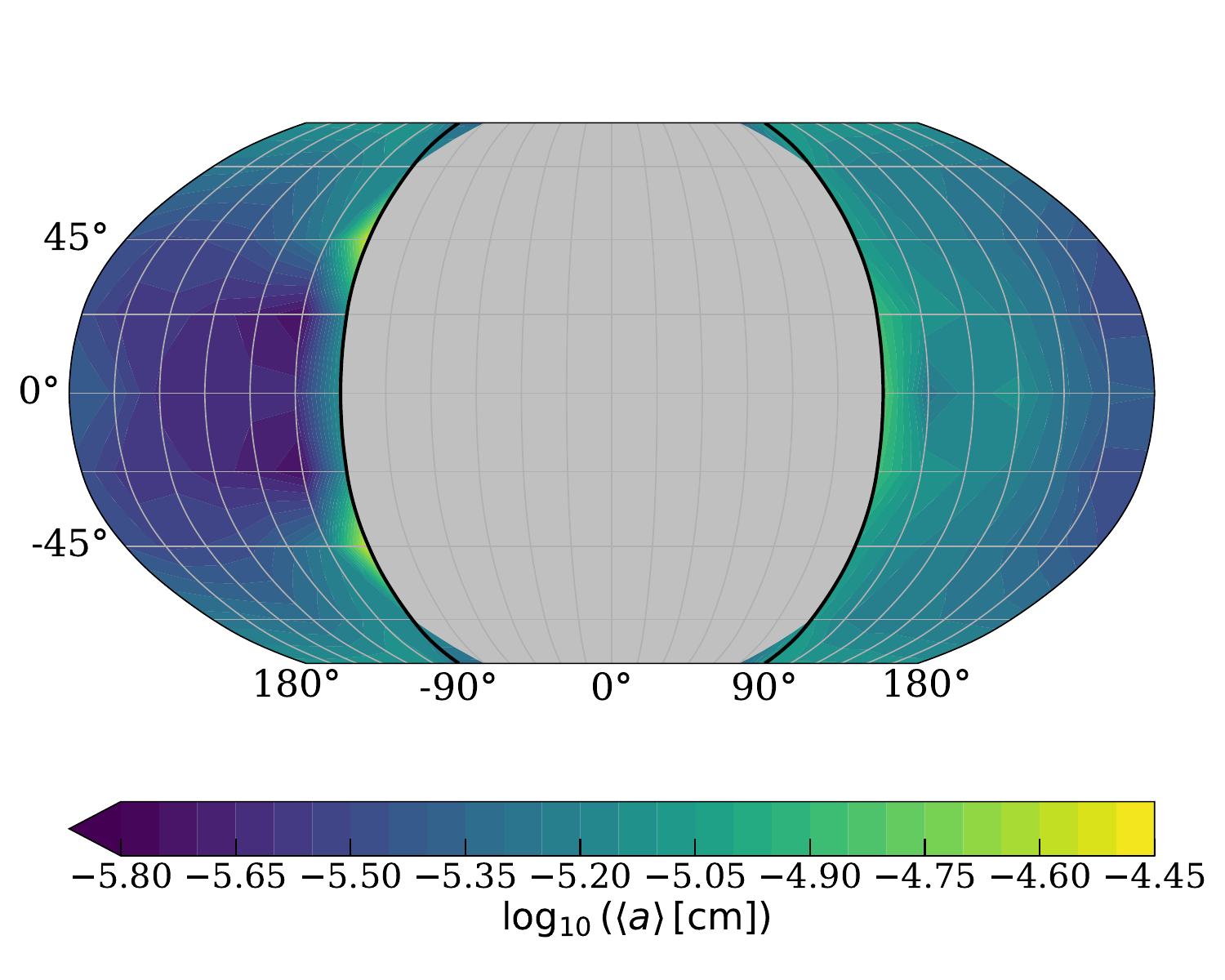}
      \includegraphics[width=0.49\textwidth]{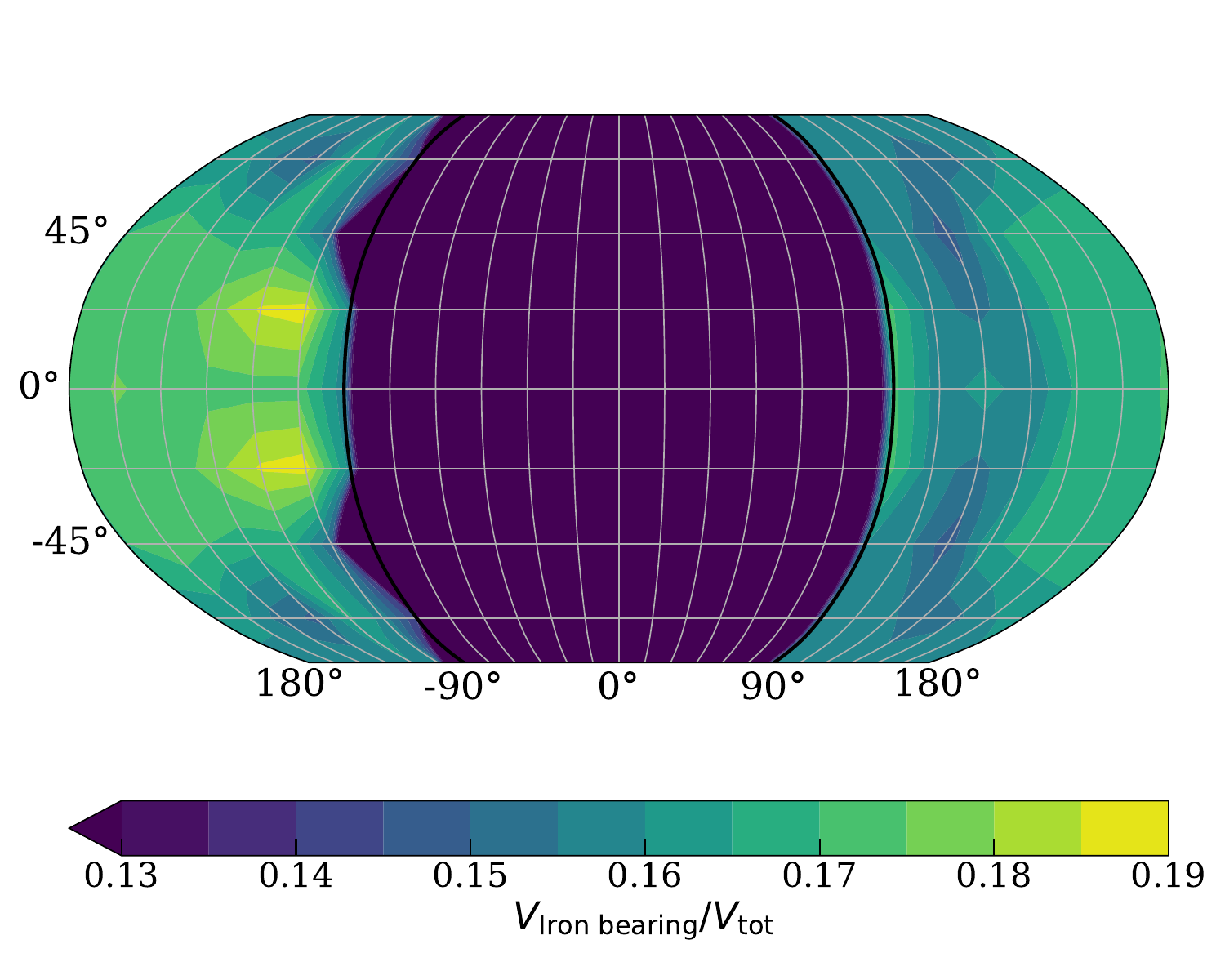}
      \caption{GCM and cloud formation results for HD~209458~b at $p_{\rm gas} = 0.1\,{\rm mbar}$. The maps are centred on the substellar point ($\phi_{\rm long} = 0\degree,\, \lambda_{\rm latt} = 0\degree$), with the terminator limbs ($\phi_{\rm long}=\pm 90\degree$) shown in bold black lines. \textbf{Top Left:} The local gas temperature from the GCM solution. \textbf{Top Right:} Cloud particle mass load in the atmosphere as the ratio of the density of cloud particle mass to the gas density. \textbf{Middle Left:} Total nucleation rate. \textbf{Middle Right:} Cloud particle number density. \textbf{Bottom Left:} Average cloud particle size. \textbf{Bottom Right:} Volume fraction of the cloud particles composed of iron-bearing condensate species.}
      \label{fig:HD209458b_01mbar_maps}
  \end{figure*}

    \begin{figure*}
      \centering
      \includegraphics[width=0.49\textwidth]{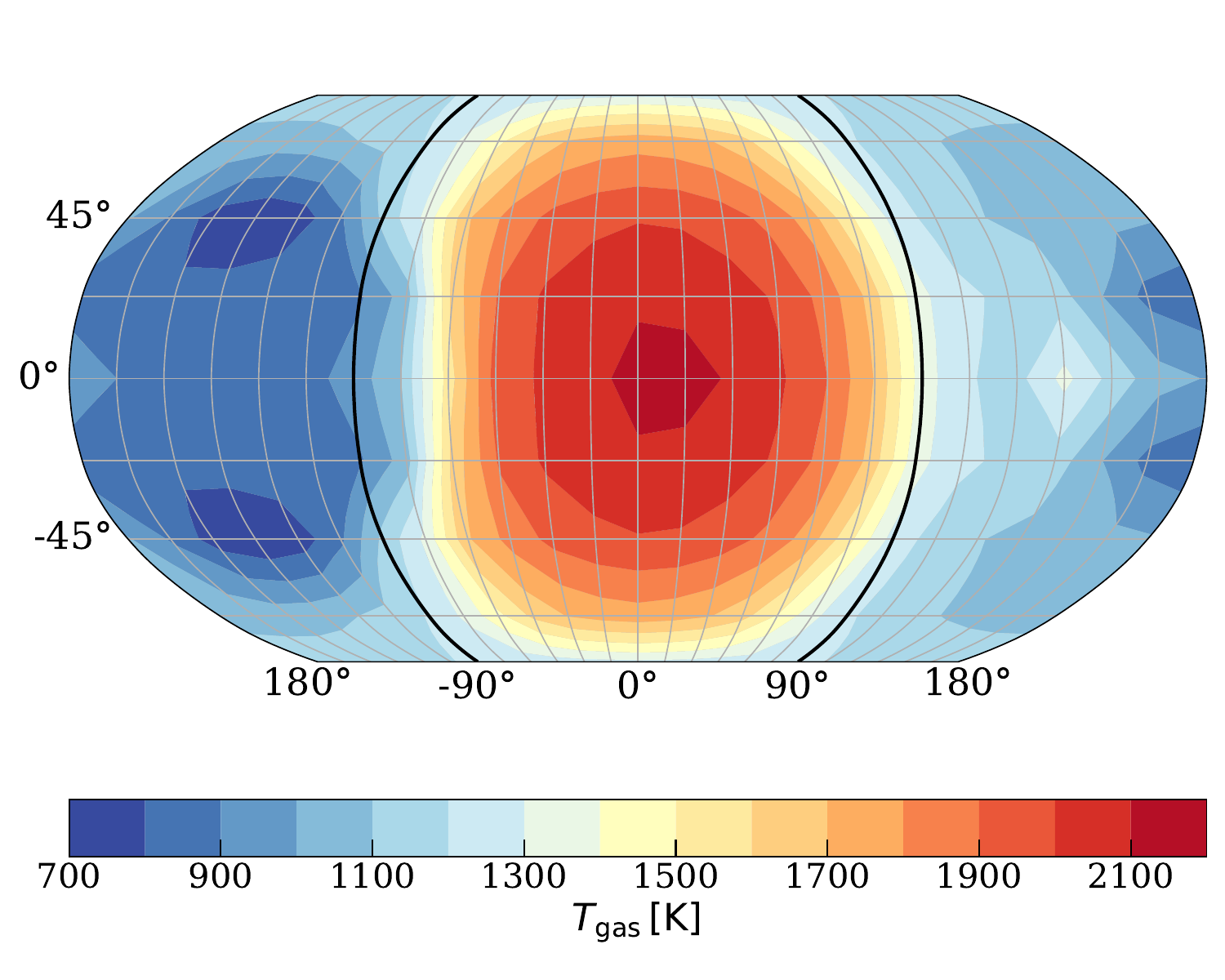}
      \includegraphics[width=0.49\textwidth]{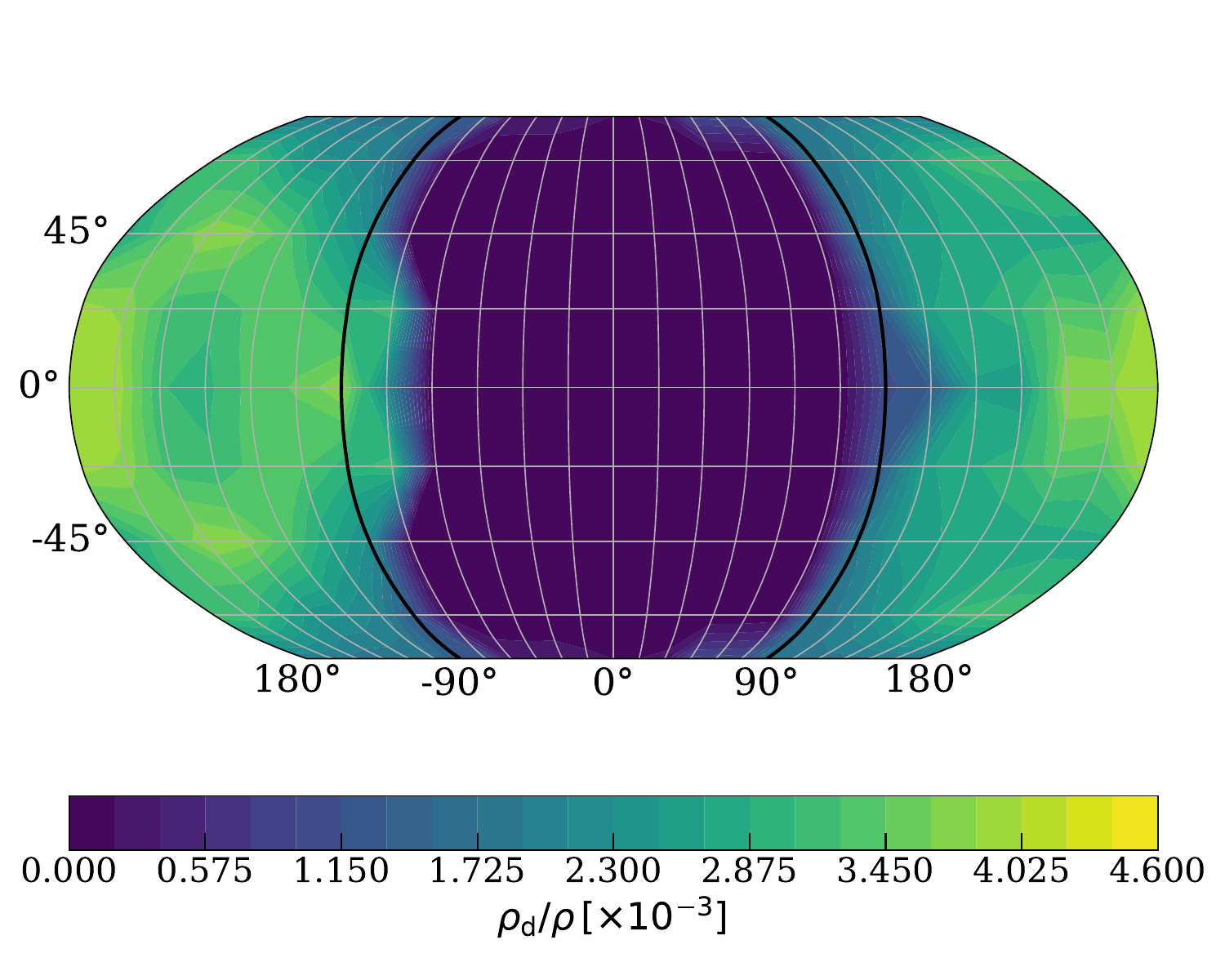}\\
      \includegraphics[width=0.49\textwidth]{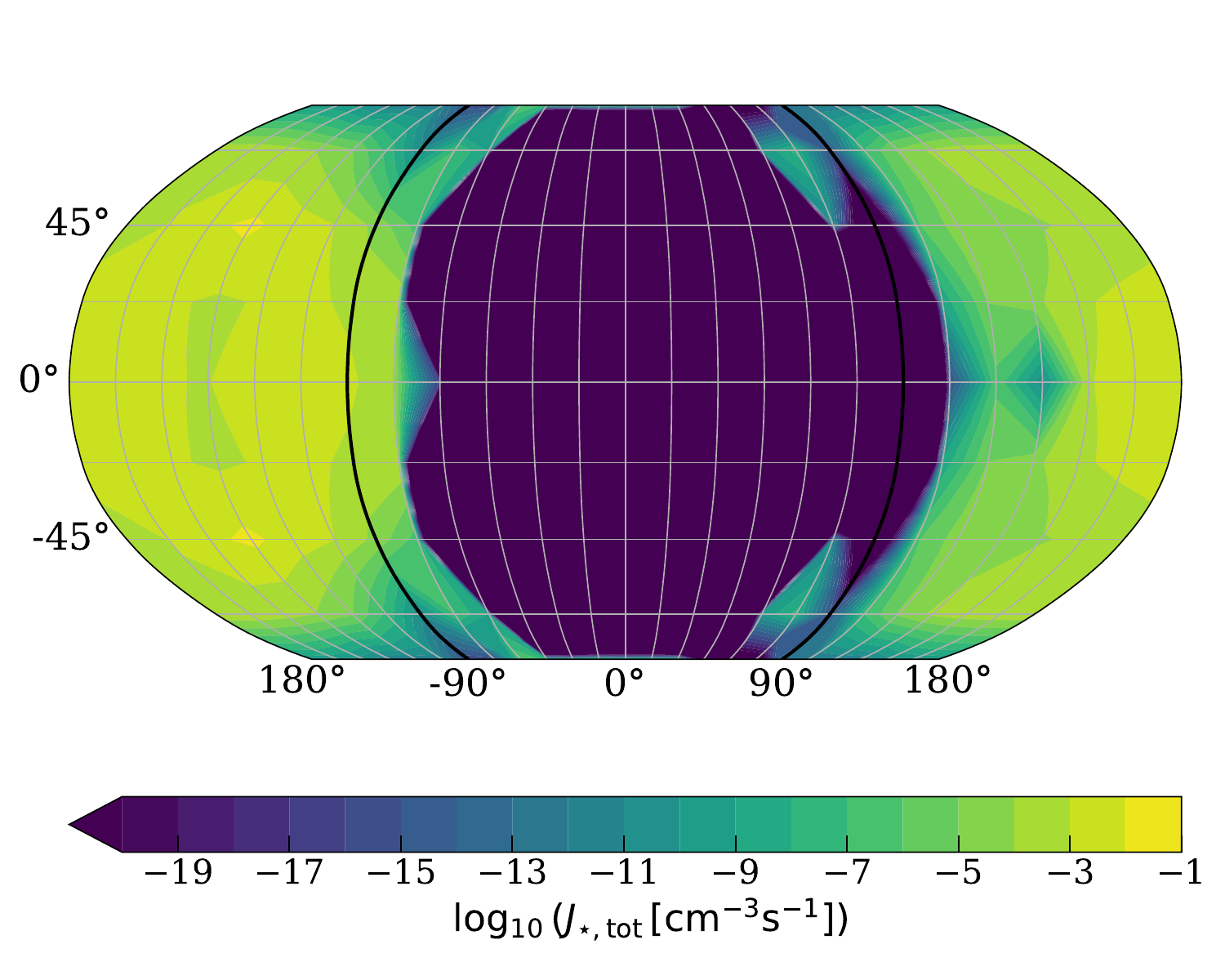}
      \includegraphics[width=0.49\textwidth]{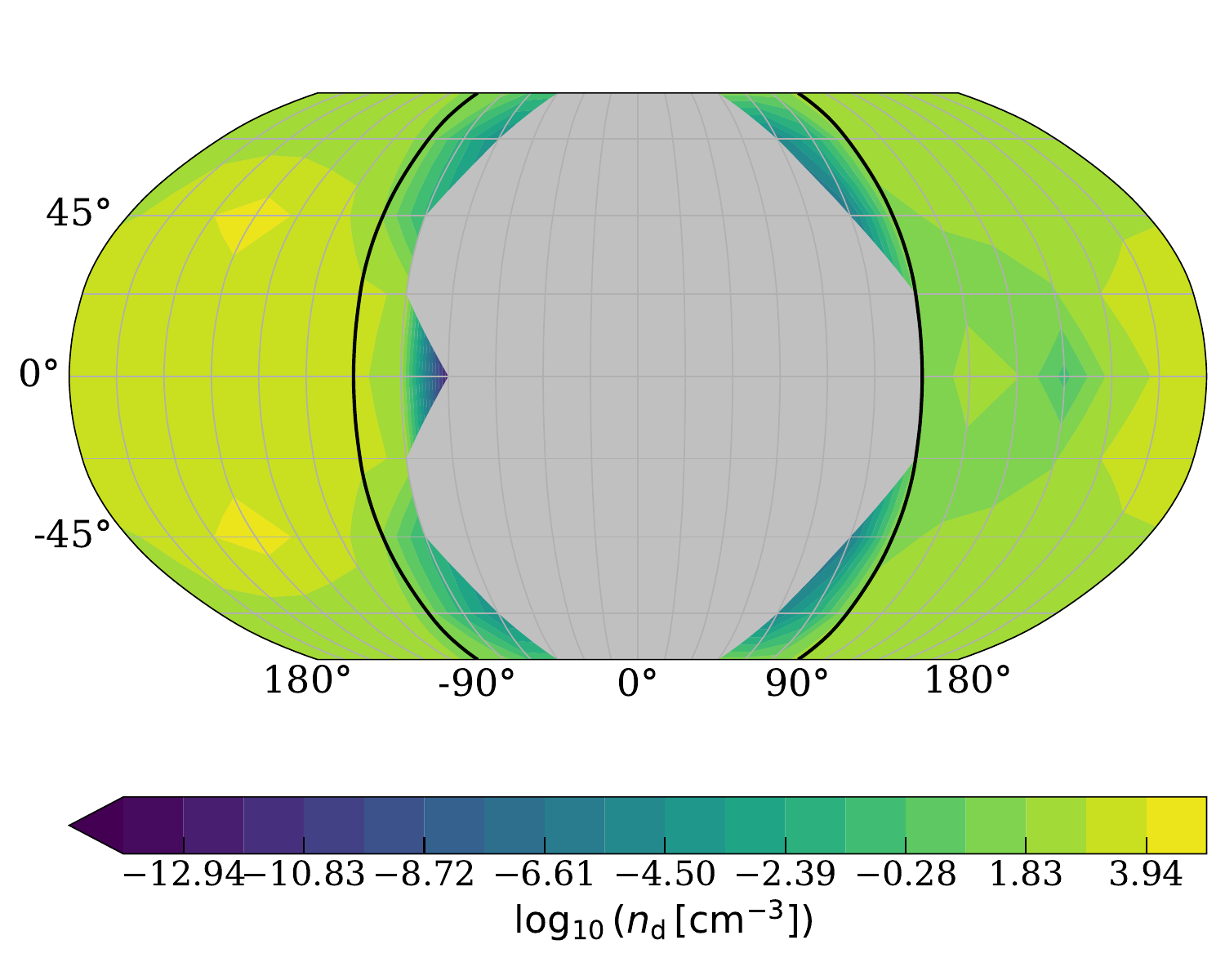}\\
      \includegraphics[width=0.49\textwidth]{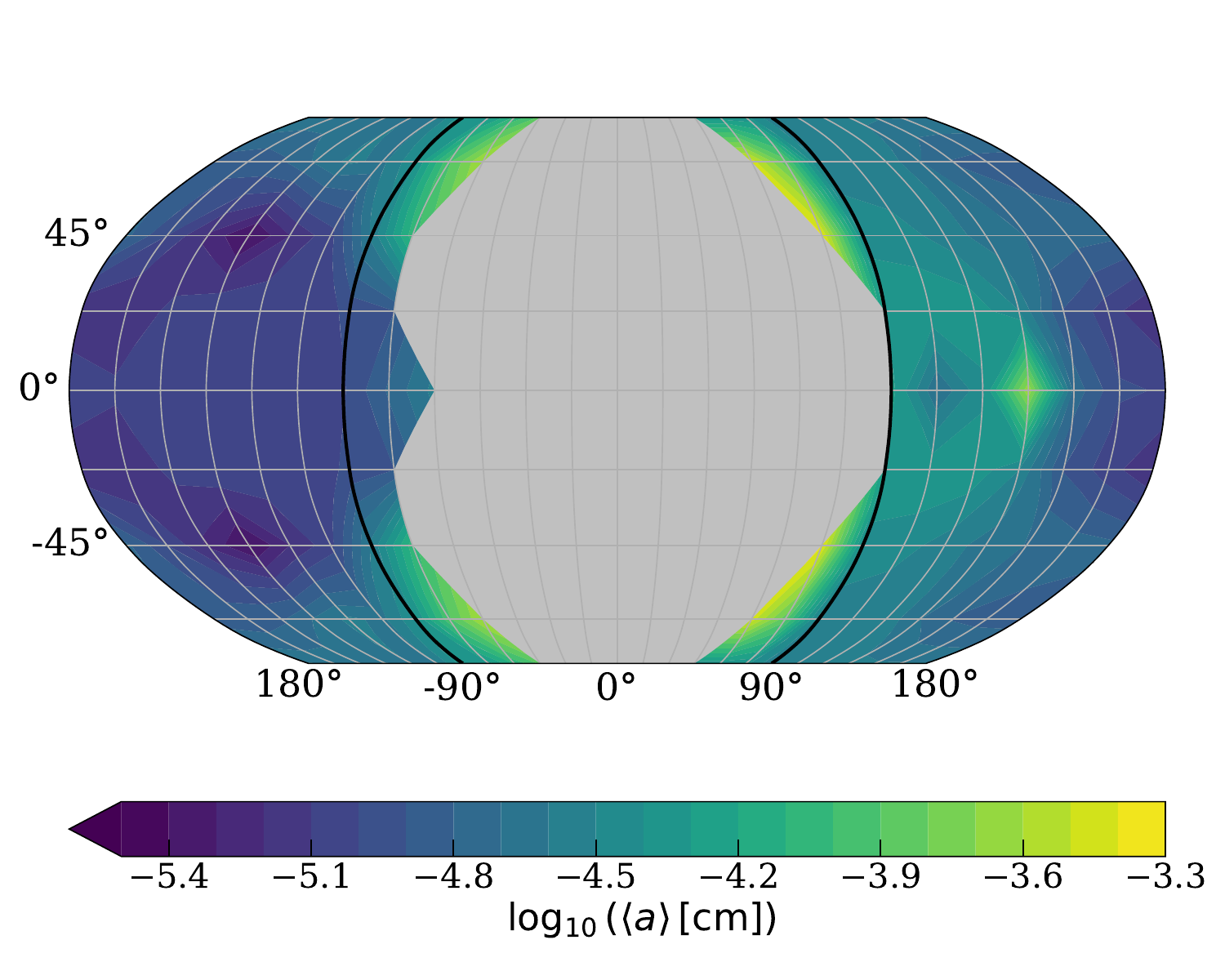}
      \includegraphics[width=0.49\textwidth]{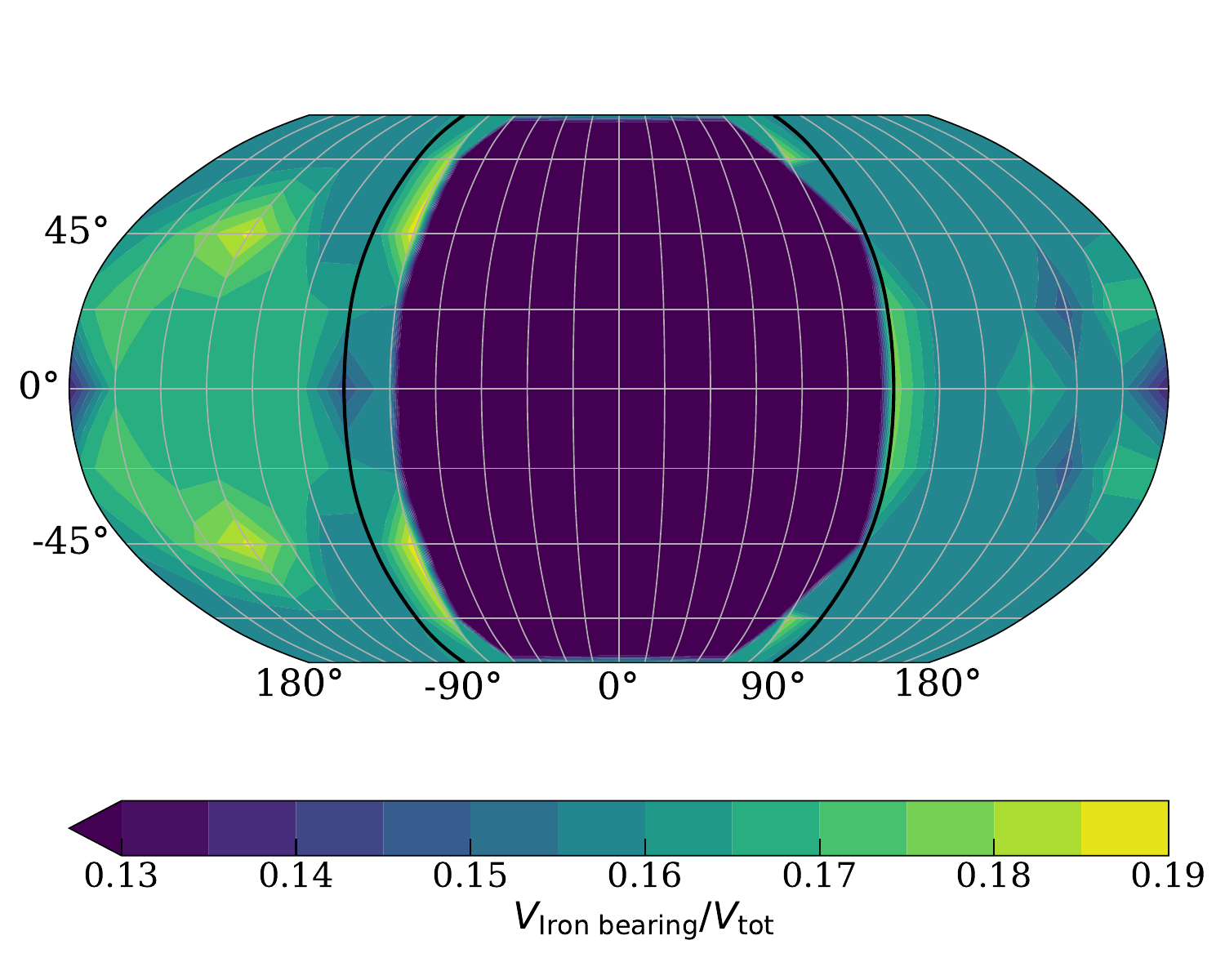}
      \caption{GCM and cloud formation results for HD~209458~b at $p_{\rm gas} = 1\,{\rm mbar}$. The maps are centred on the substellar point ($\phi_{\rm long} = 0\degree,\, \lambda_{\rm latt} = 0\degree$), with the terminator limbs ($\phi_{\rm long}=\pm 90\degree$) shown in bold black lines. \textbf{Top Left:} The local gas temperature from the GCM solution. \textbf{Top Right:} Cloud particle mass load in the atmosphere as the ratio of the density of cloud particle mass to the gas density. \textbf{Middle Left:} Total nucleation rate. \textbf{Middle Right:} Cloud particle number density. \textbf{Bottom Left:} Average cloud particle size. \textbf{Bottom Right:} Volume fraction of the cloud particles composed of iron-bearing condensate species.}
      \label{fig:HD209458b_mbar_maps}
  \end{figure*}

\begin{figure*}
	\centering    
\includegraphics[width=0.45\textwidth]{Figures/PolHEx/cloud_files_overleaf/cloudstructure_mixingratios_out3_dust_WASP43b_solar_scaledmixing_all_species_log_A_colours.pdf}
\includegraphics[width=0.45\textwidth]{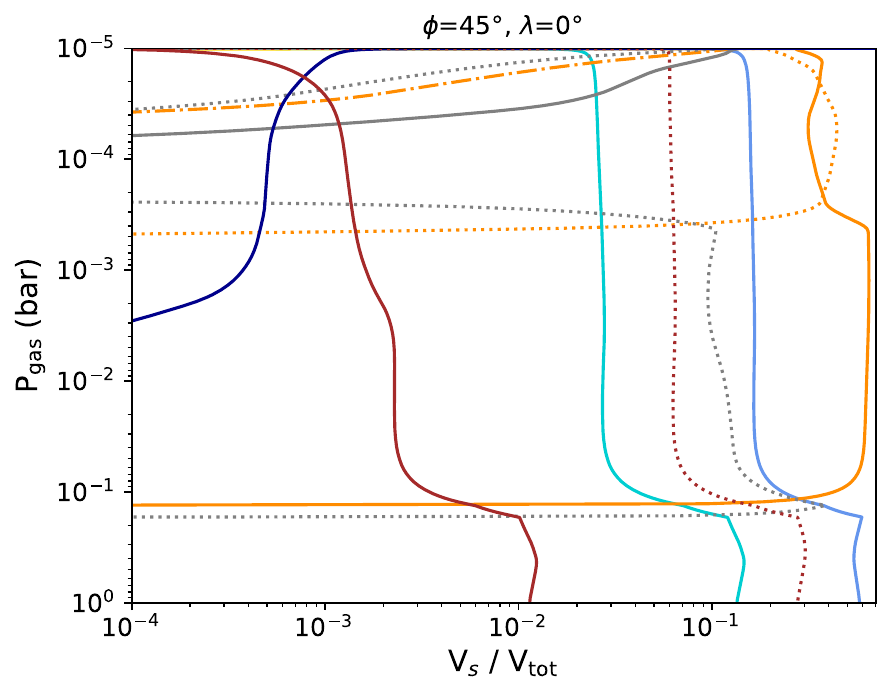}
\includegraphics[width=0.45\textwidth]{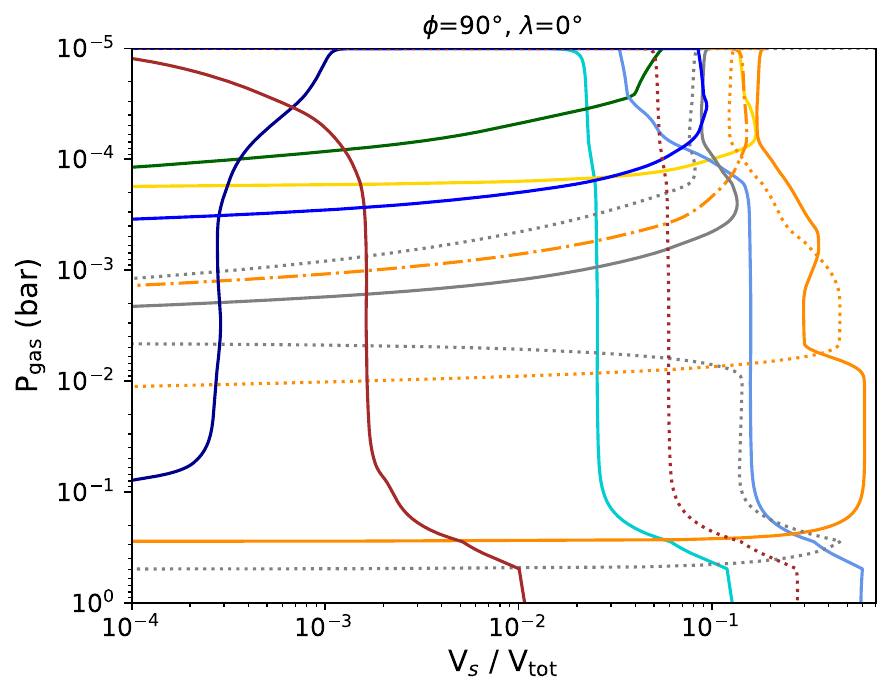}
\includegraphics[width=0.45\textwidth]{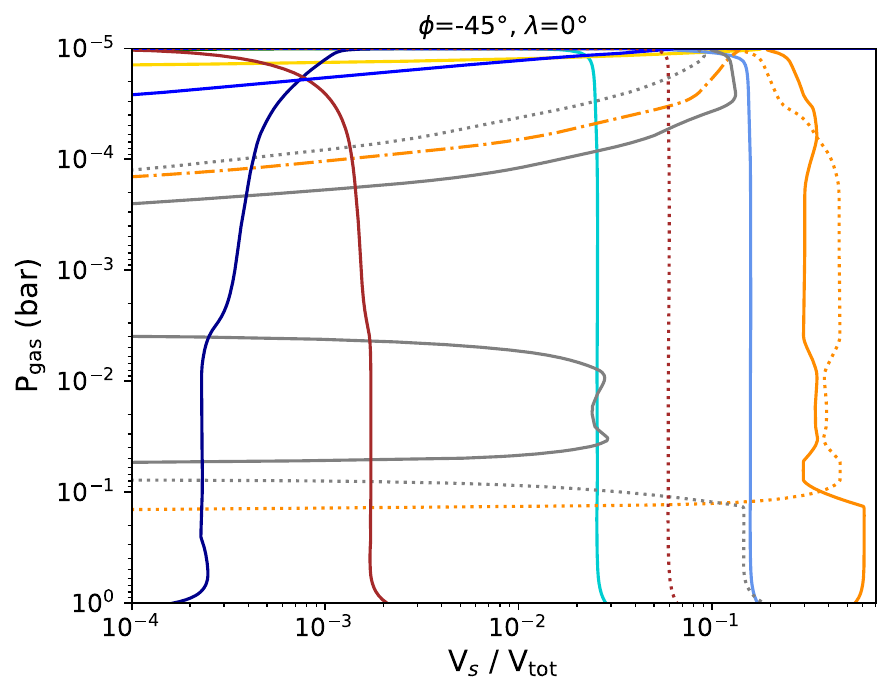}
\includegraphics[width=0.45\textwidth]{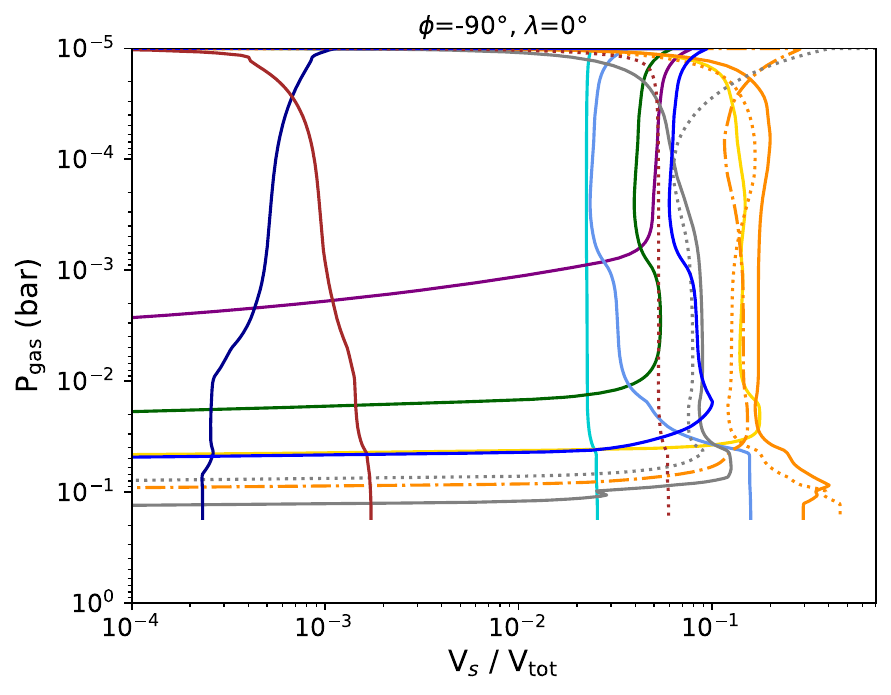}
\includegraphics[width=0.45\textwidth]{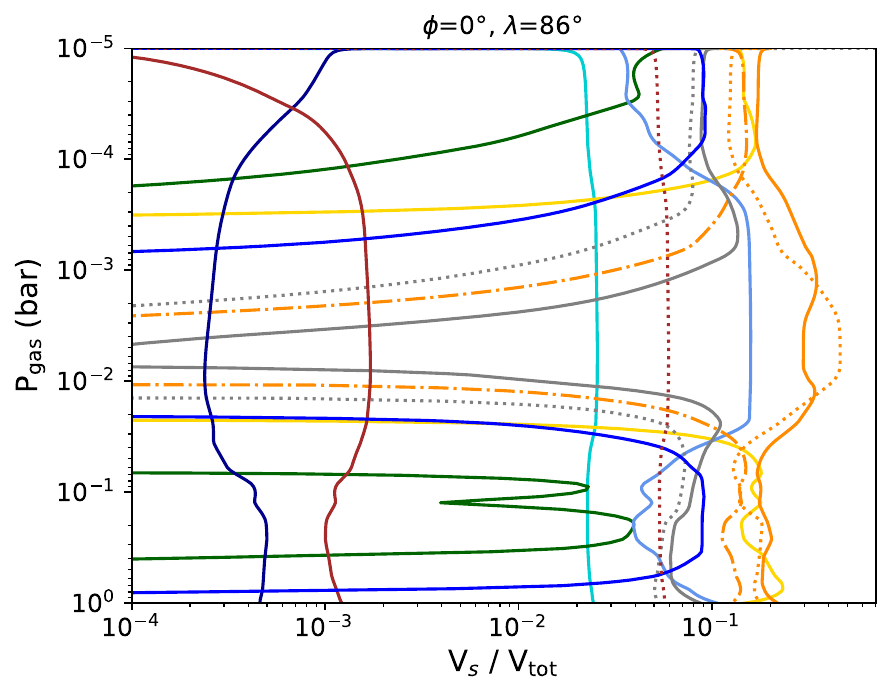}
   \includegraphics[width=0.49\textwidth]{Figures/PolHEx/cloud_files_overleaf/cloud_legend_only.pdf}
    \caption{Material volume fractions V$_{s}$~/~V$_{\rm tot}$ of the different materials forming the mixed-material cloud particles for WASP-43~b at the longitude, latitude specified in each panel. 
    %{\bf Cut plots at $10^{-1}$ bars}
    }\label{fig:cloud_W43b}
\end{figure*}

\begin{figure*}
	\centering    
\includegraphics[width=0.45\textwidth]{Figures/PolHEx/cloud_files_overleaf/cloudstructure_mixingratios_out3_dust_HD209458b_scaledmixing_all_species_log_A_colours.pdf}
\includegraphics[width=0.45\textwidth]{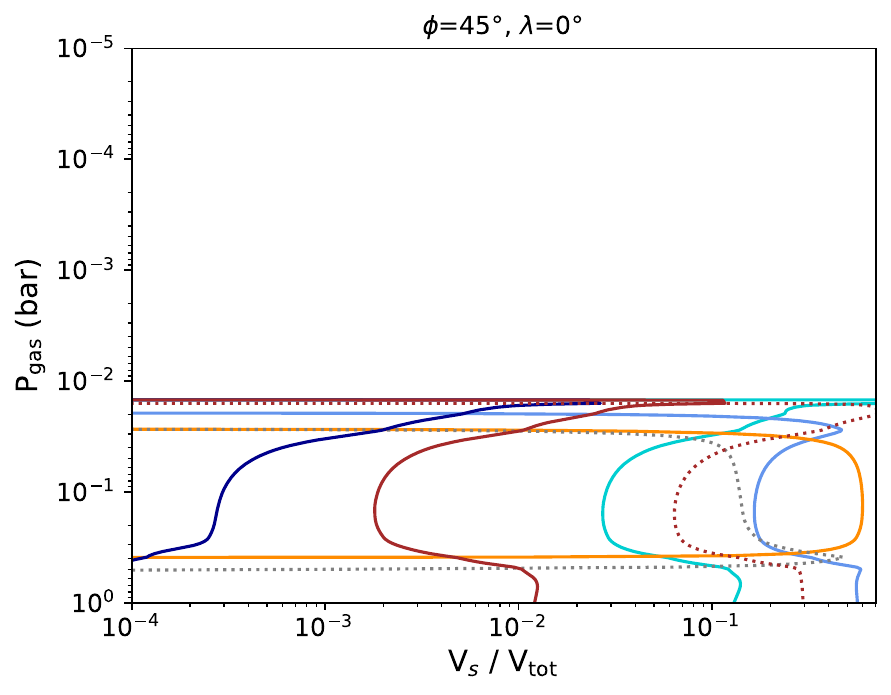}
\includegraphics[width=0.45\textwidth]{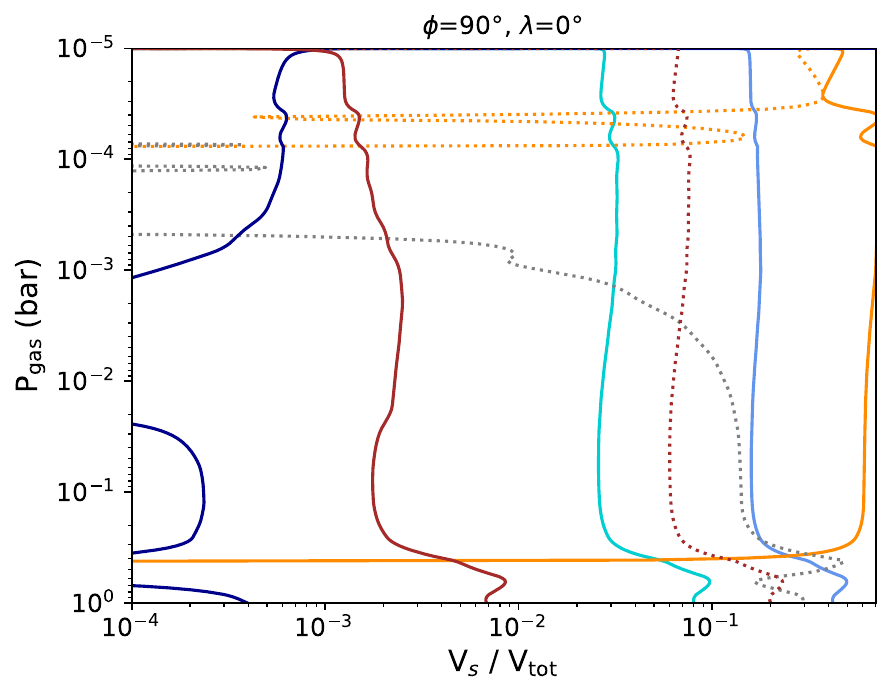}
\includegraphics[width=0.45\textwidth]{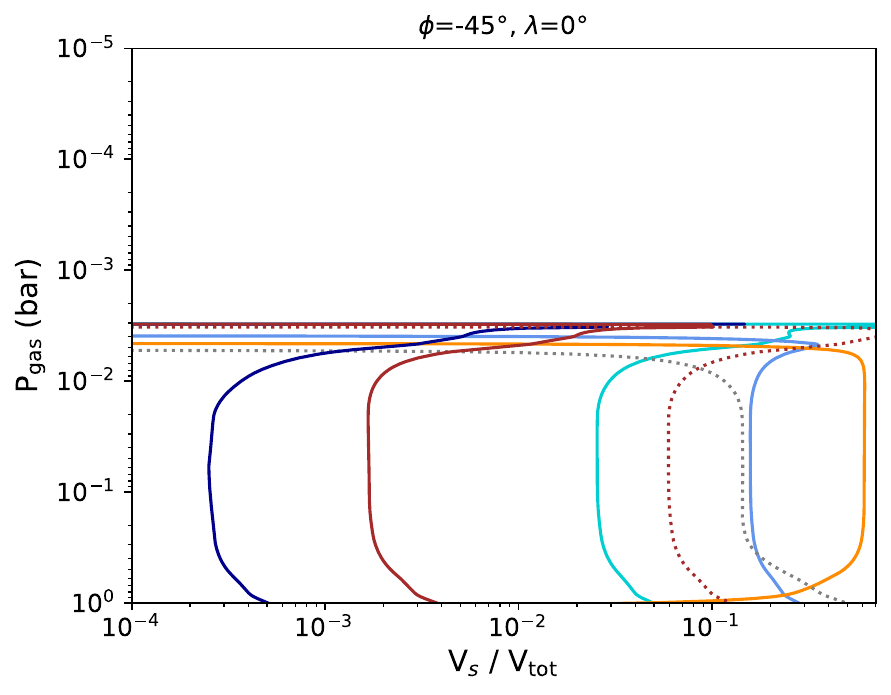}
\includegraphics[width=0.45\textwidth]{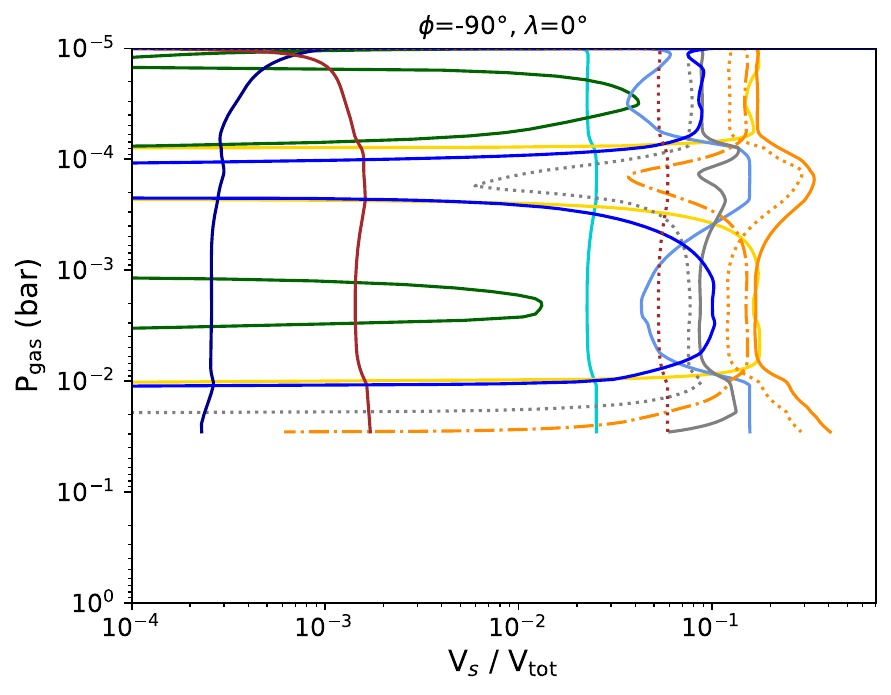}
\includegraphics[width=0.45\textwidth]{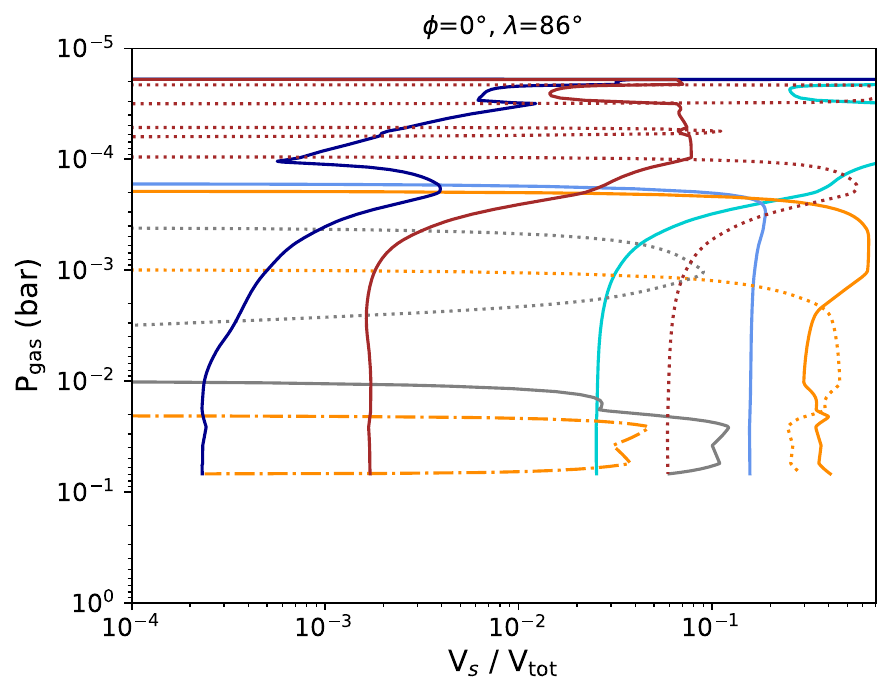}
       \includegraphics[width=0.49\textwidth]{Figures/PolHEx/cloud_files_overleaf/cloud_legend_only.pdf}
    \caption{Material volume fractions V$_{s}$~/~V$_{\rm tot}$ of the different materials forming the mixed-material cloud particles for HD~209458~b at the longitude, latitude specified in each panel. }\label{fig:cloud_HD209}
\end{figure*}

\begin{figure*}
	\centering    
\includegraphics[width=0.6\textwidth]{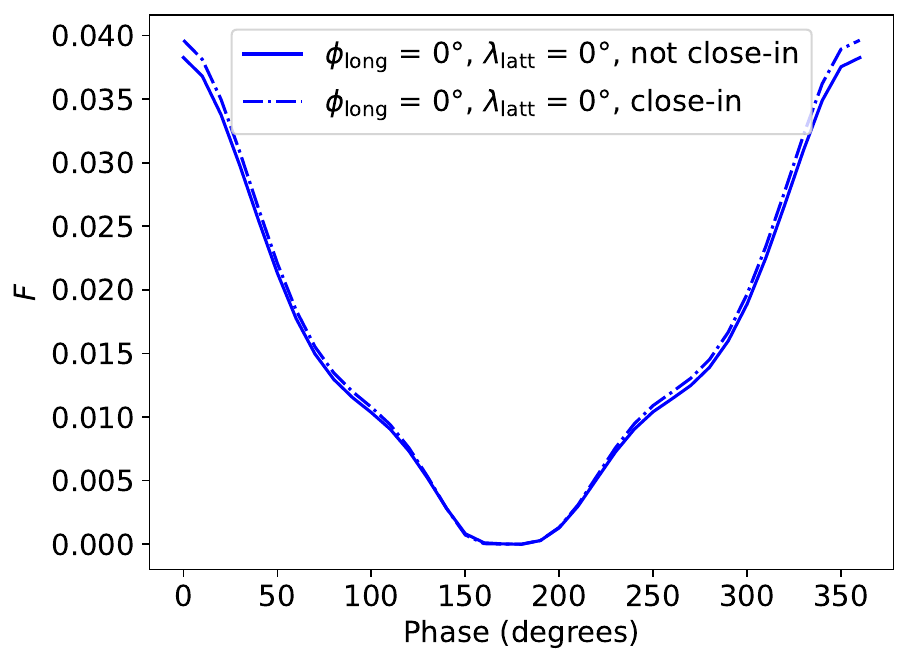}
    \caption{A comparison of before the amendments for close-in exoplanets were made (solid line) to after (dash-dotted line) for a horizontally homogeneous atmosphere of WASP-43~b modelled using the atmosphere at the sub-solar point. 
    %Although the geometric albedo $A_g$ is defined at a phase angle of 0$^\circ$, we show the whole phase curve here, at $\lambda$~=~0.35~$\mu$m.
    } \label{fig:closein}
\end{figure*}

\begin{figure*}
    \centering
    \includegraphics{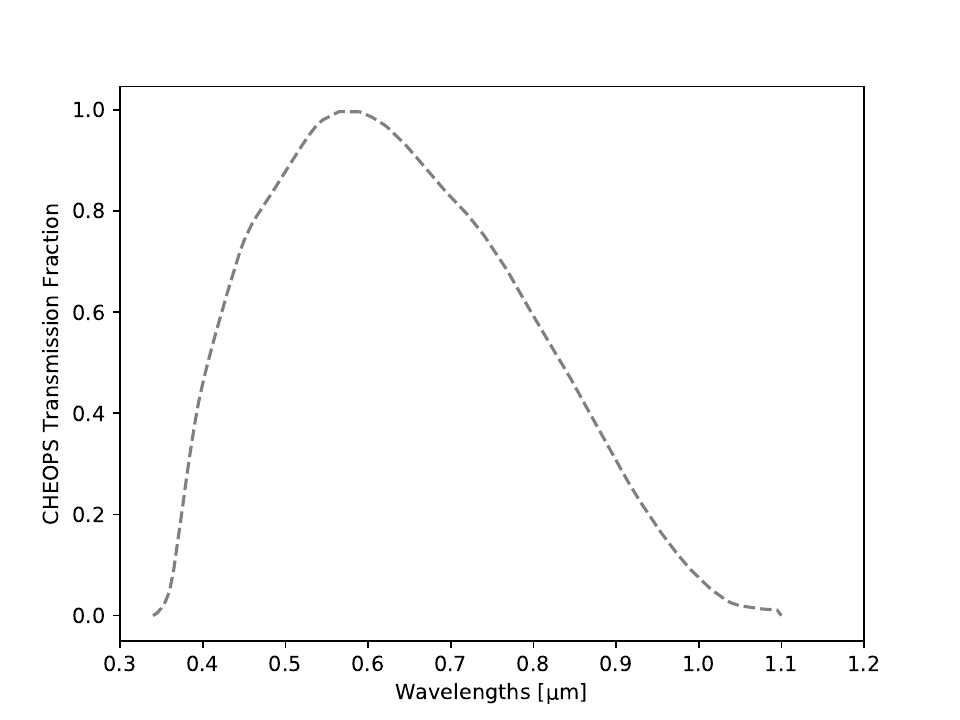}
    \caption{Transmission fraction for the CHEOPS bandpass taken from 
    %\url{http://svo2.cab.inta-csic.es/theory/fps/index.php?id=CHEOPS/CHEOPS.band\&\&mode=browse\&gname=CHEOPS&gname2=CHEOPS\#filter}
    \url{http://svo2.cab.inta-csic.es/theory/fps/}, taking into account both filter throughput and CCD sensitivity.}
    \label{fig:CHEOPS_SVO_bandpass}
\end{figure*}

\begin{figure*}
    \centering    
    \includegraphics[width=\textwidth]{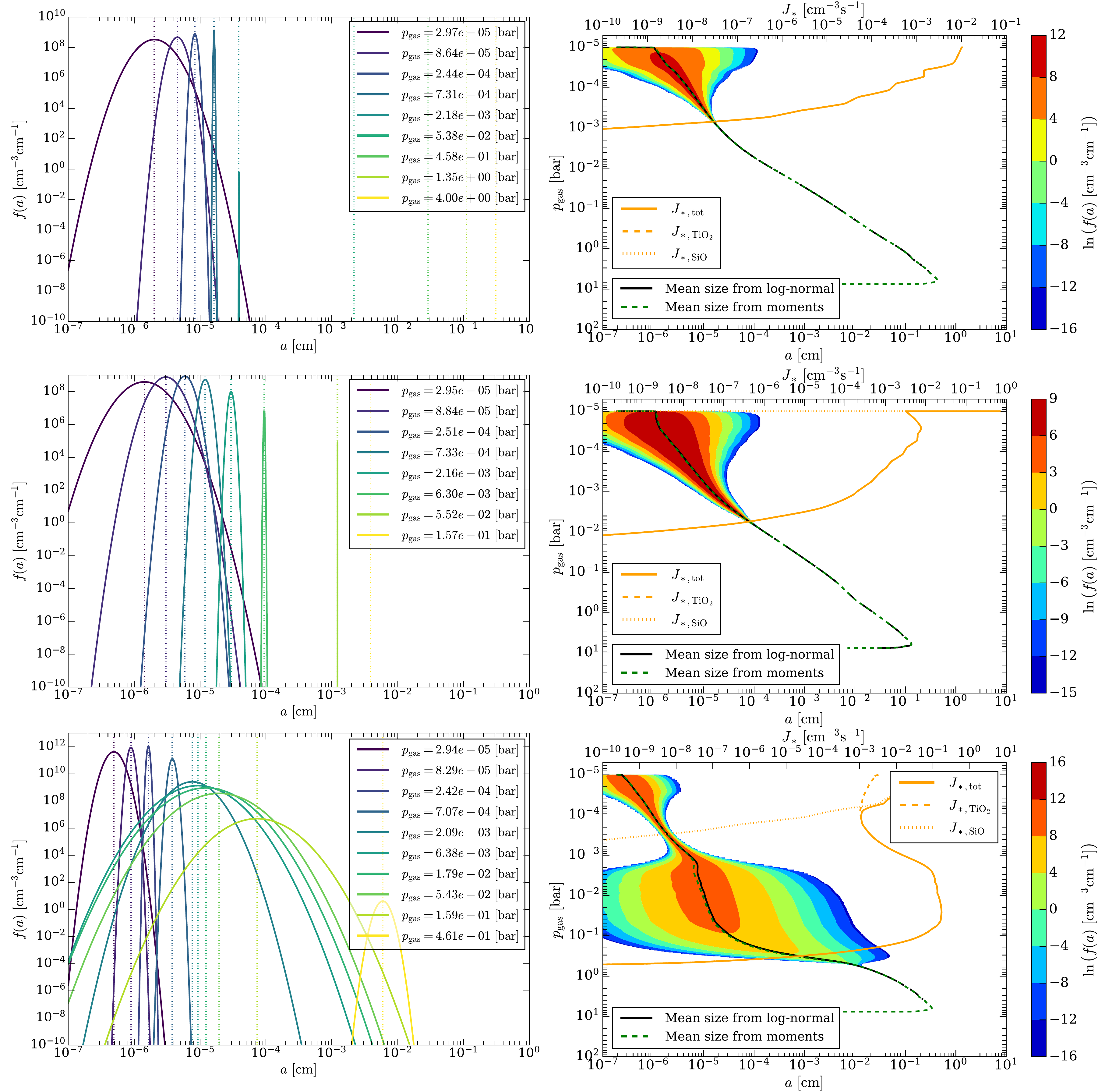}
    \caption{Log-normal distribution for WASP~43b for points at the equator ($\lambda = 0\degree$) from top to bottom these show the substellar point, evening terminator, and the morning terminator. % and the antistellar point, 
    \textbf{Left:} The derived log-normal size distribution function for individual pressure levels through the atmosphere, dotted lines indicate the average particle size, in the case that the distribution becomes monodisperse (i.e. that no valid solution for the log-normal parameters exists), only the average particle size is shown. \textbf{Right:} Cloud Particle size distribution as contours against pressure. The average particle size from the log-normal is shown as a black line. The average particle size derived from the moments ($\langle a \rangle = (3/4\pi)^{1/3}L_1/L0$) is shown in dashed green, showing good agreement at all pressures. Also over-plotted in orange are the nucleation rates, indicating that the log-normal size distribution broadens in regions with efficient nucleation, in order to capture the presence of the population of small cloud particles resulting `recently nucleated' particles that have not yet undergone substantial particle growth.}
\end{figure*}

\begin{figure*}
    \centering    
        \includegraphics[width=\textwidth]{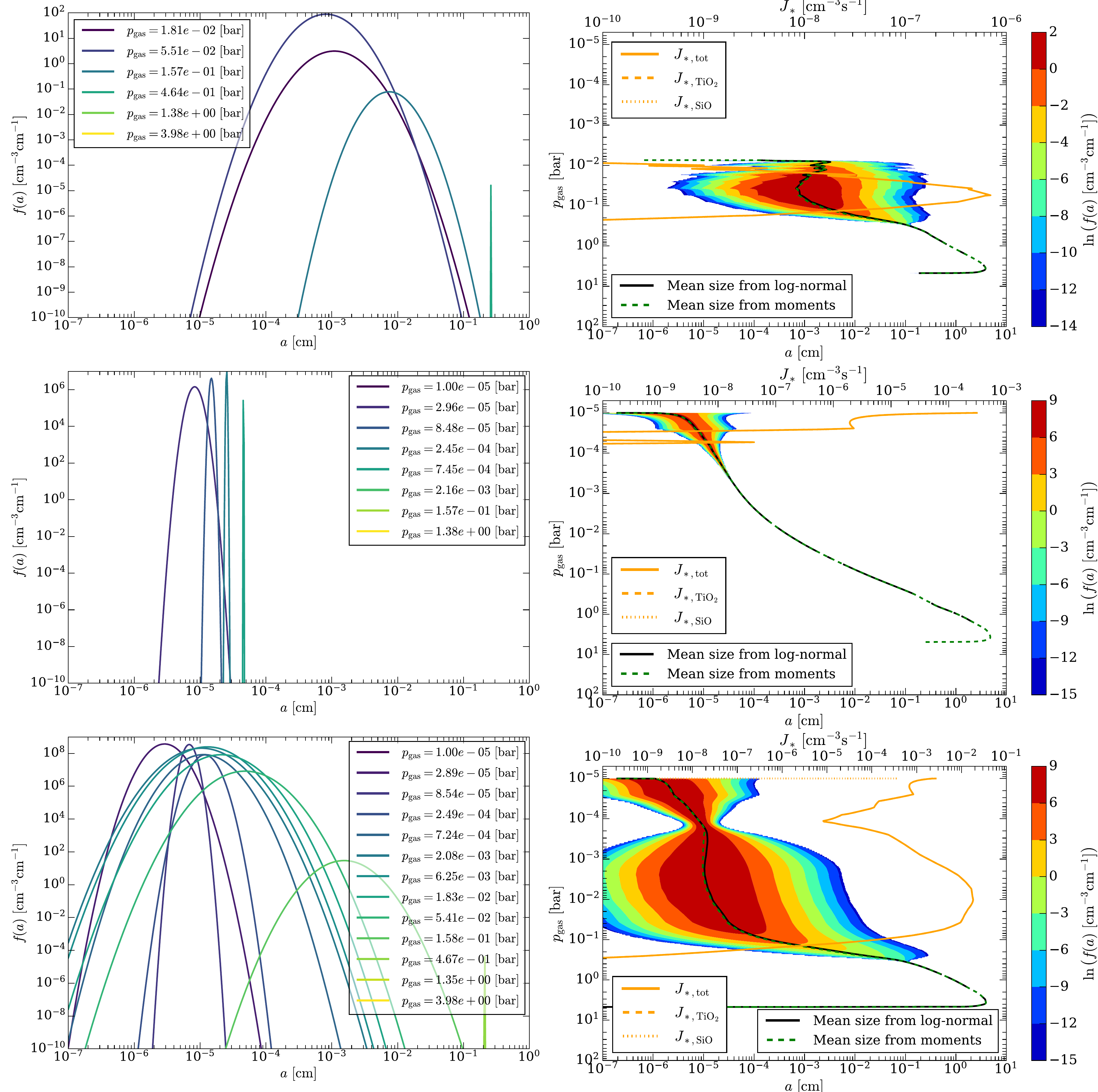}
    \caption{Log-normal distribution for HD209458~b for points at the equator ($\lambda = 0\degree$) from top to bottom these show the substellar point, evening terminator, and the morning terminator. % and the antistellar point, 
    \textbf{Left:} The derived log-normal size distribution function for individual pressure levels through the atmosphere, dotted lines indicate the average particle size. \textbf{Right:} Cloud Particle size distribution as contours against pressure. The average particle size from the log-normal is shown as a black line. The average particle size derived from the moments is shown in dashed green. Also over-plotted in orange are the nucleation rates.}
\end{figure*}

		% Don't change these lines
		\bsp	% typesetting comment
		\label{lastpage}
	\end{document}